\documentclass[prodmode]{acmsmall} 

\usepackage[ruled]{algorithm2e}

\SetAlFnt{\small}
\SetAlCapFnt{\small}
\SetAlCapNameFnt{\small}
\SetAlCapHSkip{0pt}
\IncMargin{-\parindent}

\usepackage[T1]{fontenc}
\usepackage{amsmath}
\usepackage{tikz}
\usetikzlibrary{automata}
\usetikzlibrary{chains}
\usetikzlibrary{arrows}
\usetikzlibrary{calc}
\usepackage{url}
\usepackage{amsmath} 
\usepackage{subfigure} 
\usepackage{upquote}
\usepackage{multirow}

\usepackage{fancyvrb}
\DefineVerbatimEnvironment{verbatim}{Verbatim}{xleftmargin=-1cm}

\newcommand{\np}{{\sc NP}\xspace}
\newcommand{\exptime}{{\sc EXPTIME}\xspace}
\newcommand{\pspace}{{\sc PSPACE}\xspace}

\newcommand{\uri}[1]{\text{\scriptsize {\tt\vphantom{yI}#1}}}
\newcommand{\urit}[1]{\text{{\tt\vphantom{yI}#1}}}

\newcommand{\urie}[1]{\uri{#1\,=\,}}

\newcommand{\ttt}[1]{{\texttt{#1}}} 

\newcommand{\uriq}[1]{\uri{#1}}
\newcommand{\uriqt}[1]{\urit{#1}}

\renewcommand{\arraystretch}{0.97}

\newcommand{\al}[2]{
\addtolength{\jot}{-1ex} 
\begin{minipage}{#1} 
\vspace{1pt}
\begin{center}
$\begin{aligned}
#2
\end{aligned}$
\end{center}
\end{minipage}
\addtolength{\jot}{1em} 
}

\newcommand{\alt}[2]{
\addtolength{\jot}{-1ex} 
\begin{minipage}{#1} 
\vspace{2pt}
\begin{center}
$\begin{aligned}
#2
\end{aligned}$
\end{center}
\end{minipage}
}

\usetikzlibrary{arrows,positioning,backgrounds} 
\tikzset{
    rt/.style={
		rectangle,
		fill = white,
		draw=black, 
		text centered,
		inner sep=0.5ex
		},
    rtt/.style={ 
    	rt,
    	inner sep=0.1ex
    	},
    ert/.style={ 
     	rt,
     	dashed
     	}, 
    ertt/.style={ 
        rtt,
        dashed
        }, 
    rect/.style={ 
        rectangle,
        fill = white,
        rounded corners,
        draw=black, 
        text centered,
        inner sep=0.8ex
        },
    rectw/.style={
        rect,
        draw=white
        },
    erect/.style={ 
    	rect,
    	dashed
    	},
    erectw/.style={ 
     	rectw,
     	dashed
     	},
    arrout/.style={
           ->,
           -latex,
           },
    arrin/.style={
           <-,
           latex-,
           },
    arrb/.style={
           <->,
           >=latex,
           }
}

\doi{0000001.0000001}

\issn{1234-56789}

\begin{document}

\markboth{R. Angles et al.}{Foundations of Modern Graph Query Languages}

\title{Foundations of Modern Query Languages for Graph Databases\footnote{Work funded 
by the Millennium Nucleus Center for Semantic Web Research under Grant NC120004.}}
\author{
RENZO ANGLES
\affil{Universidad de Talca \& Center for Semantic Web Research}
MARCELO ARENAS
\affil{Pontificia Universidad Cat\'olica de Chile \& Center for Semantic Web Research}
PABLO BARCEL\'O
\affil{DCC, Universidad de Chile \& Center for Semantic Web Research}
AIDAN HOGAN
\affil{DCC, Universidad de Chile \& Center for Semantic Web Research}
JUAN REUTTER
\affil{Pontificia Universidad Cat\'olica de Chile \& Center for Semantic Web Research}
DOMAGOJ VRGO\v{C}
\affil{Pontificia Universidad Cat\'olica de Chile \& Center for Semantic Web Research}}

\begin{abstract}
We survey foundational features underlying modern graph query languages. We first discuss two popular graph data models: {\em edge-labelled graphs}, 
where nodes are connected by directed, labelled edges; and {\em property graphs}, 
where nodes and edges can further have attributes. Next we discuss the two most fundamental graph querying functionalities: 
{\em graph patterns} and {\em navigational expressions}. We start with graph patterns, in which a graph-structured query is matched against the data. Thereafter we discuss navigational expressions, in which patterns can be matched recursively against the graph to navigate paths of arbitrary length; we give an overview of what kinds of expressions have been proposed, and how they can be combined with graph patterns. We also discuss several semantics under which queries using the previous features can be evaluated, what effects the selection of features and semantics has on complexity, and offer examples of such features in three modern languages that are used to query graphs: SPARQL, Cypher and Gremlin. We conclude by discussing the importance of formalisation for graph query languages; a summary of what is known about SPARQL, Cypher and Gremlin in terms of expressivity and complexity; and an outline of possible future directions for the area.
\end{abstract}

%
%

\ccsdesc[500]{Information systems~Query languages}
\ccsdesc[500]{Theory of computation~Database query languages (principles)}

\newcommand{\bL}{\text{\it Lab}}
\newcommand{\bP}{\text{\it Prop}}
\newcommand{\bV}{\text{\it Val}}
\newcommand{\bVr}{\text{\it Var}}
\newcommand{\bC}{\text{\it Const}}

%
%


\keywords{Property graphs, graph databases, query languages, graph patterns, navigation, 
aggregation}

\acmformat{}


\maketitle

\section{Introduction}\label{sec:intro}

The last decade has seen a resurgence in interest in \textit{graph databases}, wherein entities from the domain of interest are represented by \textit{nodes} and relationships between them by \textit{edges}. Part of this resurgence stems from the growing realisation that there are a variety of domains for which graph databases offer a more intuitive conceptualisation than their more well-established relational cousins. For example, one can view a social network as a graph of people who know each other. One may likewise view transport networks, biological pathways, citation networks, and so on, as a graph. 
%
%
Although graphs can still be (and sometimes still are) stored in relational databases, the choice to use a graph database for certain domains has significant benefits in terms of querying, where the emphasis shifts from joining various tables to specifying graph patterns and navigational patterns between nodes that may span arbitrary-length paths.
To support these new types of queries, a variety of graph database engines~\cite{virtuoso,bigdata,cypher}, graph data models~\cite{sparql11,cypher} and graph query languages~\cite{cypher,sparql11,tinkerpop} have been released over the past few years. 


\paragraph{Scope} Our goal in this survey is to give an in-depth discussion of the main conceptual features found in modern graph query languages, as both supported by graph database engines, and studied in the theoretical literature. By organising our survey at the level of query features, rather than languages, we provide a foundational introduction to the area, which helps to understand, and even define, individual query languages as the composition of features.
We consider two high-level categories of query features: graph patterns and path expressions. 
These features collectively form the core of a variety of modern graph query languages~\cite{cypher,sparql11,pgql}, and form the core of what has been studied in the theoretical literature~\cite{Woo,B13}.

After introducing and defining the graph query features in each category, 
we list various semantics under which such features can be evaluated, 
provide examples of how such features are applied in a selection of modern query languages, 
discuss the computational complexity of key problems underlying such features, and 
present some of their most important extensions as implemented in modern graph database engines.  

We wrap up by drawing all of the foundational discussion together into a summary of the types of features we have covered, how these features can be used to understand the complexity and expressivity of modern query languages, the importance of formalisation for such languages, the key challenges underlying their implementation and optimisation in practical engines, and possible ways in which they might evolve. 

\paragraph{Survey structure} The survey is structured as follows:

\begin{itemize}
\item We first discuss two graph \textit{data models} in Section~\ref{sec:prelim}: \textit{edge-labelled graphs}, which is the foundational model considered in the graph database literature; and \textit{property graphs}, which is a model commonly employed in practice, where nodes and edges in labelled graphs can be annotated with additional meta-information. 

\item In Section~\ref{gp}, we discuss \textit{graph patterns}, where a graph-structured query is matched against the graph database. We also discuss the extension of such graph patterns with additional operators, such as projection, difference, union, etc. 
\item Section~\ref{sec:nav} then introduces \textit{navigational expressions}, which, unlike graph patterns, can match paths of arbitrary length. We study different types of expressions, including path expressions, expressions that additionally allow checking branches from a path, and expressions that are based on recursively matching graph patterns. 
\item In Section \ref{sec:final} we present our final remarks. 
\end{itemize}

\paragraph{Online Appendix} An online appendix for this paper discusses additional features that can be found in graph query languages -- or have been proposed for such languages -- including \textit{aggregation}, where results can be grouped, counted, etc.; \textit{path unwinding}, where elements can be projected from paths to be further processed by the query; \textit{graph-to-graph} queries, where the evaluation of a query over a graphs can (recursively) form new graphs; as well as further extensions that can be considered. 

\paragraph{Proviso} Throughout the survey, following the conventions of theoretical papers, we will use the phrase ``graph database'' to refer to a specific data model or an instance of that data model. We will use the phrase ``graph database engine'' to specify an implementation for executing queries over graph databases.

\paragraph{Intended audience} The main ambition of this survey is to bridge theory and practice, relating theoretical notions of querying graphs to three modern query languages that are popular in practice. The survey is thus primarily aimed at both theoretical and applied researchers interested in graph databases. For a theory-oriented researcher, the survey outlines a practical context for proposals in the theoretical literature, providing concrete examples of how practical languages instantiate or relate to theoretical proposals, discussing choices of semantics, highlighting aspects of such languages not yet well understood in theory, etc. For a practice-oriented researcher, the survey shows how the core of various graph query languages can be understood and compared from a more foundational perspective, the possible semantics that can be chosen, the effects on complexity and the practicality of a language by changing certain features and/or semantics, etc. Aside from researchers, practitioners -- i.e., developers, database administrators, engineers, consultants -- may also be interested in this survey, particularly those involved in the development of graph database engines or query languages.

To keep the paper accessible to a broad audience, we keep formal definitions only for core notions where it is important to be precise. Throughout the survey, we provide a wide variety of examples, including examples in three concrete query languages: SPARQL, Cypher and Gremlin.

\paragraph{Previous surveys} A number of surveys have been published 
in recent years in the area of graph databases. \citeN{AG08} provide a survey of graph database models. More recently,~\citeN{90525} presents a systematic analysis of the functionalities of current graph database engines. Neither of these surveys covers querying graph databases in depth, rather focusing on models and engines.

\citeN{Woo} and \citeN{B13} study several graph query languages from a theoretical point of view, focusing on their expressive power and the computational complexity of associated problems. Given the theoretical focus of both surveys, neither covers practical aspects of modern graph query languages in detail.

Particular aspects of graph querying have also been surveyed; 
for example, works by \citeN{Bunke00}, \citeN{G06}, \citeN{RJB10}, \citeN{LR13}, and \citeN{YYLDZY16} deal with 
particular aspects of graph pattern matching, while \citeN{YC10} 
concentrate on graph reachability queries. Again, however, all such works have a narrower focus than our survey.

Our survey complements these previous works in two novel aspects:

\begin{enumerate}
\item Instead of surveying the myriad of different graph data models available, 
we build our presentation in terms of two popular such data models; namely, edge-labelled and property graphs. In spite of their simplicity, these models are flexible enough to express most practical graph database scenarios. In addition, the most fundamental issues related to querying graphs are already present for these models. 
\item Though we discuss semantics and complexity, we do not focus only on the theoretical aspects of graph query languages. Instead, we identify and explain in detail the basic features that appear in such languages, providing examples of how they are applied in a selection of practical query languages. In summary, our paper bridges the theory and practice of graph query languages in a novel manner; as previously discussed, our survey thus targets a broader audience than previous works.
\end{enumerate} 

Specific novel aspects of this survey include a new formalisation of the property graph model; discussion of how this model can be understood through the lens of existing theory; comparisons of practical aspects of the SPARQL, Gremlin and Cypher query languages and the semantics they adopt; and examples of how the design of such languages influences the complexity of query evaluation.
\section{Graph data models}\label{sec:prelim}



Graphs can be used to encode data whereby nodes represent objects in a domain of interest, and edges represent relationships between these objects. For instance, if a graph is used to encode data about movies, nodes may be actors and movies, and a (directed) edge from a node $a$ to a node $b$ may indicate that $a$ is an actor in $b$. Note that the direction of an edge matters here: we want to say that an actor stars in a movie, but not vice-versa. A movie database can then be modelled using graphs as follows:
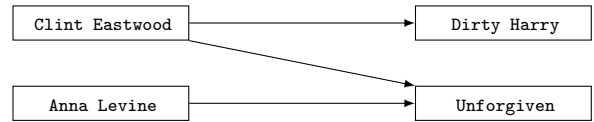
\begin{center}
\begin{tikzpicture}

\node[rtt] (n1) {
\al{60pt}{ 
\uriq{Clint Eastwood}}};

\node[rtt, right=3cm of n1] (n2) {
\al{60pt}{ 
\uriq{Dirty Harry}}};

\draw[arrout] (n1) to  (n2);

\node[rtt, below=0.6cm of n1] (n3) {
\al{60pt}{ 
\uriq{Anna Levine}}};

\node[rtt, right=3cm of n3] (n4) {
\al{60pt}{ 
\uriq{Unforgiven}}};

\draw[arrout] (n1) to  (n4);

\draw[arrout] (n3) to  (n4);

\end{tikzpicture}
\end{center}
However, it is difficult to express different types of relationships in such a simple form of graph. For instance, suppose that we wish to encode that \uriqt{Clint Eastwood} is also the director of  \uriqt{Unforgiven}. We could consider adding an edge between these nodes, thus ending up with two nodes connected in the following way:
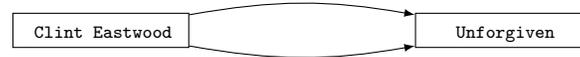
\begin{center}
\begin{tikzpicture}

\node[rtt] (n1) {
\al{60pt}{ 
\uriq{Clint Eastwood}}};

\node[rtt, right=3cm of n1] (n2) {
\al{60pt}{ 
\uriq{Unforgiven}}};

\draw[arrout, bend left=10] (n1) to  (n2);

\draw[arrout, bend right=10] (n1) to  (n2);

\end{tikzpicture}
\end{center}

But which edge here represents the fact that \uriqt{Clint Eastwood} is the director of \uriqt{Unforgiven}? And, more generally, if we have many different types of relationships between nodes, how can we distinguish between them? 

\medskip
\noindent \paragraph{{\em {\bf Edge-labelled graphs}}}

\begin{figure}[h]
\begin{center}
\begin{tikzpicture}

\node[rtt] (n1) {
\al{60pt}{ 
\uriq{Clint Eastwood}}};

\node[rtt, right=3cm of n1] (n2) {
\al{60pt}{\uriq{Unforgiven}}};

\draw[arrout, bend left=10] (n1) to 
node[ertt] (e1) 
{\al{40pt}{
\uri{acts\_in}}}
(n2);

\draw[arrout, bend right=10] (n1) to 
node[ertt] (e2) {
\al{40pt}{
\uri{directs}}}
(n2);

\node[rtt,right=3cm of n2] (n3) {
	\al{60pt}{ 
		\uriq{Anna Levine}}};

\draw[arrout] (n3) to 
node[ertt] (e3) 
{\al{40pt}{
		\uri{acts\_in}}}
(n2);

\end{tikzpicture}
\end{center}
\caption{An edge-labelled graph encoding basic movie information with dashed labels on edges}\label{fig:graphdb}
\end{figure}
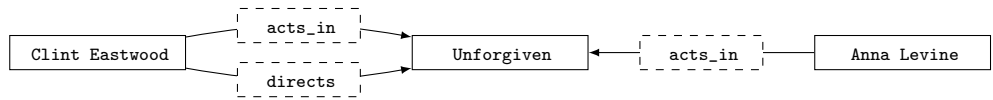

A simple and widely-adopted solution is the use of \textit{edge-labelled graphs}, where we additionally assign {\it labels} to edges that indicate the different types of relationships in the domain being described. We can see an example in Figure~\ref{fig:graphdb} where  \uriqt{Clint Eastwood} has two relations to \uriqt{Unforgiven}: one represented by the edge labelled \urit{acts\_in}, another represented by the edge labelled \urit{directs}, and where \uriqt{Anna Levine} also has an edge labelled \urit{acts\_in} to this movie.

In the following, we formalise the notion of an \textit{edge-labelled graph}.

\begin{definition}[Edge-labelled graph]\label{def-gdb}
	An edge-labelled graph $G$ is a pair $(V,E)$, where:
	\begin{enumerate} 
		\item $V$ is a finite set of {\em vertices} (or {\em nodes}).
		
		\item $E$ is a finite set of {\em edges}; formally, $E \subseteq V \times \bL \times V$ where $\bL$ is a set of \textit{labels}. \qed
	\end{enumerate}  
\end{definition}

\begin{example}
Letting $G=(V,E)$ denote the graph from Figure~\ref{fig:graphdb}, the set of vertices and edges, respectively, are: 

\begin{center}
	\small
\begin{tabular}{rcl@{\,}l}
$V$ & $=$ & $\{$ & $\uriqt{Clint Eastwood}, \uriqt{Unforgiven}, \uriqt{Anna Levine}\,\}$	\\[1ex]
$E$ & $=$ & $\{$ & $(\uriqt{Clint Eastwood}, \urit{acts\_in}, \uriqt{Unforgiven}),$ \\
 & & & $(\uriqt{Clint Eastwood}, \urit{directs}, \uriqt{Unforgiven}),$ \\
 & & & $(\uriqt{Anna Levine}, \urit{acts\_in}, \uriqt{Unforgiven})\,\,\}$ \\	
\end{tabular}
\end{center}	

\noindent
The labels \urit{acts\_in} and \urit{directs} are taken from the set $\bL$. \qed
\end{example}

Edge-labelled graphs are widely adopted in practice where, for example, 
they form the basis of the Resource Description Framework (RDF) 
standard used for encoding machine-readable content on the Web \cite{RDF11}. 
An \textit{RDF graph} is simply a set of triples analogous to the edges in a 
graph database, but with some further detailing: in the case of RDF, the set $V$ 
can be partitioned into disjoint sets of IRIs, literals and blank nodes, and the set 
$\bL$ is a subset of IRIs (not necessarily disjoint from $V$). But for our purposes, 
we require no special consideration on the types of nodes,\footnote{We will 
not consider the existential semantics of blank nodes nor the interpretation of datatype values nor other special vocabularies. 
These issues are orthogonal to our goal of introducing query features for graphs.} and for simplicity, we consider an RDF graph as simply a special type of edge-labelled graph.

Note that the definition of an edge-labelled graph does not impose any particular restriction on the topology of graphs. For example, although Figure \ref{fig:graphdb} does not contain a cycle, one can be obtained if we also add an edge labelled \texttt{directedBy} between Unforgiven and Clint Eastwood, signifying that the movie was directed by Clint Eastwood. For more involved cycles please refer to our social network example in Figure \ref{fig:sn-sample} below.

%
%
%
%
%
%
%
%
%

\begin{figure}  
	\begin{center}
		\begin{tikzpicture}
		
		\node[rtt] (n1) {
			\al{60pt}{ 
				\uriq{Peter Sellers}}};
		
		\node[rtt,above=0.9cm of n1,xshift=3.5cm] (n2) {
			\al{65pt}{ 
				\uriq{Lionel Mandrake}}};	
		
		\draw[arrout] (n1) to 
		node[ertt] (e1) 
		{\al{40pt}{
				\uri{plays}}}
		(n2);
		
		\node[rtt,below=0.9cm of n1,xshift=3.5cm] (n3) {
			\al{65pt}{ 
				\uriq{Merkin Muffley}}};	
		
		\draw[arrout] (n1) to 
		node[ertt] (e2) 
		{\al{40pt}{
				\uri{plays}}}
		(n3);
		
		\node[rtt,right=5cm of n1] (n4) {
			\al{60pt}{ 
				\uriq{Dr. Strangelove}}};    
		
		\draw[arrout] (n2) to 
		node[ertt] (e3) 
		{\al{40pt}{
				\uri{movie}}}
		(n4);
		
		\draw[arrout] (n3) to 
		node[ertt] (e4) 
		{\al{40pt}{
				\uri{movie}}}
		(n4);
		
		\node[rtt,right=4cm of n2] (n5) {
			\al{60pt}{ 
				\uriq{18 minutes}}};
		
		\node[rtt,right=4cm of n3] (n6) {
			\al{60pt}{ 
				\uriq{34 minutes}}};
		
		\draw[arrout] (n2) to 
		node[ertt] (e5) 
		{\al{40pt}{
				\uri{screentime}}}
		(n5);
		
		\draw[arrout] (n3) to 
		node[ertt] (e6) 
		{\al{40pt}{
				\uri{screentime}}}
		(n6); 
		\end{tikzpicture}
	\end{center}
	\caption{An edge-labelled graph encoding information about actors that have acted in movies under different roles}\label{fig:reified}
\end{figure}
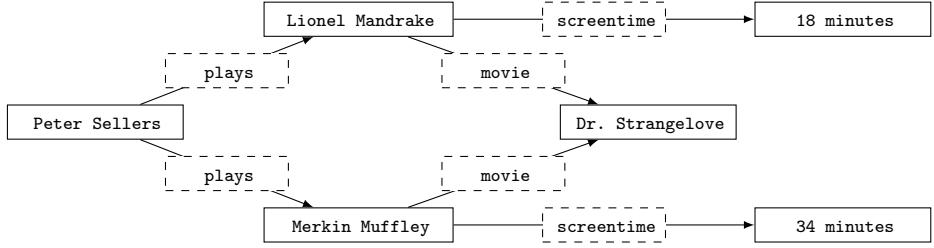

Although edge-labelled graphs have a simple structure, they can encode complex information. 
For example, when describing certain movies in a graph database, we may wish to encode that an actor has acted 
multiple times in the same movie under different roles. At first, this may seem incompatible with our definition of a graph database $G = (V,E)$ since 
$E$ is defined as a \textit{set} of edges: we cannot have multiple edges with the same label between the same two nodes. 
However, with some lateral thinking, we can model such information as an edge-labelled graph, as per Figure \ref{fig:reified}. 
Here we see that by using a node (rather than an edge) to represent each role played by the actor in the movie, we can not only encode cases where an actor plays multiple roles in a movie, but we can also encode additional information about the role, in this case the total on-screen time for the character in question. With this principle of using nodes to represent \textit{$n$-ary relations} (where $n > 2$), it becomes feasible to encode increasingly more complex information in an edge-labelled graph, such as, for example, to encode that the same character can be portrayed by different actors, and so forth. 


\medskip
\noindent \paragraph{{\em {\bf Property graphs}}} In edge-labelled graphs, we use labels to indicate the type of edge, where multiple edges may have the same type. In a similar way, we could consider labelling nodes.\footnote{Such graphs are often called \emph{heterogeneous information networks}, see e.g. \cite{SHYYW11}.} For example, in the movie graph of Figure~\ref{fig:graphdb}, we could label the nodes \texttt{Clint Eastwood} and \texttt{Anna Levine} as \texttt{Person}, and the node \texttt{Unforgiven} as \texttt{Movie}; we may even add multiple labels to a node, for example to label \texttt{Clint Eastwood} as \texttt{Director} and \texttt{Actor}. While this information can be represented in edge-labelled graphs -- for example, as done in RDF, a new node is created for \texttt{Movie} with an edge labelled \texttt{type} extended to it from \texttt{Unforgiven} -- having node labels as part of the model can offer a more direct abstraction that is easier for users to query and understand.

In the same way, it is often cumbersome to add information about the edges to an edge-labelled graph.
For example, let's say that to Figure \ref{fig:graphdb}, we wished to add the source of information, e.g., that the \urit{acts\_in} relations were sourced from the web-site IMDb; for this, we cannot simply add edges to the graph. Instead, we would need to start again from the graph in Figure \ref{fig:graphdb}, and create a new type of $n$-ary relation with the information we need: the facts in the \urit{acts\_in} relation together 
with their source of information.\footnote{A more generic technique involves applying ``reification'' where edges are represented as nodes (for a more detailed discussion see \cite{HHK15}).} 
Adding new types of information to edges in an edge-labelled graph may thus require a major change to the graph's structure, entailing a significant cost.

Thus, for scenarios where various new types of meta-information may regularly 
need to be added to edges 
or nodes,
the most general and 
widely adopted alternative is to use an extension of an edge-labelled graph called a \textit{property graph}. 
This model is currently adopted by some major graph database engines, such as Neo4j \cite{70101}, 
and has been recently standardised by a working group of the Linked Data Benchmark Council (LDBC) formed by 
members of academia and industry \cite{ldbc}. 

In property graphs, both edges \textit{and} nodes can be labelled. Each edge and node is additionally associated with a unique identifier that can be used as a ``hook'' to associate additional meta-information -- in the form of a set of property--value pairs called \textit{attributes} -- directly to that edge or node. Again, while it would be possible to instead encode attributes and labels as additional edges, in practice, such features allow one to directly annotate the graph without modifying its overall structure.

\begin{figure}[t]
\begin{center}
\resizebox{\textwidth}{!}{
\begin{tikzpicture}
\node[rect] (n1) {
\alt{84pt}{ 
 \urie{name} & \uriq{Clint Eastwood}\\
\urie{gender} & \uriq{male}}};

\node[rt] (ln1) at (n1.north) {\uri{$n_1$ $:$ Person}};

\node[rect, right=2.6cm of n1] (n2) {
\alt{66pt}{ 
\urie{title} & \uriq{Unforgiven}}};

\node[rt] (ln2) at (n2.north) {\uri{$n_2$ $:$ Movie}};

\draw[arrout, bend left=10] (n1) to 
node[above, erect] (e1) {
\alt{50pt}{
\urie{role} & \uriq{Bill}\\
\urie{ref} & \uriq{IMDb}}}
(n2);

\node[ert] (le1) at (e1.north) {\uri{$e_1$ $:$ acts\_in}};

\draw[arrout, bend right=10] (n1) to 
node[below, ert] (e2) {\uri{$e_2$ $:$ directs}}
(n2);

\node[rect, right=2.8cm of n2] (n3) {
\alt{70pt}{
\urie{name} & \uriq{Anna Levine}\\
\urie{gender} & \uriq{female}}};

\node[rt] (ln3) at (n3.north) {\uri{$n_3$ $:$ Person}};

\draw[arrin] (n2) to
node[above, erect] (e3) 
{
\alt{50pt}{
\urie{role} & \uriq{Delilah}\\
\urie{ref} & \uriq{IMDb}}}
(n3);

\node[ert] (le3) at (e3.north) {\uri{$e_3$ $:$ acts\_in}};
\end{tikzpicture}
}
\end{center}
\caption{A property graph with attribute values storing information about movies. \label{fig-movies}}
\end{figure}
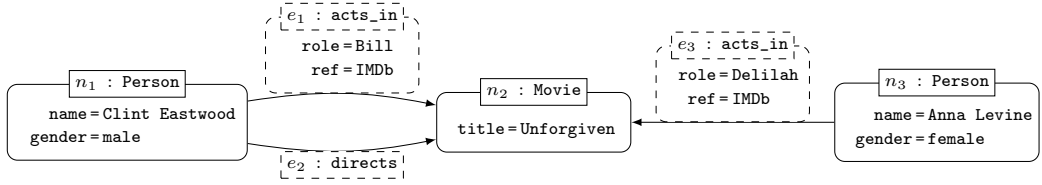

For example, in Figure \ref{fig-movies} we show a graph for our movie database that includes labels and attributes on nodes and edges. In this figure, the attributes for a node are shown in the round rectangle below it. Thus, for example, the attributes associated to the node with identifier $n_1$ are {\it name} and {\it gender}, and their values are \uriqt{Clint Eastwood} and \urit{male}, respectively. On the other hand, the edge with identifier $e_2$ does not have any attribute. In this model, we can directly encode multiple edges (having different identifiers) with the same label between the same two nodes, and can extend the graph with additional attributes on edges without having to remodel complex relations as nodes.

We now provide a formal definition of the notion of a property graph.

\begin{definition}[Property graph]\label{def-pg}
A property graph $G$ is a tuple $(V,E,\rho,\lambda,\sigma)$, where: 
\begin{enumerate}
	\item $V$ is a finite set of {\em vertices} (or {\em nodes}).
		
	\item $E$ is a finite set of {\em edges} such that $V$ and $E$ have no elements in common.
		
	\item $\rho : E \to (V \times V)$ is a total function. Intuitively, $\rho(e) = (v_1,v_2)$ indicates that $e$ is a directed edge {\em from} node  $v_1$ {\em to} node $v_2$ in $G$.   
		
	\item $\lambda : (V \cup E) \to \bL$ is a total function with $\bL$ a  set of labels. Intuitively, if $v \in V$ (resp., $e \in E$) and $\rho(v) = \ell$ (resp., $\rho(e) = \ell$) , then $\ell$ is the label of node $v$ (resp., edge $e$) in $G$. 
	
	\item $\sigma : (V \cup E) \times \bP \to \bV$ is a partial function with $\bP$ a finite set of properties and $\bV$ a set of values. 
	Intuitively, if $v \in V$ (resp., $e \in E$), $p \in \bP$ and $\sigma(v,p) = s$ (resp., $\sigma(e,p) = s$), then $s$ is the value of property $p$ for node $v$ (resp., edge $e$) in the 
	property graph $G$. \qed
\end{enumerate} 
\end{definition} 

\begin{example}
For the property graph $G$ shown in Figure \ref{fig-movies}, we have that $G = (V,E,\rho,\lambda,\sigma)$, where
$V$, $E$, $\rho$, $\lambda$, and $\sigma$ are as shown in Figure \ref{fig:components}. \qed
\end{example}

\begin{figure} 
	\small
	\begin{center}
		 \begin{tabular}{r@{~}c@{~}l@{~~~~~~~}r@{~}c@{~}l@{~~~~~~~~}r@{~}c@{~}l}
		 	$V$ & $=$ & $\{n_1, n_2, n_3\}$ & $E$ & $=$ & $\{e_1, e_2, e_3\}$ & 
		 	$\sigma(n_1,\urit{name})$ & $=$ & $\uriqt{Clint Eastwood}$\\ 
		 	
		 	 & & & & & & $\sigma(n_1,\urit{gender})$ & $=$ & $\uriqt{male}$\\
		 	 
		 	$\rho(e_1)$ & $=$ & $(n_1,n_2)$ & $\rho(e_2)$ & $=$ & $(n_1,n_2)$ & 
		 	$\sigma(n_2,\urit{title})$ & $=$ & $\uriqt{Unforgiven}$\\
		 	
		 	$\rho(e_3)$ & $=$ & $(n_3,n_2)$ & & & & 
		 	$\sigma(n_3,\urit{name})$ & $=$ & $\uriqt{Anna Levine}$\\
		 	
		 	 & & & & & & $\sigma(n_3,\urit{gender})$ & $=$ & $\uriqt{female}$\\
		 	
		 	$\lambda(n_1)$ & $=$ & $\urit{Person}$ & $\lambda(n_2)$ & $=$ & $\urit{Movie}$ & 
		 	$\sigma(e_1,\urit{role})$ & $=$ & $\uriqt{Bill}$\\
		 	
		 	$\lambda(n_3)$ & $=$ & $\urit{Person}$ & $\lambda(e_1)$ & $=$ & $\urit{acts\_in}$ & 
		 	$\sigma(e_1,\urit{ref})$ & $=$ & $\uriqt{IMDb}$\\
		 	
		 	$\lambda(e_2)$ & $=$ & $\urit{directs}$ & $\lambda(e_3)$ & $=$ & $\urit{acts\_in}$ & 
		 	$\sigma(e_3,\urit{role})$ & $=$ & $\uriqt{Delilah}$\\
		 	
		 	 & & & & & & $\sigma(e_3,\urit{ref})$ & $=$ & $\uriqt{IMDb}$\\
		 \end{tabular}
	\end{center} 
\caption{The components of the graph $G$ shown in Figure \ref{fig-movies}} \label{fig:components}
\end{figure}

In our definition of a property graph, each node and edge is associated with a single label, and at most one value for each attribute property. In some applications, it may be useful to have multiple values in these positions. We could thus consider a variant of property graphs, which we call \textit{multi-valued property graphs}, to allow multiple labels and multi-valued attribute properties within the property graph model: in Definition~\ref{def-pg}, the mapping 
$\lambda$ would then return a set of labels and $\sigma$ would return a set of values. In practice, engines may have custom policies; for example, Neo4j~\cite{cypher} -- a popular engine implementing the property graph model that we will introduce later -- allows only one label on each edge, multiple labels on nodes, and one value on each attribute property (albeit potentially a list). In any case, we focus on the single-valued variant of a property graph as given in Definition~\ref{def-pg}; whether or not $\lambda$ or $\sigma$ return a single label/value or sets of labels/values is not exigent for us.

We conclude our discussion about property graphs by presenting 
a second real-world example of how connected data can
be modelled by using this class of graphs. 

\begin{example}
\label{exa:sn} 
%
A property graph representation of a (fictitious) social network is shown in Figure \ref{fig:sn-sample}. 
Each node is labelled either as \urit{Person}, \urit{Post}, or \urit{Tag}, and each edge is labelled either as \urit{dislikes}, \urit{knows}, \urit{likes}, \urit{hasFollower} or \urit{hasTag}. Nodes with label \urit{Person} may have attributes for \urit{firstName}, \urit{lastName}, \urit{gender} and \urit{country}; nodes with
label \urit{Tag} may have an attribute for \urit{name}; nodes with label \urit{Post} may have attributes for \urit{content} and \urit{language}; and edges with label \urit{dislikes} or \urit{likes} may have an attribute for \urit{date}. We highlight that edge-sets $\{ e_1, e_2 \}$, $\{ e_3, e_4, e_5 \}$, etc., form directed cycles.
\qed
\end{example}


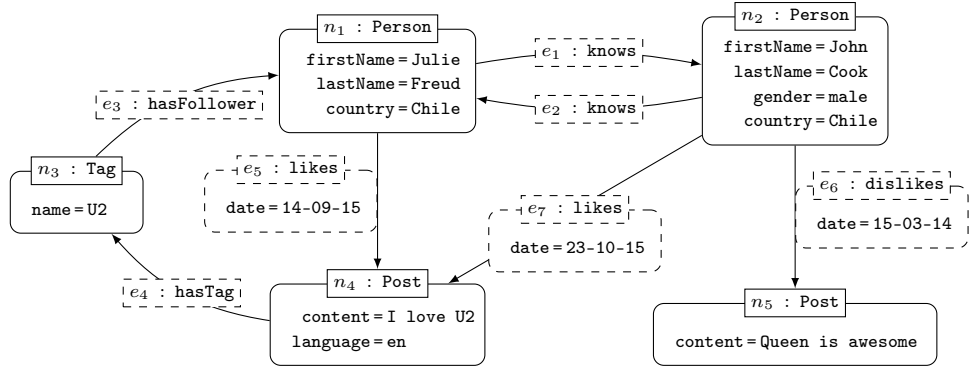
\begin{figure}[t]
\begin{center}
\begin{tikzpicture}
\node[rect] (n1) {
\alt{64pt}{
\urie{firstName} & \uriq{Julie}\\
\urie{lastName} & \uriq{Freud}\\
\urie{country} & \uriq{Chile}
}};

\node[rt] (ln1) at (n1.north) {\uri{$n_1$ $:$ Person}};

\node[rect, right=3cm of n1] (n2) {
\alt{60pt}{ 
\urie{firstName} & \uriq{John}\\
\urie{lastName} & \uriq{Cook}\\
\urie{gender} & \uriq{male}\\
\urie{country} & \uriq{Chile}
}};

\node[rt] (ln2) at (n2.north) {\uri{$n_2$ $:$ Person}};

\draw[arrout,bend left=10] (n1) to 
node[below, rectw] (e1) {} 
(n2);

\node[ert] (le1) at (e1.north) {\uri{$e_1$ $:$ knows}};

\draw[arrout,bend left=10] (n2) to 
node[below, erectw] (e2) {} 
(n1);

\node[ert] (le2) at (e2.north) {\uri{$e_2$ $:$ knows}};

\node[rect, below=0.5cm of n1,xshift=-4cm] (n3) {
\alt{40pt}{
\urie{name} & \uriq{U2}}};

\node[rt] (ln3) at (n3.north) {\uri{$n_3$ $:$ Tag}};

\draw[arrin, bend right=20] (n1) to
node[below, erectw] (e3) 
{}
(ln3);

\node[ert,yshift=-0.1cm] (le3) at (e3.north) {\uri{$e_3$ $:$ hasFollower}};

\node[rect,below=2cm of n1] (n4) {
\alt{71pt}{ 
\urie{content} & \uriq{I love U2}\\
\urie{language} & \uriq{en}}};

\node[rt] (ln4) at (n4.north) {\uri{$n_4$ $:$ Post}} ;

\draw[arrout, bend left=20] (n4) to
node[below, erectw] (e4) 
{}
(n3);

\node[ert] (le4) at (e4.north) {\uri{$e_4$ $:$ hasTag}};

\node[rect,anchor=south] at (n4.south -| n2) (n5) {
\alt{97pt}{
\urie{content} & \uriq{Queen is awesome}}};

\node[rt] (ln5) at (n5.north) {\uri{$n_5$ $:$ Post}} ;

\draw[arrout] (n1) to
node[left, erect] (e5) {
\alt{55pt}{
\urie{date} & \uriq{14-09-15}}}
(ln4);

\node[ert] (le5) at (e5.north) {\uri{$e_5$ $:$ likes}};

\draw[arrout] (n2) to
node[right, erect] (e6) {
\alt{55pt}{ 
\urie{date} & \uriq{15-03-14}}}
(ln5);

\node[ert] (le6) at (e6.north) {\uri{$e_6$ $:$ dislikes}};

\draw[arrout] (n2) to
node[below, erect] (e7) {
\alt{55pt}{ 
\urie{date} & \uriq{23-10-15}}}
(n4);

\node[ert] (le7) at (e7.north) {\uri{$e_7$ $:$ likes}};

\end{tikzpicture}
\end{center}
\caption{A property graph storing social network data. \label{fig:sn-sample}}
\end{figure}

Per the proviso in the introduction, in the following, we refer to edge-labelled graphs and property graphs generically as \textit{graph databases}. We refer to systems implementing such a data model as \textit{graph database engines}.



\section{Graph Patterns}
\label{gp} 

A variety of practical, declarative query languages have emerged in the past ten years for interrogating instances of graph data models presented in the previous section. One of the earliest such languages to be adopted by multiple vendors -- for the purposes of querying RDF graphs -- was \textit{SPARQL} (SPARQL Protocol and RDF Query Language), which was initially standardised by the W3C in 2008~\cite{sparql}, with an updated version called SPARQL 1.1 published in 2013~\cite{sparql11}. With respect to property graphs, perhaps the most well-known implementation thereof is the Neo4j engine, whose development team released a declarative query language called Cypher~\cite{cypher}. Another query language for property graphs is Gremlin~\cite{tinkerpop}, which forms an important part of the Apache TinkerPop3 graph computing framework.\footnote{Although SPARQL (1.1) has been officially standardised, Cypher and Gremlin have not and are subject to change. This survey is based on Cypher/Neo4j v.3 and Gremlin/TinkerPop v.3. Issues we discuss relating to these languages may thus change in future versions. However, given that many such systems now rely on these languages, significant (non-backwards-compatible) changes to the core features covered here would incur major migration costs. In revising the recent change-logs of these languages, we informally note that the core features and semantics discussed in this survey have not changed in recent years.}

Although these three query languages vary significantly in terms of style, purpose, expressivity, implementation, etc., they share a common conceptual core, which consists of two natural operations that one could imagine in the context of querying graphs: \textit{graph pattern matching} and \textit{graph navigation}. In this section, we focus on the former operation; navigation will be covered in detail in Section~\ref{sec:nav}.
\medskip

The simplest form of graph pattern is a \textit{basic graph pattern}, which is a graph-structured query that should be matched against the graph database. 
Additionally, basic graph patterns can be augmented with other (relational-like) features, such as projection, union, optional and difference. These allow for refining what sorts of matches are allowed and, ultimately, what results are returned. We call basic graph patterns augmented with such features \textit{complex graph patterns}. Graph pattern matching is then the evaluation of graph patterns over graph databases; it forms part of the conceptual core of SPARQL, Cypher and Gremlin; it has also found use in a variety of practical applications, including chemical structure analysis, machine learning, planning, semantic networks, and pattern recognition (see, e.g., \cite{Bunke00,aggarwal,OFGK00,MAYU03,motifs}).

We begin by introducing basic graph patterns and complex graph patterns, discuss different semantics used to evaluate them, and present concrete examples of graph patterns in SPARQL, Cypher and Gremlin. Thereafter, we make some general remarks on the complexity of graph pattern matching. 


\subsection{Basic graph patterns}  
\label{sec:bgp}  

At the core of query answering over graph databases is basic graph pattern matching.\footnote{In the context of query answering over graphs, basic graph patterns are equivalent to \textit{conjunctive queries} \cite{AHV} without projection (which will be added later).} 
Basic graph patterns (bgps) follow the same structure as the type of graph database they are intended to query but instead of only allowing constants, basic graph patterns also permit variables. In other words, a bgp for querying an edge-labelled graph is just an edge-labelled graph where variables can now appear as nodes or edge labels; a bgp for querying property graphs is just a property graph where variables can appear in place of any constant. A \textit{match} for a bgp is a mapping from variables to constants such that when the mapping is applied to the bgp, the result is, roughly speaking, contained within the original graph database. The results for a bgp are then all mappings from variables in the query to constants in the database that comprise a match.

We start with an example of a bgp for an edge-labelled graph; later we will give a more complex example involving a bgp for a property graph.

\begin{figure}[t]
	\centering
	\subfigure[Example for an edge-labelled graph]{
		\begin{tikzpicture}[->,>=stealth',auto,
		thick, scale = 1.0]

		\tikzstyle{rectl}=[rt,minimum width=5mm,minimum height=5mm,inner sep=0.5ex]
		
		\node [rectl] (z) {\uri{$x_{3}$}};
		\node [rectl,below left of=z,node distance=3cm] (x){\uri{$x_{1}$}};
		\node [rectl,below right of=z,node distance=3cm] (y) {\uri{$x_{2}$}};

		\draw[arrout] (x) to node[ertt,anchor=center] (e1) {\al{40pt}{ \uri{acts\_in}}} (z);
		\draw[arrout] (y) to node[ertt,anchor=center] (e2) {\al{40pt}{ \uri{acts\_in}}} (z);

		\end{tikzpicture} 
		\label{fig-bgp-a}
	}
	\hfill
	\subfigure[Example for a property graph]{
		
		\begin{tikzpicture}
		\node[rect] (n1) {
			\alt{60pt}{
				\urie{firstName} & \uri{$x_{1}$}\\
				\urie{lastName} & \uri{$x_{2}$}}};
		
		\node[rt] (ln1) at (n1.north) {\uri{$x_{10}$ $:$ Person}};
		
		\node[rect, right=3cm of n1] (n2) {
			\alt{60pt}{ 
				\urie{firstName} & \uri{$x_{3}$}\\
				\urie{lastName} & \uri{$x_{4}$}}};
		
		\node[rt] (ln2) at (n2.north) {\uri{$x_{11}$ $:$ Person}};
		
		\draw[arrout,bend left=10] (n1) to 
		node[below,erectw] (e1) {} 
		(n2);
		
		\node[ert] (le1) at (e1.north) {\uri{$x_{12}$ $:$ knows}};
		
		\draw[arrout,bend left=10] (n2) to 
		node[below, erectw] (e2) {} 
		(n1);
		
		\node[ert] (le2) at (e2.north) {\uri{$x_{13}$ $:$ knows}};

		\node[rect,below=2.6cm of le1] (n4) {
			\alt{71pt}{ 
				\urie{$x_{8}$} & \uri{$x_{9}$}}};
		
		\node[rt] (ln4) at (n4.north) {\uri{$x_{16}$ $:$ $x_7$}} ;
		
		\draw[arrout] (n1) to
		node[erect] (e5) {
			\alt{55pt}{
				\urie{date} & \uri{$x_{5}$}}}
		(ln4);
		
		\node[ert] (le5) at (e5.north) {\uri{$x_{14}$ $:$ likes}};

		\draw[arrout] (n2) to
		node[erect] (e7) {
			\alt{55pt}{ 
				\urie{date} & \uri{$x_{6}$}}}
		(ln4);
		
		\node[ert] (le7) at (e7.north) {\uri{$x_{15}$ $:$ likes}};
		
		\end{tikzpicture}
		
		\label{fig-bgp-b}
	}
	\caption{Two example basic graph patterns: \ref{fig-bgp-a} applies to the graph database depicted in Figure~\ref{fig:graphdb} while \ref{fig-bgp-b} applies to the property graph depicted in Figure~\ref{fig:sn-sample}}
	\label{fig-bgps}
\end{figure}
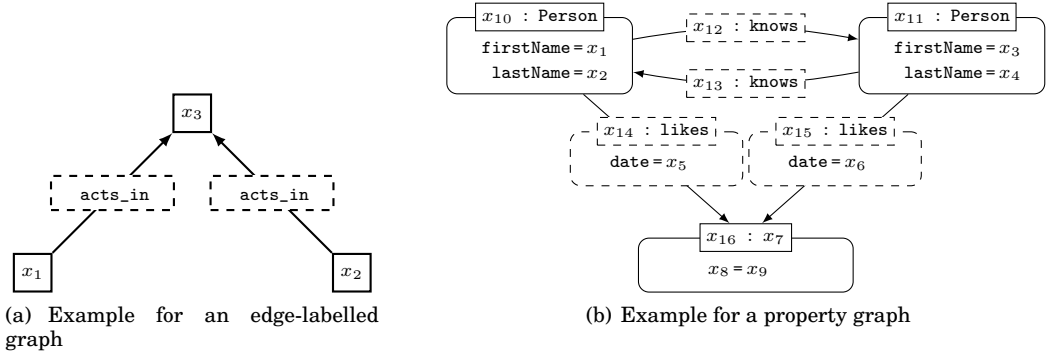

\begin{example}\label{ex-bgp-gdb}
	Let $G$ be the graph in Figure~\ref{fig:graphdb}. Assume we wish to find all co-stars in this graph. We can do this by matching the bgp in Figure~\ref{fig-bgp-a}, which we shall call $Q$, against $G$. In $Q$, we use terms $x_i$ as variables that will match any term in the database. On the other hand, \urit{acts\_in} is a constant from the set $\bL$ that will only match edges with the corresponding label in the original graph. The results of evaluating the bgp $Q$ against the graph $G$, which we denote as $Q(G)$, will thus be as follows:
	
	\begin{center}
		\begin{tabular}{|l|l|l|}
			\hline
			\uri{$x_1$} & \uri{$x_2$} & \uri{$x_3$} \\\hline
			\uriq{Clint Eastwood} & \uriq{Anna Levine} & \uriq{Unforgiven}\\
			\uriq{Anna Levine} & \uriq{Clint Eastwood} & \uriq{Unforgiven}\\
			\uriq{Clint Eastwood} & \uriq{Clint Eastwood} & \uriq{Unforgiven}\\
			\uriq{Anna Levine} & \uriq{Anna Levine} & \uriq{Unforgiven}\\\hline
		\end{tabular}
	\end{center}
	
	\noindent
	Taking the first mapping as an example, in the original bgp, if we replace variable $x_1$ by \uriqt{Clint Eastwood}, $x_2$ by \uriqt{Anna Levine} and $x_3$ by \uriqt{Unforgiven}, we get a sub-graph of the original graph database; thus we call this mapping a \textit{match} for $Q$ against $G$. The results then consist of all such valid matches. \qed
\end{example}

Though not shown in the prior example, we may also refer to specific nodes in the bgp; for example, to find the co-stars of Clint Eastwood, we could replace the variable $x_1$ (or $x_2$) with the term \uriqt{Clint Eastwood}. Basic graph patterns may also contain cycles, where, for example, we could also query for co-stars who are siblings.

We now look at an example of a bgp for a property graph.

\begin{example}\label{ex-bgp-pg}
	Let $G$ be the property graph in Figure~\ref{fig:sn-sample}. Assume we wish to query for things that (mutual) friends in the social network both like, where we wish to view the first and last name of the users in question, all the details of the item(s) they both like, and the date on which they both liked the item(s) in question. We can achieve this by matching the bgp in Figure~\ref{fig-bgp-b}, which we shall call $Q$, against the graph $G$. Again, we use terms $x_i$ as variables that will match any term in the graph database. In this case, the results $Q(G)$ will be as follows:
	
	\begin{center}
		\begin{tabular}{|l|l|l|l|l|l|l|l|l|l|l@{}}
			\cline{1-10}
			\uri{$x_1$} & \uri{$x_2$} & \uri{$x_3$} & \uri{$x_4$} & \uri{$x_5$} & \uri{$x_6$} & \uri{$x_7$} & \uri{$x_8$} & \uri{$x_9$} & \uri{$x_{10}$} & $\ldots$\\\cline{1-10}
			\uriq{Julie} & \uriq{Freud} & \uriq{John} & \uriq{Cook} & \uriq{14-09-15} & \uriq{23-10-15} & \uri{Post}  & \uri{content} & \uriq{I love U2} &  \uri{$n_1$} & $\ldots$\\
			\uriq{John} & \uriq{Cook} & \uriq{Julie} & \uriq{Freud} & \uriq{23-10-15} & \uriq{14-09-15} & \uri{Post}  & \uri{content} & \uriq{I love U2} &  \uri{$n_2$} & $\ldots$\\
			\uriq{Julie} & \uriq{Freud} & \uriq{John} & \uriq{Cook} & \uriq{14-09-15} & \uriq{23-10-15} & \uri{Post}  & \uri{language} & \uriq{en} &  \uri{$n_1$} & $\ldots$\\
			\uriq{John} & \uriq{Cook} & \uriq{Julie} & \uriq{Freud} & \uriq{23-10-15} & \uriq{14-09-15} & \uri{Post}  & \uri{language} & \uriq{en} & \uri{$n_2$} & $\ldots$\\\cline{1-10}		
		\end{tabular}
	\end{center}
	
	We omit the columns for variables $x_{11}$--$x_{16}$ for space reasons: these variables will simply match the corresponding node ids and edge ids in a manner analogous to $x_{10}$. Please note that in the expression $x_{8} =  x_{9}$, the equality sign refers to a mapping from the attribute name to its value (not equality between variables).
	
	As for the previous example, if we replace the variables in $Q$ per any of the mappings in the results above, we find that the corresponding property graph is contained within $G$, where $Q(G)$ is again defined to contain all (and only) such matches. \qed
\end{example}

\paragraph{\bf Definition}
More formally, let us refer collectively to the sets of terms $V$ and $\bL$ from Definition~\ref{def-gdb} and the sets of terms $V$, $E$, $\bL$, $\bP$ and $\bV$ from Definition~\ref{def-pg} as \textit{constants}, denoted $\bC$. Let $\bVr$ denote a set of \textit{variables}. We could then define bgps for graph databases in relation to Definition~\ref{def-gdb} by allowing $V$ and $\bL$ to contain variables, and likewise we could define bgps for property graphs in relation to Definition~\ref{def-pg} by allowing $V$, $E$, $\bL$, $\bP$ and $\bV$ to contain variables. For brevity, we skip repetitive definitions and instead continue with some quick examples.

\begin{example}
	For the bgp $Q$ shown in Figure~\ref{fig-bgp-a}, as per Definition~\ref{def-gdb}, we can denote $Q = (V,E)$, where:
    \begin{center}	
    		\small
	\begin{tabular}{c}
		$V = \{ x_1,x_2,x_3 \}\,,\,  E$ = $\{ (x_1,\urit{acts\_in},x_3) , (x_2,\urit{acts\_in},x_3) \}$ \\
	\end{tabular}
	\end{center}	
	In this case, $x_i \in \bVr$ for $1 \leq i \leq  3$, while 
	$\urit{acts\_in} \in \bC$. \qed
\end{example}

\begin{example}
	For the bgp $Q$ shown in Figure~\ref{fig-bgp-b}, as per Definition~\ref{def-pg}, we can denote $Q = (V,E,\rho,\lambda,\sigma)$, where:
	
	\begin{center}
		\small
		\begin{tabular}{r@{~}c@{~}l@{~~~~~~~}r@{~}c@{~}l@{~~~~~~~~}r@{~}c@{~}l}
			$V$ & $=$ & $\{x_{10}, x_{11}, x_{16}\}$ & $E$ & $=$ & $\{x_{12}, x_{13}, x_{14}, x_{15}\}$
			& $\sigma(x_{10},\urit{firstName})$ & $=$ & $x_1$\\
			          
			 & & & & & & $\sigma(x_{10},\urit{lastName})$ & $=$ & $x_2$ \\

			$\rho(x_{12})$ & $=$ & $(x_{10},x_{11})$ & $\rho(x_{13})$ & $=$ & $(x_{11},x_{10})$ 
			& $\sigma(x_{11},\urit{firstName})$ & $=$ & $x_3$ \\
			
			$\rho(x_{14})$ & $=$ & $(x_{10},x_{16})$ & $\rho(x_{15})$ & $=$ & $(x_{11},x_{16})$
			& $\sigma(x_{11},\urit{lastName})$ & $=$ & $x_4$\\
			
			 & & & & & & $\sigma(x_{14},\urit{date})$ & $=$ & $x_5$\\
			 
			$\lambda(x_{10})$ & $=$ & $\urit{Person}$ & $\lambda(x_{12})$ & $=$ & $\urit{knows}$
			& $\sigma(x_{15},\urit{date})$ & $=$ & $x_6$\\
			
			$\lambda(x_{11})$ & $=$ & $\urit{Person}$ & $\lambda(x_{13})$ & $=$ & $\urit{knows}$
			& $\sigma(x_{16},x_8)$ & $=$ & $x_9$ \\
			
			$\lambda(x_{16})$ & $=$ & $x_7$ & $\lambda(x_{14})$ & $=$ & $\urit{likes}$
			& $\sigma(x_{16},x_8)$ & $=$ & $x_9$ \\			
			
			 & & & $\lambda(x_{15})$ & $=$ & $\urit{likes}$
			& & & \\

		\end{tabular}
	\end{center} 
	\noindent
	As before, $x_i \in \bVr$ for $1 \leq i \leq 16$, and all other domain terms are in $\bC$.\hfill\qed
\end{example}

\paragraph{\bf Evaluation}
Evaluating a bgp $Q$ against a graph database $G$ corresponds to listing all possible \textit{matches} of $Q$ with respect to $G$ (as per Examples~\ref{ex-bgp-gdb}~\&~\ref{ex-bgp-pg}). More formally, we can define a match as follows.

\begin{definition}[Match]\label{def-match-gdb}
	Given an edge-labelled graph $G = (V,E)$ and a bgp $Q = (V',E')$, a {\em match} $h$ of $Q$ in $G$ is a mapping from $\bC \cup \bVr$ to $\bC$ such that:
	
	\begin{enumerate}
		
		\item for each constant $a \in \bC$, it is the case that $h(a) = a$; that is, the mapping maps constants to themselves; and
		
		
		\item for each edge $(b,l,c) \in E'$, it holds that $(h(b),h(l),h(c)) \in E$; this condition imposes that (a) each edge of $Q$ is mapped to an edge of $G$, and (b) the structure of $Q$ is preserved in its image under $h$ in $G$ (that is, when $h$ is applied to all the terms in $Q$, the result is a sub-graph of $G$). \qed
		
	\end{enumerate} 
\end{definition}

We leave implicit the analogous definition for property graphs since the principle is the same: a mapping $h$ maps constants to themselves and variables to constants; if the image of $Q$ under $h$ is contained within $G$, then $h$ is a match (see Example~\ref{ex-bgp-pg}).
\medskip

In technical terms, a match $h$ corresponds to a {\em homomorphism} from $Q$ to $G$ (see, e.g., \cite{B13}), whereby multiple variables in $Q$ can map to the same term in $G$, as was the case in Example~\ref{ex-bgp-gdb} where the latter two matches mapped variables $x_1$ and $x_2$ to the same term. In some cases, however, it may be desirable to require that variables map to distinct terms, where these latter two matches would be dropped; in other words it may be desirable to restrict $h$ to be an injective (i.e., one-to-one) mapping, in which case the matching process corresponds to the well-known notion of {\em subgraph isomorphism} (see, e.g., \cite{Ullmann76,F12}). But this may be too strict in certain applications, where, for example, it may be desirable to allow multiple label variables to match one label, but to enforce that node and/or edge ids are kept distinct (with the intuition that nodes and edges represent the structure of the graph, and labels are simply annotations on that structure). These preferences lead to different semantics for the {\em evaluation of a bgp $Q$ over a graph database $G$}, as explained next:

\begin{enumerate}
	
	\item {\em Homomorphism-based semantics:} This is the unconstrained semantics: no additional restriction is imposed on the matches $h$ of $Q$ in $G$ other than the base conditions from Definition~\ref{def-match-gdb}. The evaluation of $Q$ against $G$ then consists of all possible homomorphisms from $Q$ to $G$. Since homomorphism-based approach corresponds to the familiar semantics of select-from-where queries in relational databases, and since it forms the basis for the other, more restrictive semantics of bgps, it is often studied in the theoretical community (see, e.g., \cite{CGLVkr,Woo,B13,BLR14,RRV15}). 
There are also several papers that study implementation issues related to this semantics (see, e.g., \cite{CYDYW08,ZCO09,FLMWW10}) and it is currently used, for example, by the SPARQL query language~\cite{sparql11}.
	
	\item {\em Isomorphism-based semantics:} Under this type of semantics, the structure of the query (in some potentially application-dependent sense) should be preserved under the image of the permitted mappings; in more practical terms, certain types of variables are restricted to match distinct constants in the database. Since the precise type of isomorphism -- i.e., the precise type of structure preserved -- may depend on the application, this leaves us with a variety of different possible isomorphism-based semantics, where we can highlight:
	
	\begin{itemize} 
		\item {\em No-repeated-anything semantics:} Only injective mappings are allowed, meaning that no two variables can be bound to the same term in a given match.
	    \item {\em No-repeated-node semantics:} The injective restriction only applies to variables that map to nodes (or node ids). In edge-labelled graphs, for example, it is common to only require mappings of node variables to be injective, meaning that multiple variables can still be mapped to the same edge labels. This ``no-repeated-node'' semantics is often preferred in graph matching applications (see, e.g., \cite{Bunke00}) where no nodes in the query graph should be ``collapsed'' as it would change the structure of the query graph. 
		\item {\em No-repeated-edge semantics:} The injective restriction only applies to variables that map to edges: in other words, ``edge variables'' (variables that map to edge ids in $E$) must be mapped one-to-one, whereas other types of variables (for nodes, labels, attribute properties and values) need not be injective. This semantics is currently used by the Cypher query language \cite{cypher}.   
		\end{itemize}
\end{enumerate}
		

\begin{example} \label{ex:bgp-eval} In order to illustrate the differences between these semantics, let $G$ be the property graph of Figure~\ref{fig-movies} and $Q$ the following basic graph pattern:

	
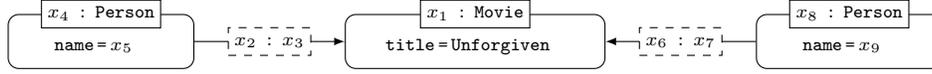
\begin{center}
		\begin{tikzpicture}
		\node[rect] (n1) {
			\alt{88pt}{ 
				\urie{title} & \uriq{Unforgiven}}};
		
		\node[rt] (ln1) at (n1.north) {\uri{$x_1$ $:$ Movie}};
		
		\node[rect,left=2cm of n1] (n2) {
			\alt{60pt}{ 
				\urie{name} & \uri{$x_5$}}};
		
		\node[rt] (ln2) at (n2.north) {\uri{$x_4$ $:$ Person}};
		
		\draw[arrout] (n2) to 
		node[ert] (e2) {\uri{$x_2$ $:$ $x_3$}}
		(n1);
		
		\node[rect,right=2cm of n1] (n3) {
			\alt{60pt}{ 
				\urie{name} & \uri{$x_9$}}};
		
		\node[rt] (ln3) at (n3.north) {\uri{$x_8$ $:$ Person}};
		
		\draw[arrout] (n3) to 
		node[ert] (e2) {\uri{$x_6$ $:$ $x_7$}}
		(n1);
		\end{tikzpicture}
	\end{center}
	
\noindent
From evaluating $Q(G)$, we have the following (unrestricted) results:

	\begin{center}
		\begin{tabular}{|l|l|l|l|l|l|l|l|l|}\cline{1-9}
			\uri{$x_1$} & \uri{$x_2$} & \uri{$x_3$} & \uri{$x_4$} & \uri{$x_5$} & \uri{$x_6$} & \uri{$x_7$} & \uri{$x_8$} & \uri{$x_9$} \\\cline{1-9}
			
			\uri{$n_2$} & \uri{$e_2$} & \uri{directs} & \uri{$n_1$} &  \uriq{Clint Eastwood} & \uri{$e_3$} & \uri{acts\_in} & \uri{$n_3$} & \uriq{Anna Levine} \\
			
			\uri{$n_2$} & \uri{$e_3$} & \uri{acts\_in} & \uri{$n_3$} &  \uriq{Anna Levine} & \uri{$e_2$} & \uri{directs} & \uri{$n_1$} & \uriq{Clint Eastwood} \\
			
			\uri{$n_2$} & \uri{$e_1$} & \uri{acts\_in} & \uri{$n_1$} &  \uriq{Clint Eastwood} & \uri{$e_3$} & \uri{acts\_in} & \uri{$n_3$} & \uriq{Anna Levine} \\
			
			\uri{$n_2$} & \uri{$e_3$} & \uri{acts\_in} & \uri{$n_3$} &  \uriq{Anna Levine} & \uri{$e_1$} & \uri{acts\_in} & \uri{$n_1$} & \uriq{Clint Eastwood} \\
			
			\uri{$n_2$} & \uri{$e_2$} & \uri{directs} & \uri{$n_1$} &  \uriq{Clint Eastwood} & \uri{$e_1$} & \uri{acts\_in} & \uri{$n_1$} & \uriq{Clint Eastwood} \\
			
			\uri{$n_2$} & \uri{$e_1$} & \uri{acts\_in} & \uri{$n_1$} & \uriq{Clint Eastwood} & \uri{$e_2$} & \uri{directs} & \uri{$n_1$} &  \uriq{Clint Eastwood} \\
			
			\uri{$n_2$} & \uri{$e_1$} & \uri{acts\_in} & \uri{$n_1$} & \uriq{Clint Eastwood}  & \uri{$e_1$} & \uri{acts\_in} & \uri{$n_1$} & \uriq{Clint Eastwood} \\		
			
			\uri{$n_2$} & \uri{$e_2$} & \uri{directs} & \uri{$n_1$} & \uriq{Clint Eastwood}  & \uri{$e_2$} & \uri{directs} & \uri{$n_1$} & \uriq{Clint Eastwood} \\
			
			\uri{$n_2$} & \uri{$e_3$} & \uri{acts\_in} & \uri{$n_1$} & \uriq{Anna Levine}  & \uri{$e_3$} & \uri{acts\_in} & \uri{$n_1$} & \uriq{Anna Levine} \\	
			\hline
		\end{tabular}
	\end{center} 
	
All matches are valid under the homomorphism-based semantics. Only the first two matches would be permitted under the no-repeated-anything semantics since the latter seven matches all map multiple variables to the same term. Only the first four matches would be valid under the no-repeated-node semantics since in the latter five matches, the ``node variables'' $x_4$ and $x_8$ map to the same node. Only the first six matches would be valid under the no-repeated-edge semantics since in the latter three matches, the ``edge variables'' $x_2$ and $x_6$ map to the same edge.

As the example suggests, the appropriate selection of semantics may vary from application to application: no one semantics fits all. \qed
\end{example}


There are two main criticisms of the above semantics. First, as we discuss in more detail later, the computational complexity of key problems associated with these semantics can be quite high since they directly capture notions of graph homomorphism and subgraph isomorphism \cite{Ullmann76} (which are both known to have NP-complete decision problems). 
Second, the matches defined by the above semantics are rigid, in a sense that they require the entire query to be matched onto the graph continuously. That is, even when all parts of the query can be matched to (possibly different parts of) the graph, they may return zero answers. To remedy the situation one can deploy the more flexible notion of {\em graph-simulations}~\cite{M89simulation} when defining a match, which gives rise to an additional semantics.

\begin{enumerate}
	\setcounter{enumi}{2}
	\item {\em Simulation-based semantics:}	
	A generalisation of the notion of a homomorphism-based match has been proposed in the form of {\em graph-simulations} \cite{M89simulation}, which, intuitively speaking, allow matching one node of a pattern to several nodes in the graph, as long as the structure of the pattern is preserved. 
	Given an edge-labelled graph $G = (V,E)$ and a bgp $Q = (V',E')$, a simulation between $Q$ and $G$ is a	relation $S \subseteq V' \times V$ such that: (i) for every node $n'$ in $V'$ there is a node $n$ in $V$ such that 
	the pair $(n',n)$ is in $S$, and (ii) for every pair $(n',n) \in S$ and every edge $(n',r',m')$ in $E'$, there exists an edge $(n,r,m)$ in $E$ such that $(m',m) \in S$ and $r=r'$ if $r'\in \bC$ (and can be any value when $r'\in \bVr$). Then an answer to $Q$ over $G$ under the simulation-based semantics is  any simulation $S$ between $Q$ and $G$. 
	As shown in the literature, simulation-based semantics is computationally lighter for certain problems \cite{HenzingerHK95,FanLMTW11} and is more versatile when handling large graphs that might contain incomplete information \cite{FLMTWW10,F12,FLMWW10}.
	
\end{enumerate}  

Simulation-based semantics can be naturally extended to property graphs. In this case, if a query node (or edge) uses constants in its attributes, we also require that it matches a graph node with equal values in the corresponding attributes. That is, when $(v',v)$ belongs to our simulation, we also need the following conditions: (iii) if $\lambda(v') = r$ with $r \in \bC$,  then $\lambda(v)=r$; and (iv) $\sigma(v',e') = a'$, then $\sigma(v,e)=a$, for some $e$ and $a$, with $e=e'$ when $e'\in \bC$ and $a=a'$ when $a'\in \bC$. A similar condition is also required for edge properties when specified in the query.

\begin{example}\label{ex:bgp-sim}
Consider again the graph $G$ from Figure~\ref{fig:graphdb}, and let $Q$ be the following BGP:	
	
\begin{center}
\begin{tikzpicture}

\node[rtt] (n1) {
\al{60pt}{ 
\uriq{$x_1$}}};

\node[rtt, right=3cm of n1] (n2) {
\al{60pt}{\uriq{$x_2$}}};

\draw[arrout] (n1) to 
node[ertt] (e1) 
{\al{40pt}{
\uri{acts\_in}}}
(n2);
\end{tikzpicture}
\end{center}

One simulation between the query $Q$ above and the graph $G$ is given by the relation $S=\{(x_1,\uri{Clint Eastwood}),(x_2,\uri{Unforgiven})\}$. 
Another simulation is given by the relation $S' = \{(x_1,\uri{Clint Eastwood}),(x_2,\uri{Unforgiven}),(x_1,\uri{Anna Levine})\}$, which in a sense contains matches for both Clint Eastwood and for Anna Levine. This exemplifies the fact that simulation-based semantics can capture multiple homomorphic matches in a single relation, which is one of the reasons why it can be evaluated more efficiently. \qed
\end{example}

The idea of matching the same query node to multiple graph nodes may be counter intuitive, as it captures ``too much" information in a single relation. For this reason simulation-based semantics is often viewed as a base semantics for defining a set of ``candidate matches'' that can be further restricted and refined for particular use-cases, as has been explored recently by \citeN{MCFHW14}.


While simulation-based semantics offer an interesting, more flexible alternative to homomorphism- or isomorphism-based semantics, the query languages we include in our survey do not support such simulation-based partial matches over bgps; henceforth, we will thus focus on the latter two ``complete match'' semantics for bgps.
\medskip

While some of the previous semantics may restrict the duplication of terms within a single match -- namely the isomorphism-based semantics -- we can also consider an orthogonal choice of semantics with respect to duplicate matches in the result of evaluating a bgp $Q$ over a graph database $G$, as follows:

\begin{itemize}
	\item {\em Set semantics:} $Q(G)$ is defined as a \textit{set} of matches; in other words, the result of evaluating $Q$ over $G$ cannot contain duplicate matches.
	\item {\em Bag semantics:} $Q(G)$ is defined as a \textit{bag} of matches; more specifically, the number of times a match appears in the result corresponds with the number of unique mappings that witness the match.
\end{itemize}

In fact, on the level of bgps, duplicate matches cannot occur, and hence the set and bag semantics are equivalent. However, when we later extend bgps with features such as projection, union, etc., duplicate matches can occur, distinguishing both semantics.

We can then consider, for example, homomorphism-based set semantics, or isomorphism-based bag semantics, and so forth. Since in much of our discussion it will be inessential which underlying semantics we use for evaluating bgps, we may refer to $Q(G)$ as the evaluation of bgp $Q$ over a graph database $G$ in a generic manner, where we assume a homomorphism-based set semantics unless otherwise stated.

\subsection{Complex graph patterns} \label{sec:cgp} 

In terms of traditional relational operations, basic graph patterns (bgps) cover the natural join, and selection based on equality (since constants can be embedded into a bgp). Complex graph patterns (cgps) extend bgps with further traditional relational operations -- namely projection, union, difference, optional (aka. left-outer-join) and filter (which covers selection). We will now go through each of these features in turn.




\noindent 
\paragraph{{\em {\bf Projection}}}
We call the set of variables for which $Q(G)$ potentially returns matches the \textit{output variables} of the graph pattern $Q$ (which is independent of $G$). For a bgp, this is always the set of all variables in a query. However, projection allows for selecting a subset of the output variables of a graph pattern as the new output variables: it allows for stating which variables are deemed relevant in the evaluation of a cgp. 
For instance, in Example~\ref{ex:bgp-eval}, to retrieve only the names of actors who starred together in \texttt{Unforgiven} -- e.g., for a user who is uninterested in node or edge ids -- we can project variables $x_5$ and $x_9$; other columns will then be simply omitted from the results. 
As expected, this operator is present in all practical query languages for graphs, often using the projection keyword \texttt{SELECT} as used by SQL.

\noindent 
\paragraph{{\em {\bf Join}}}
While the join of two bgps can be easily expressed as another bgp (under homomorphism-based semantics), more complex graph patterns or different semantics require the explicit use of this
operator. This corresponds to the usual relational join (more specifically, a \textit{natural join}) over the queries that are defined by two graph patterns $Q_1$ and $Q_2$. The output variables of this join corresponds to the union of the output variables of $Q_1$ and $Q_2$, and its evaluation contains all matches that can be obtained by joining a match in the evaluation of $Q_1$ with a match in the evaluation of $Q_2$. More specifically, two such matches can be joined when they take the same values for the variables that are shared by the output variables of $Q_1$ and $Q_2$; in this case, we say that the matches are \textit{compatible}. An explicit join is essential in any query language that goes beyond bgps to combine results from different operations.

\noindent
\paragraph{{\em {\bf Union and difference}}}
Let $Q_1$ and $Q_2$ be two graph patterns. 
The union of $Q_1$ and $Q_2$ is a complex graph pattern whose evaluation is defined
as the union of the evaluations of $Q_1$ and
$Q_2$; for example, in a movie database such as the one from Figure~\ref{fig-movies}, one could use union to find the movies in which Clint Eastwood acted \textit{or} which he directed. 
The difference of $Q_1$ and $Q_2$ is also a complex graph pattern whose evaluation is defined 
as the set of matches in the evaluation of $Q_1$ that do not belong to the evaluation of $Q_2$; 
for example, one could use difference to find the movies in which Clint Eastwood acted but did not direct.
Computing the union of two sets of matches is rather simple in computational terms, and as expected most systems implement the union operator. However, difference is computationally more difficult for certain evaluation problems and as such some systems prefer to leave its implementation out. In some other cases, the implementation of the difference operator has been delayed for future revisions of the language, as was the case for SPARQL, where an explicit difference operator, called \texttt{MINUS}, was only introduced in SPARQL 1.1~\cite{sparql11}.\footnote{We briefly note that SPARQL supports a variant of \texttt{UNION} and \texttt{MINUS} where, if the output of the base patterns $Q_1$ and $Q_2$ differ, then \textit{compatible} matches are unioned or removed, respectively~\cite{PAG09,AnglesG16}. In the case of union, this may create partial matches.}

\noindent 
\paragraph{{\em {\bf Optional}}}
This operator is based on the join of two graph patterns $Q_1$ and $Q_2$, but instead of dismissing those matches in the evaluation of $Q_1$ that cannot be joined with a match in the evaluation of $Q_2$, it keeps them in the result in order to maximise the amount of information retrieved. This feature is particularly useful when dealing with incomplete information, or in cases where the user may not know what information is available. For example, in the context of Figure~\ref{fig:sn-sample}, information relating to the gender of users is incomplete but may still be interesting to the client, where available. Let us assume that the client wishes to retrieve users that follow the U2 tag, where available, to find out what their genders are. Using a natural join, users such as Julie Freud that do not have an explicit gender would be excluded from the results. But instead by using optional, users without a gender will be returned and the value for gender in the corresponding match will simply be left undefined/blank. This operation, then, supports partial answers over incomplete data. 
In relational terms, the optional operator corresponds to the {\em left-outer join} \cite{GR97}. 
The optional operator has been present in SPARQL since the original version \cite{sparql,PAG09}, and is also included, for example, in the Cypher query language \cite{cypher}.

\noindent 
\paragraph{{\em {\bf Filter}}}
Users may wish to restrict the matches of a cgp over a graph database $G$ based on some of the intermediate values returned using, e.g., inequalities, or other types of expressions. For instance, with respect to Example~\ref{ex-bgp-pg}, a client may be interested in finding things that mutual friends both liked during \textit{October 2015}, in which case, the client could apply a filter on the cgp of the following form: 
\[ \uriqt{01-10-15} \leq x_5 \leq \uriqt{31-10-15} \ \ \textsc{and} \ \ \uriqt{01-10-15} \leq x_6 \leq \uriqt{31-10-15}  \]
\noindent
Applying a filter over a graph pattern does not change its output variables. In general, the filter expression covers the usual conditions permitted by the selection relational operator, including inequalities; boolean connectives such as \textsc{and}, \textsc{or} or \textsc{not}; etc. However, while basic filter operators are present in some form for all practical graph-based query languages, in certain languages a wide range of expressions is provided to support complex filtering criteria, including regular expressions over strings, arithmetic operators, casting, etc. We give some examples in the following section. 

\subsection{Graph patterns in practice}
\label{subsec-bgp-practice}

\begin{figure}[t]
	\centering
	\subfigure[Example for an edge-labelled graph]{ \label{fig-dh-gdb}
		\begin{tikzpicture}
		
		\node[rtt] (n1) {
			\al{12pt}{ 
				\uri{$x_1$}}};
		
		\node[rtt, right=4cm of n1] (n2) {
			\al{12pt}{ 
				\uri{$x_3$}}};
		
		\node[rtt, right=4cm of n2] (n3) {
			\al{12pt}{ 
				\uri{$x_2$}}};
		
		\draw[arrout] (n1) to 
		node[ertt] (e1) 
		{\al{28pt}{
				\uri{acts\_in}}}
		(n2);
		
		\draw[arrout] (n3) to 
		node[ertt] (e2) {
			\al{28pt}{
				\uri{acts\_in}}}
		(n2);
		
		\node[rtt, below=1.4cm of n1] (n4) {
			\al{40pt}{ 
				\uri{Person}}};
		
		\draw[arrout] (n1) to 
		node[ertt] (e3) {
			\al{25pt}{
				\uri{type}}}
		(n4);		
		
		\node[rtt, below=1.4cm of n3] (n5) {
			\al{40pt}{ 
				\uri{Person}}};
		
		\draw[arrout] (n3) to 
		node[ertt] (e4) {
			\al{25pt}{
				\uri{type}}}
		(n5);	
		
		\node[rtt, below=1.4cm of n2,xshift=2cm] (n6) {
			\al{50pt}{ 
				\uriq{Unforgiven}}};
		
		\draw[arrout] (n2) to 
		node[ertt] (e5) {
			\al{25pt}{
				\uri{title}}}
		(n6);
		
		\node[rtt, below=1.4cm of n2,xshift=-2cm] (n7) {
			\al{40pt}{ 
				\uri{Movie}}};
		
		\draw[arrout] (n2) to 
		node[ertt] (e6) {
			\al{25pt}{
				\uri{type}}}
		(n7);	
		\end{tikzpicture}
	}
	\subfigure[Example for a property graph]{	\label{fig-dh-pg}
		\begin{tikzpicture}[->,>=stealth',auto,
		thick, scale = 0.5]

		\tikzstyle{every node}=[minimum width=2mm,minimum height=2mm]
		
		
		\node [rect] (y) {
			\alt{80pt}{ 
				\urie{title} & \uri{Unforgiven}\\}
		};
		\node[rt,left of=y,node distance=5cm] (x) {\uri{$x_1$ $:$ Person}};
		
		\node [rt,right of=y, node distance=5cm] (z) {\uri{$x_2$ $:$ Person}};
		
		
		
		\node[rt] (ly) at (y.north) {\uri{$x_3$ $:$ Movie}};
		
		\node[above=0.5ex of ly.north] {};
		
		\path[->]
		(x) edge  node[rectw,below] (e1) {} (y)
		(z) edge  node[rectw] (e2) {} (y);
		
		\node[ert] (le1) at (e1.north) {\uri{$x_4$ $:$ acts\_in}};
		\node[ert] (le2) at (e2.north) {\uri{$x_5$ $:$ acts\_in}};
		
		\end{tikzpicture}
	}
	\caption{Two versions of a basic graph pattern to retrieve all pairs of co-stars for the movie Unforgiven: one for a graph database and one for a property graph}
	\label{fig-bgp-practice}
\end{figure}
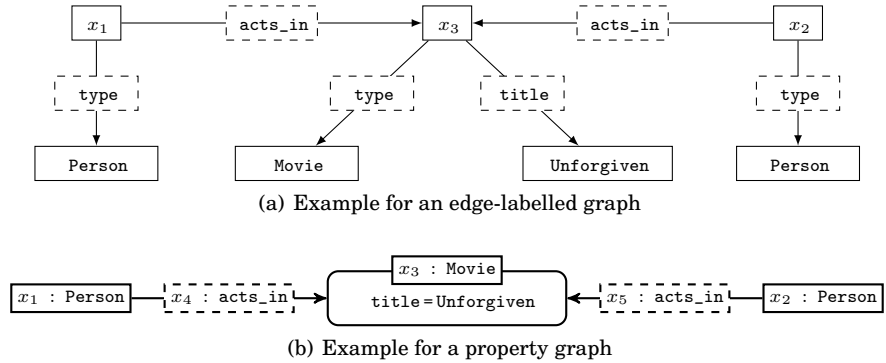

We now take a closer look at how graph patterns are applied in three practical query languages: SPARQL, Cypher and Gremlin. We choose these languages because they are the most widely-used 
query languages in practice but offer significant differences: SPARQL operates over 
RDF graphs;
Cypher is designed to operate over property graphs as defined previously; meanwhile, Gremlin is more imperative in nature than the other two, and is geared more towards graph traversal than graph pattern matching. Given that each of these three languages is associated with lengthy documentation, in the following our goal is not to be complete in discussing the graph pattern matching features of all three engines, but rather to give a quick comparative impression of each language through examples (for which we will use the bgps depicted in Figure \ref{fig-bgp-practice}) and to highlight and contrast some important aspects.

\begin{figure}
	\centering
	\begin{tikzpicture}
	
	\node[rtt] (n1) {
		\al{60pt}{ 
			\uri{:Clint\_Eastwood}}};
	
	\node[rtt, right=3cm of n1] (n2) {
		\al{48pt}{ 
			\uri{:Unforgiven}}};
	
	\node[rtt, right=3cm of n2] (n3) {
		\al{50pt}{ 
			\uri{:Anna\_Levine}}};
	
	\draw[arrout,bend left=10] (n1) to 
	node[ertt] (e1) 
	{\al{33pt}{
			\uri{:acts\_in}}}
	(n2);
	
	\draw[arrout,bend right=10] (n1) to 
	node[ertt] (e1a) 
	{\al{33pt}{
			\uri{:directs}}}
	(n2);
	
	\draw[arrout] (n3) to 
	node[ertt] (e2) {
		\al{33pt}{
			\uri{:acts\_in}}}
	(n2);
	
	\node[rtt, below=1.4cm of n1] (n4) {
		\al{42pt}{ 
			\uri{:Person}}};
	
	\draw[arrout] (n1) to 
	node[ertt] (e3) {
		\al{25pt}{
			\uri{:type}}}
	(n4);		
	
	\node[rtt, below=1.4cm of n3] (n5) {
		\al{42pt}{ 
			\uri{:Person}}};
	
	\draw[arrout] (n3) to 
	node[ertt] (e4) {
		\al{25pt}{
			\uri{:type}}}
	(n5);	
	
	\node[rtt, below=1.4cm of n2,xshift=1.5cm] (n6) {
		\al{50pt}{ 
			\uriq{"Unforgiven"}}};
	
	\draw[arrout] (n2) to 
	node[ertt] (e5) {
		\al{25pt}{
			\uri{:title}}}
	(n6);
	
	\node[rtt, below=1.4cm of n2,xshift=-1.5cm] (n7) {
		\al{40pt}{ 
			\uri{:Movie}}};
	
	\draw[arrout] (n2) to 
	node[ertt] (e6) {
		\al{25pt}{
			\uri{:type}}}
	(n7);	
	\end{tikzpicture}
	\caption{RDF graph extending the edge-labelled graph of Figure~\ref{fig:graphdb}}\label{fig:rdf}
\end{figure}
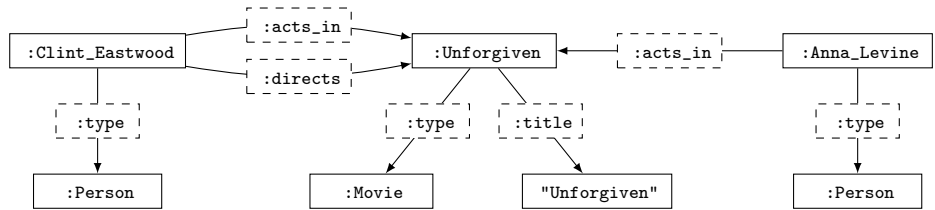

\medskip
\noindent \paragraph{{\em {\bf SPARQL}}} SPARQL is a declarative language recommended by the W3C for querying RDF graphs \cite{sparql,sparql11}. The basic building blocks of SPARQL queries are \emph{triple patterns}, which are RDF triples where the subject, object or predicate may be a variable (variables in SPARQL typically start with the symbol `\verb+?+'). Several triple patterns can be combined (conjunctively) into a basic graph pattern. On top of basic graph patterns, SPARQL also supports all of the complex graph pattern features discussed previously (and more besides). The evaluation of bgps in SPARQL is done following homomorphism-based bag semantics. In the following, we will use Figure~\ref{fig:rdf} as our example RDF data.

\begin{example}
The following SPARQL query represents a complex graph pattern that combines the basic graph pattern of Figure~\ref{fig-dh-gdb} with a projection that asks to only return the co-stars and not the movie identifier:

\begin{center}
	\small
	\begin{verbatim}
	PREFIX : <http://ex.org/#>
	SELECT ?x1 ?x2 
	WHERE { 
	  ?x1 :acts_in ?x3 . ?x1 :type :Person .  
	  ?x2 :acts_in ?x3 . ?x2 :type :Person . 
	  ?x3 :title "Unforgiven" . ?x3 :type :Movie .
	  FILTER(?x1 != ?x2)
	}
	\end{verbatim}
\end{center}

\noindent
Recalling that constants in RDF graphs can be IRIs, the purpose of the \texttt{PREFIX} statement is to define a shortcut for a namespace under which constants appear; since prefixes are inessential to our discussion, we will henceforth leave them implicit. In the \texttt{SELECT} clause, we specify the variables we wish to project as output. The \texttt{WHERE} clause then captures the basic graph pattern of Figure~\ref{fig-dh-gdb}: it contains six triple patterns (delimited by periods) that correspond to the edges of Figure~\ref{fig-dh-gdb}. Additionally, since the semantics of SPARQL evaluation is homomorphism-based, we add a \texttt{FILTER} to ensure that we do not match cases where \texttt{?x1} and \texttt{?x2} map to the same person.

Applied to Figure~\ref{fig:rdf}, this query would thus return:

\begin{center}
	\small
	\begin{tabular}{|l|l|}\hline
		\uri{?x1} & \uri{?x2}\\\hline
		\uri{:Clint\_Eastwood} & \uri{:Anna\_Levine} \\		
		\uri{:Anna\_Levine} & \uri{:Clint\_Eastwood} \\\hline
	\end{tabular}
\end{center}

\noindent
Other matches for the bgp are removed by the filter and \texttt{?x3} is projected away.\qed

\end{example}

The previous example shows how bgps, projection and filter are supported in SPARQL. We now look at some brief examples for the remaining cgp features that are all based on the graph database of Figure~\ref{fig:rdf}.

\begin{example}
We start with an example of a union to find movies that Clint Eastwood has acted \textit{or} directed in.

\begin{center}
	\small
	\begin{verbatim}
	SELECT ?x
	WHERE {{ :Clint_Eastwood :acts_in ?x . } UNION { :Clint_Eastwood :directs ?x . }}
	\end{verbatim}
\end{center}

\noindent
Both patterns to the left and right of the \texttt{UNION} will be evaluated \textit{independently} and their results unioned. This will return \ttt{:Unforgiven}; in fact, this result will be returned twice since SPARQL, by default, adopts a bag semantics.\qed
\end{example}

\begin{example}
We could use difference to ask for people who acted in the movie Unforgiven but who did not (also) direct it:

\begin{center}
	\small
	\begin{verbatim}
	SELECT ?x 
	WHERE {{ ?x :acts_in :Unforgiven . } MINUS { ?x :directs :Unforgiven . }}
	\end{verbatim}
\end{center}

\noindent
Any match for the left side of the \texttt{MINUS} that is compatible with a match from the right side will be removed. Hence, this query will return \ttt{:Anna\_Levine}.\qed
\end{example}

\begin{example}
Using optional, we could ask for movies that actors have appeared in, and any other participation they had with the movie besides acting in it:

\begin{center}
	\small
	\begin{verbatim}
	SELECT ?x1 ?x2 ?x3
	WHERE {{ ?x1 :acts_in ?x2 .} OPTIONAL { ?x1 ?x3 ?x2 . FILTER(?x3 != :acts_in) }}
	\end{verbatim}
\end{center}

\noindent
This will return:

\begin{center}
	\small
\begin{tabular}{|l|l|l|}\hline
\uri{?x1} & \uri{?x2} & \uri{?x3}\\\hline
\uri{:Clint\_Eastwood} & \uri{:Unforgiven} & \uri{:directs}\\
\uri{:Anna\_Levine} & \uri{:Unforgiven} & \\\hline
\end{tabular}
\end{center}

\noindent
A result is still returned for \texttt{:Anna\_Levine} even though she had no other participation in the movie; instead the relevant column is left blank for that result.\qed
\end{example}

In the latter example, we show how optional and filter can be combined. Of course, it is also possible to combine these features in other ways to form increasingly more complex graph patterns, for example, to find movies Clint Eastwood has neither acted nor directed in, or to find his co-stars in those movies he did not also direct, etc. 
\medskip

Here we have provided a few brief examples of the most notable features for graph patterns that SPARQL supports. However, the list of graph pattern features we cover is far from complete, where for example SPARQL 1.1 now supports a wide range of \texttt{FILTER} expressions, variable assignments, arithmetic operations, conditionals, federation, and so forth. Likewise, rather than operate over a single graph, SPARQL operates over collections of graphs, called ``Named Graphs'', which allow for selecting customised partitions of the data over which queries should be executed. We refer to the official standard for more details~\cite{sparql11}. Other SPARQL features such as property paths will be covered in later sections.

\medskip
\noindent \paragraph{{\em {\bf Cypher}}} Cypher is a declarative language for querying property graphs that uses ``patterns'' as its main building blocks~\cite{cypher}. Patterns are expressed syntactically following a ``pictorial'' intuition to encode nodes and edges with arrows between them. 
Unlike SPARQL, Cypher uses isomorphism-based no-repeated-edges bag semantics.
We again give a quick flavour of Cypher in some examples, where this time we will consider evaluation against the property graph of Figure~\ref{fig-movies}.

\begin{example}\label{ex-cyp} The pattern in Figure \ref{fig-dh-pg} would be written in Cypher as: 
\label{exa-cypher-match}
\begin{center}
	\small
	\begin{verbatim}
	MATCH (x1:Person) -[:acts_in]-> (:Movie {title:"Unforgiven"})
	                    <-[:acts_in]- (x2:Person)
	RETURN x1,x2
	\end{verbatim}
\end{center}

\noindent
The \texttt{MATCH} clause specifies the bgp in question. Nodes are written inside ``\verb+( )+'' brackets and edges inside ``\verb+[ ]+'' brackets. 
Filters for labels can be written after the node separated with 
a ``\verb+:+'' symbol, such that \verb+(x1:Person)+ represents a node \verb+x1+ that must match to a node labelled \verb+Person+. Specific values for properties can be specified within 
``\verb+{ }+'' brackets; for instance \verb+(:Movie {title:"Unforgiven"})+ represents a node that must match to a node labelled \verb+Movie+ and that must have value \verb+Unforgiven+ for the property 
\verb+title+. The \verb+RETURN+ clause can be used to project the output variables. Implicit projection is also allowed 
inside the pattern itself by simply omitting some of the variables; we have done this for the edges and the node with label \verb|Movie|.

Cypher implements a no-repeated-edge semantics, and thus the evaluation of this query against the movie graph of Figure \ref{fig-movies} would not include the match that sends both \verb|x1| and \verb|x2| to the node of \uriqt{Clint Eastwood} (that is, \texttt{n1}) since it would require mapping to the same edge $e_1$ twice in a single match (and likewise for the match that sends \verb|x1| and \verb|x2| to the node of \uriqt{Anna Levine}). One possibility to overcome this restriction is to use the explicit (natural) join operation of Cypher, which is invoked by simply including additional \verb+MATCH+ commands. For example, if we wanted to construct a pattern that retrieves all pairs of actors who act in the same movie, including pairs that repeat the same actor, we would use the following Cypher statement: 

\begin{center}
	\small
	\begin{verbatim}
	MATCH (x1:Person) -[:acts_in]-> (x3:Movie {title:"Unforgiven"}) 
	MATCH (x2:Person) -[:acts_in]-> (x3) 
	RETURN x1,x2
	\end{verbatim}
\end{center}

\noindent
This is equivalent to the natural join of the evaluations of the two patterns given by the two \texttt{MATCH} statements. In this case, we would also get the matches that send both \verb|x1| and \verb|x2| to the node of \uriqt{Clint Eastwood} (and likewise to the node of \uriqt{Anna Levine}).
\qed
\end{example}

If a variable \urit{x} stores a node or edge id, Cypher offers an ``\urit{.}'' operator to refer to the value of some property of \urit{x}. For instance, in our previous example we can refer to the value of the property \urit{name} for the variable \urit{x1} by using notation ``\urit{x1.name}'', and thus use ``\urit{RETURN x1.name,x2.name}" to return the actors' names (rather than their node ids).

Cypher 
supports union, difference, optional, and filter. We now provide similar example queries as for SPARQL, this time against the property graph of Figure~\ref{fig-movies}.

\begin{example}
In the following query, we use union to ask for the titles of movies that \uriqt{Clint Eastwood} either acted in or directed: 

\begin{center}
	\small
	\begin{verbatim}
	MATCH (:Person {name:"Clint Eastwood"}) -[:acts_in]-> (x3:Movie) 
	RETURN x3.title 
	UNION ALL MATCH (:Person {name:"Clint Eastwood"}) -[:directs]-> (x3:Movie) 
	RETURN x3.title 
	\end{verbatim}
\end{center}

\noindent
Both patterns will be evaluated independently and their results unioned. The ``\ttt{ALL}''  keyword indicates that duplicates should be returned; in this case, the title \uriqt{Unforgiven} will be returned twice. Omitting the ``\ttt{ALL}''  keyword, the title would appear once.\qed
\end{example}

\begin{example}
We can use difference to return the people who acted in but did not direct the movie \uriqt{Unforgiven}:
	
\begin{center}
	\small
	\begin{verbatim}
	MATCH (x1:Person) -[:acts_in]-> (x3:Movie {title:"Unforgiven"})
	WHERE NOT  (x1) -[:directs]-> (x3)
	RETURN x1.name
	\end{verbatim}
\end{center}	
	
\noindent The ``\ttt{NOT}'' keyword indicates the difference operator: any match for the initial pattern that is compatible with a match for the pattern indicated after ``\ttt{NOT}'' will be removed. In this case, \uriqt{Anna Levine} will be returned.\qed
\end{example}	

\begin{example}
We now use optional and filter to find movies in which people have acted and other ways they participated in the movie, if any.

\begin{center}
	\small
	\begin{verbatim}
	MATCH (x1:Person) -[:acts_in]-> (x3:Movie)
	OPTIONAL MATCH (x1) -[x4]-> (x3) 
	WHERE type(x4) <> "acts_in"
	RETURN x1.name AS name, x3.title AS movie, type(x4) as part
	\end{verbatim}
\end{center}

\noindent
In this query, the \ttt{WHERE} clause is a true filter expression: ``\ttt{<>}'' denotes inequality and ``\ttt{type}'' is a built-in function to return the label of an edge. The first match will retrieve all pairs of actors and movies, where the second optional match will check the other edges between each such pair matching edges where the label is not \ttt{acts\_in}. 

Cypher allows the use of the operator \urit{AS} in the \urit{RETURN} clause to indicate that the results of the query should be displayed under some specific names for columns. For instance, the use of ``\urit{x3.title AS movie}'' indicates that the values of the property \urit{title} of the nodes stored in the variable \urit{x3} will be displayed in a column with name \urit{movie}. Hence, the query in this example returns:
\begin{center}
	\small
\begin{tabular}{|l|l|l|}\hline
	\uri{name} & \uri{movie} & \uri{part}\\\hline
	\uri{Clint Eastwood} & \uri{Unforgiven} & \uri{directs}\\
	\uri{Anna Levine} & \uri{Unforgiven} & \\\hline
\end{tabular}
\end{center}

\noindent
Given that we use the optional matching functionality, we see that the result for the \urit{Anna Levine} node is preserved even though she only acted in the movie. \qed
\end{example}

Once again, here we only provide examples of the \textit{core} matching features supported by Cypher to give a flavour of the language; we refer the interested reader to the online documentation for further details~\cite{cypher}.

\medskip
\noindent \paragraph{{\em {\bf Gremlin}}} The last language we review is Gremlin: the query language of the Apache TinkerPop3 graph Framework \cite{tinkerpop}. Although 
Gremlin is also specified with the property graph model in mind, it differs quite significantly from the previous two declarative languages and has a more ``functional'' feel: while SPARQL and Cypher have obvious influences from SQL for example, Gremlin feels more like a programming language interface.\footnote{Strictly speaking, Gremlin is a functional language that includes several operators that are out of the scope of this survey. We concentrate on querying functionalities, denoted as ``graph traversals" in the documentation \cite{tinkerpop}. There are various versions of Gremlin for integration with different programming languages. Here we stick with Gremlin-Groovy.} 
Likewise, its focus is on  navigational queries rather than matching patterns; however, amongst the ``graph traversal'' operations that it defines, we can find familiar graph pattern matching features. Similarly to SPARQL, Gremlin also uses the homomorphism-based bag semantics.

\begin{example}
Intuitively, Gremlin traversals give explicit instructions as to how the graph is to be navigated. For example, to retrieve all 
movies where \uriqt{Clint Eastwood} is an actor, we first load a ``graph traversal'' object (labelled \texttt{G} here) and write:

\begin{center}
	\small
	\begin{verbatim}
	G.V().hasLabel('Person').has('name','Clint Eastwood')
	  .out('acts_in').hasLabel('Movie')
	\end{verbatim}
\end{center}

\noindent
The call \verb+G.V()+ will return the set of all nodes in the graph (\verb+V+ stands for ``vertex").  We then apply two selections on the set of nodes, where the sequence of calls \verb+G.V().hasLabel('Person').has('name','Clint Eastwood')+ retrieves precisely those nodes with label \urit{Person} and name \uriqt{Clint Eastwood}. The command \verb+out('acts_in')+ retrieves all nodes that can be reached from these latter nodes with an edge labelled \urit{acts\_in}. Finally \verb+hasLabel('Movie')+ filters nodes not labelled with \urit{Movie}. \qed
\end{example}

Gremlin is most natural when expressing paths because all such patterns can be simulated by a traversal on the graph. 

\begin{example}\label{ex:g-ce-ca}
The following Gremlin traversal allows us to obtain all co-actors of \uriqt{Clint Eastwood}: 

\begin{center}
	\small
	\begin{verbatim}
	G.V().hasLabel('Person').has('name','Clint Eastwood')
	  .out('acts_in').hasLabel('Movies')
	    .in('acts_in').hasLabel('Person')
	\end{verbatim}
\end{center}

\noindent
This query navigates through the movies of \uriqt{Clint Eastwood} as before, but then continues: the command \verb+in('acts_in')+ looks for nodes that are connected by an edge labelled \urit{acts\_in} in the opposite direction as the traversal, and then \verb+hasLabel('Person')+ again filters out any nodes that are not of label \uriqt{Person}. \qed
\end{example}

Nevertheless, Gremlin does include a way of specifying more general bgps (including branches and cycles): traversals are used to encode the structure, but nodes can be cross-referenced at different points using variables specified by means of the \verb+as+ command, while the pattern is then evaluated using the \verb+match+ command.


\begin{example} \label{ex:gremlinBGP}
To illustrate a more complex example, we show how the bgp in Figure \ref{fig-dh-pg} can be expressed in Gremlin. The following example additionally includes an explicit filter to ensure that  \texttt{x1} does not map to the same constant as \texttt{x2} in any match, and also adds a projection to return only results for the \texttt{x1} and \texttt{x2} variables (in this case returning only the co-stars, not the movie they co-starred in).


\begin{center}
	\small
	\begin{verbatim}
	G.V().match(
	  __.as('x1').hasLabel('Person').out('acts_in').hasLabel('Movies').as('x3'),
	  __.as('x3').has('title','Unforgiven').in('acts_in').hasLabel('Person').as('x2'),
	  .where('x1', neq('x2'))
	).select('x1','x2')
	\end{verbatim}
\end{center}

\noindent
Again \texttt{G.V()} returns all vertices in the graph. The \verb+match+ command then takes a list of arguments; in this case, the command takes three arguments that specify two inner traversals and a filter. The `\verb+__+' operator means that the subsequent operation is applied on the parent traversal one level up, meaning that, for example, ``\verb+__as('x1')+'' will apply over all nodes in \texttt{G.V()}. The `\verb+as+' command declares a variable; however, rather than ``\verb+__as('x1')+'' binding all nodes to variable \texttt{x1}, the entire traversal acts as a bgp, meaning that subsequent steps from a node in \texttt{x1} must be satified for the variable to match that node. Each inner traversal can thus be seen as a tree-shaped bgp. These inner traversals are then joined to create a more complex bgp that may contain cycles. In the above example, the two inner traversals are accompanied by a \verb+where+ command that calls a not-equals (\verb+neq+) filter to ensure that \ttt{x1} and \ttt{x2} are not bound to the same result. The \verb+select+ command (outside \verb+match+) then performs a projection to select the output of the query: only the co-stars, not the movie.\qed
\end{example}

While Gremlin supports bgps, filters and projection, its main focus is on navigational queries, which will be discussed in Section~\ref{sec:nav}. The current version has limited support for declarative-style operators for complex graph patterns. While a ``\texttt{union}'' command exists, and difference can be emulated by the ``\ttt{drop}'' command, the current version does not have explicit support for optional. We will not go into details but instead refer the interested reader to the online documentation~\cite{tinkerpop}.

\subsection{The complexity of evaluating graph patterns}\label{ssec:complex}

To understand the computational complexity of working with a query language we consider the following evaluation problem: given a query $Q$ in this language, a possible answer $h$ and a graph database or property graph $G$, verify whether $h$ is an answer to $Q$ over $G$; that is, verify whether $h \in Q(G)$. 
The most basic fragment of graph query languages that needs to be considered is the fragment consisting of bgps and projection, which corresponds to conjunctive queries in relational databases~\cite{AHV}. The evaluation problem for this fragment is NP-complete for the homomorphism-based semantics and the three versions of the isomorphism-based semantics considered in Section \ref{sec:bgp}; NP-hardness can be proven for the former by reduction from the graph homomorphism problem \cite{HN04}, while for the later it can be established by reduction from the subgraph isomorphism problem \cite{Ullmann76}.
On the other hand, the evaluation problem for the fragment consisting of bgps and projection can be solved in polynomial time for the case of the simulation-based semantics considered in Section \ref{sec:bgp} \cite{FLMTWW10}. All of these results hold under set or bag semantics since the question of ``$h \in Q(G)$?'' is not affected.

In practice the size of the query $Q$ is typically much smaller than the size of the database $G$, so it is common practice to assign different roles to the two when analysing query evaluation.
This motivated the introduction of the notion of {\em data complexity} \cite{Vardi82}, in which $Q$ is assumed to be fixed and the input is given by $G$ only; this is in contrast to the more general notion of {\em combined complexity}, which is defined with respect to the input query and the database (as in the previous paragraph). Under data complexity, evaluation of queries consisting of bgps and projection can not only be solved in polynomial time, but also can be carried out in logarithmic space for all the semantics considered in Section \ref{sec:bgp}
\cite{AHV}. Although data complexity might seem a bit simplistic at first sight, it has proven useful for understanding the cost of evaluating small queries over datasets of moderate size. 

Furthermore, in practice one is often interested in matching simple bgps that are not necessarily that difficult to evaluate. Both the graph theory and the database communities have dedicated vast amounts of work to identifying classes of patterns for which the matching problem can be efficiently solved, even in combined complexity. One of the main results here indicates that, intuitively speaking, the more cyclical the underlying structure of the graph pattern (i.e., the less it resembles a tree), the more difficult the query is to evaluate; this notion of cyclicity is captured formally by a notion call \textit{treewidth}, where we refer the reader to, e.g., \cite{CR00,DKV02,GGLS16} for detailed discussion.
\medskip

Going beyond the fragment covering bgps and projection, the combined complexity of the evaluation of cgps has been extensively studied for SPARQL. To recap the main results, let us first consider SPARQL under set semantics. If only projection, join, union and filter are allowed in the language, then the combined complexity of the evaluation problem remains NP-complete. If difference and optional are also allowed, then SPARQL has the same operators as relational algebra, so the combined complexity is PSPACE-complete \cite{Vardi82}. Interestingly, it can be proven that the \texttt{MINUS} operator of SPARQL can be simulated using optional, filter and join~\cite{AG08b}, so the complexity of the evaluation problem remains PSPACE-complete without \texttt{MINUS}. Moreover, the same complexity bound can be obtained if only join and optional are allowed~\cite{Schmidt0L10}, but in this case the proof is not based on an expressiveness argument.
For the case of SPARQL under bag semantics, the combined complexity of the evaluation problem remains PSPACE-complete. To the best of our knowledge, the complexity of cgps has not been studied for the cases of Cypher and Gremlin, thus opening interesting opportunities for future investigation. We further discuss open questions regarding the complexity of Cypher and Gremlin in Section~\ref{sec:final}.


\section{Navigational queries}\label{sec:nav}

While graph patterns allow for querying graph databases in a bounded manner, it is often useful to provide more flexible querying mechanisms that allow to {\em navigate} the topology of the data. One example of such a query is to find all friends-of-a-friend of some person in a social network such as the one in Figure \ref{fig:sn-sample}. Here we are not only interested in immediate acquaintances of a person, but also the people she might know through other people; namely, her friends-of-a-friend, their friends, and so on. 

Queries such as the one above are called {\em path queries}, since they require us to navigate through the graph using paths of potentially arbitrary length. Path queries have found applications in areas such as the Semantic Web \cite{ABE09,PAG10,pp-primer}, provenance \cite{HBM+08} and route-finding applications \cite{Bar00}, amongst others. Of course, sometimes paths alone are not enough, and we are interested in repetitions of graph patterns inside the graph, giving rise to {\em graph motifs} which are are often used in biological networks \cite{Sc04,BIND} to discover metabolic pathways or patterns that are often repeated \cite{Leser05}. We call all such queries {\em navigational queries}, and in this section we discuss how they can be used to query graph databases. We start with path queries.

\subsection{Path Queries} \label{ss-paths}
Paths are the most basic navigational object in a graph database. 
The most fundamental type of path query is that of \emph{path existence}, which asks if there is some directed path between two nodes in a property graph, irrespective of edge labels; in some cases, one or all such paths can be additionally returned. This is a foundational notion related to the problems of reachability and transitive closure in directed graphs \cite{YC10}, and for this reason it has been well studied by the theoretical community. However, in practice, one often needs \textit{path queries} that impose additional constraints on the path that is to be computed, such as restrictions on edge labels. The transitive friend-of-a-friend relation in social networks is such an example: we are interested in paths composed only of edges labelled with \texttt{knows} (and not \texttt{likes} or any other label). 

\paragraph{\bf Definition}
We can define a {\em path query} as having the general form 
$P = x \xrightarrow{\alpha} y$, where $\alpha$ specifies conditions on the paths we wish to retrieve and $x$ and $y$ denote the endpoints of the path. 
The endpoints $x$ and $y$ can be variables, or specific nodes, or a mix of both, or even the same node (in which case we are specifying a cycle). 
For the expression $\alpha$, we can use the symbol $*$ to signify that we are only interested in the existence of a path connecting two nodes without imposing any further constraints; otherwise, there are a variety of formalisms under which $\alpha$ can express more complex \textit{path constraints}~\cite{CruzMW87,MendelzonW89,BLLW12,CGLV03,LMV16}, but probably the most famous is that of {\em regular expressions}~\cite{0011126} defined over the set $\bL$ of edge labels. When used as a path constraint, a regular expression specifies all paths whose edge labels, when concatenated, form a word in the language of the regular expression. Intuitively speaking, regular expressions allow for concatenating paths, for applying a union/disjunction of paths, and for applying a path zero or many times. Path queries specified using regular expressions are commonly known as \emph{Regular Path Queries} (RPQs).

\begin{example} \label{ex:rpqs} 
The (transitive) friend-of-a-friend relationship in our social network can be expressed via the following regular path query (RPQ):
$$P \ := \ x \xrightarrow{ \texttt{knows}^+} y.$$
\noindent
Here the symbol `+' denotes ``one-or-more'', where the regular expression $\texttt{knows}^+$ is used to specify all paths formed from a sequence of one-or-more forward-directed edges with the label $\texttt{knows}$.\footnote{Note that $\texttt{knows}^+$ is equivalent to $\texttt{knows} \cdot \texttt{knows}^*$, where `$*$' denotes the Kleene star (zero-or-more) and `$\cdot$' denotes concatenation.} Thus the endpoints $x$ and $y$ would be matched to any two nodes in the social network connected by such a path. Similarly, we can use the path query: $$P' \ := \ x \xrightarrow{\texttt{knows}^+ \cdot \texttt{likes}} y,$$
where `$\cdot$' denotes concatenation,  to 
match nodes $x$ and $y$ such that $x$ is a person and $y$ is a post that is liked by a (transitive) friend-of-a-friend of $x$. Finally we can apply a union of paths to match the liked or disliked posts of transitive friends-of-a-friend of $x$: $$P'' \ := \ x \xrightarrow{\texttt{knows}^+ \cdot (\texttt{likes}\,\mid\,\texttt{dislikes})} y,$$ where the `$|$' symbol here denotes a union.\qed
\end{example}

The features of RPQs can be combined to (implicitly) support a number of other navigational operations on graphs. For instance, the RPQ  
$P  = x \xrightarrow{\alpha} y$, with 
%
$$\alpha \ = \ \uri{knows} \mid (\uri{knows}\cdot\uri{knows}) \mid \ldots \mid
(\underbrace{\uri{knows} \cdot \uri{knows} \cdot \ldots \cdot \uri{knows}}_{\text{$k$ times}})$$
defines the friend-of-a-friend relationship up to depth $k \geq 2$. Likewise, for example, the RPQ $x\xrightarrow{\bL^*} y$, where $\bL^*$ is the regular expression that accepts all
words over $\bL$, corresponds to the path query that imposes no constraints on paths. Regardless, we will keep 
using $x \xrightarrow{*} y$ to express this query, even when talking about RPQs.

However, there are various navigational operations not supported by RPQs that seem quite natural. RPQs are sometimes thus extended to allow further expressions. One such extension is to allow an \emph{inverse} operator $a^-$ (for $a$ in $\bL$) to specify the traversal of edges in a backwards direction, giving rise to \emph{Two-way Regular Path Queries} (2RPQs), which are RPQs enhanced with inverses \cite{CGLV02,CGLV03}.

\begin{example}\label{ex:2rpqs}
Consider now a movie database such as the one in Figure \ref{fig-movies}. The following two-way regular path query (2RPQ) retrieves 
all co-stars in the database: 
$$P \ := \ x \xrightarrow{ \texttt{acts\_in} \cdot \texttt{acts\_in}^-} y.$$
The expression \texttt{acts\_in} matches a node $x$ against a person, then the path navigates to the movies that $x$ starred in, and then backwards to $x$'s co-stars (or to $x$ itself). Similarly, we can use the path query: 
$$P' \ := \ x \xrightarrow{ (\texttt{acts\_in} \cdot \texttt{acts\_in}^-)^+} y$$
 to compute the transitive closure of the 
co-star relationship; for example, if we wished to check which actors have a finite \emph{Bacon number} \cite{bacon} -- i.e., which actors have transitively co-starred in a movie with the actor Kevin Bacon -- we could use this pattern, setting $x$ to \uriqt{Kevin Bacon} and leaving $y$ as a variable.\qed
\end{example}


The need for RPQs (and their extended forms) has been long argued by the research community \cite{BunemanDHS96,Buneman97} and recently they have been implemented in various systems; for example, extensions of RPQs form the conceptual core of ``property paths'' in the SPARQL 1.1 standard~\cite{sparql11}, which have been implemented in the newest versions of various SPARQL engines~\cite{owlimswj,virtuoso,bigdata} and have been studied by numerous authors~\cite{ACP12,LM13,FPC15,KRRV15}. Likewise in the Cypher query language~\cite{cypher}, one can find RPQ-like features. We will provide examples of the use of RPQ-like features in such languages later in this section. 

\paragraph{\bf Evaluation} To define how path queries are evaluated we need to formalise the notion of a 
path over graph databases. In a property graph $G$, a path $\pi$  is a sequence 
$n_1 e_1 n_2 e_2 n_3 \dots n_{k-1} e_{k-1} n_k$, where $k\geq 1$ and with each $e_i$ being an edge in $G$ between $n_i$ and $n_{i+1}$. 
The label of the path $\pi$, denoted $\bL(\pi)$, is the concatenation of its edge labels, namely 
$\bL(\pi) \ =\ a_1 a_2 \dots a_{k-1}$, where $a_i$ is the label of $e_i$. 
For example, 
the sequence $n_1 e_1 n_2 e_6 n_5$ is a path in the property graph of Figure \ref{fig:sn-sample}. 
The label of the path is the word $\texttt{knows} \cdot \texttt{dislikes}$. 
Note that for each node $n$ of $G$ the sequence that consists exclusively of $n$ 
is also a path (of length zero). The label of such zero-length paths corresponds to the empty word, denoted by $\epsilon$.

To define paths in edge-labelled graphs we need to be more careful since we do not have edge identifiers in this model, and thus we cannot 
give the same definition as before. Instead, we define a path $\pi$ in an edge-labelled graph $G$ 
as a sequence:
 $n_1 a_1 n_2 a_2 n_3 \dots n_{k-1} a_{k-1} n_k,$ where $(n_i,a_i,n_{i+1})$ is an edge in $G$ for all $i<k$. 
In this case the label is simply $\bL(\pi) \ =\ a_1 a_2 \dots a_{k-1}$. As in the case of property graphs, a single node $n$ forms a zero-length path with the label $\varepsilon$.

The \textit{evaluation of a path query} $P = x \xrightarrow{\alpha} y$ over $G$, denoted $P(G)$, then
consists of all paths in $G$ whose label satisfies $\alpha$. 
 For instance, if $\alpha = *$, any path belongs to $P(G)$, but if $\alpha$ is the regular expression $L$, then only paths whose label belongs to $L$ appear in $P(G)$.\footnote{From a formal point of view we can treat 2RPQs (path queries with inverses) 
as standard RPQs that are evaluated over the \emph{completion} of $G$, which is constructed by adding an edge labelled $a^-$ from $v$ to $u$ for 
each edge labelled $a$ from $u$ to $v$. Hence from now on we will consider RPQs to always contain inverses.}  The set of paths matching $P(G)$ might be infinite (when $G$ has directed cycles), and thus this general definition of evaluation is not computable. Later we will see different ways in which this definition is restricted to be implemented in practice.   
 
 \begin{example}
 \label{exa-list-paths}
Let $G$ denote the property graph of Figure \ref{fig:sn-sample} and consider the RPQ 
 $P = x \xrightarrow{\texttt{knows}^+} y$. Because of the cycle between nodes $n_1$ and $n_2$ in $G$, the 
 number of paths in $P(G)$ is infinite: it contains all finite sequences of the form 
 $n_1 e_1 n_2 e_2 n_1 e_1 \cdots$ and  $n_2 e_2 n_1 e_1 n_2 e_2 \cdots$. 
For the case of the RPQ $P' = x \xrightarrow{\texttt{knows}^+ \cdot \texttt{likes} \cdot \texttt{hasTag} } y$, the following table shows a few paths in $P'(G)$: 
  \begin{center}
  	\small
		\begin{tabular}{| l l l l l l l l l l l l l l l | }
		\hline
		$n_1$ & $e_1$ & $n_2$ & $e_7$ & $n_4$ & $e_4$ & $n_3$ & & & & & & & & \\
		$n_1$ & $e_1$ & $n_2$ & $e_2$ & $n_1$ & $e_5$ & $n_4$ & $e_4$ & $n_3$ & & & & & & \\
		$n_1$ & $e_1$ & $n_2$ & $e_2$ & $n_1$ & $e_1$ & $n_2$ & $e_7$ & $n_4$ & $e_4$ & $n_3$ & & & & \\
		$n_1$ & $e_1$ & $n_2$ & $e_2$ & $n_1$ & $e_1$ & $n_2$ & $e_2$ & $n_1$ & $e_5$ & $n_4$ & $e_4$ & $n_3$ & & \\
		$\vdots$ & & & & & & $\vdots$ & & & & & & $\vdots$ & & \\
		\hline
 		\end{tabular}
\end{center} 
The number of paths in $P'(G)$ is also infinite.\qed
 \end{example}

As in the case of graph patterns, different practical considerations -- for example, the possibility of having paths involving cycles -- give rise to different semantics for the evaluation for path queries, or more specifically, for which paths are included in $P(G)$. Next we describe the most common such forms of evaluation in practice:
\begin{enumerate}
\item {\em Arbitrary path semantics:}
All paths are considered. More specifically, all paths in $G$ that satisfy the constraints of $P$ are included in $P(G)$. As per Example~\ref{exa-list-paths}, under this semantics, $P(G)$ may contain an infinite number of paths. However, while it may not be feasible to enumerate all paths under this semantics, a user may only be interested in whether or not such a path exists, or in the (finite) pairs of nodes connected by such paths, etc., in which case such a semantics can be practical~\cite{CGLV03,Woo,BLLW12}.
 

\item {\em Shortest path semantics:} 
In this case, $P(G)$  is defined in terms of {\em shortest paths} only, i.e., paths of minimal length that satisfy the constraint specified by $P$. We may use this semantics when we want to find pairs of nodes that are linked by some path and, for each such pair, a minimal path (or set of minimal paths of equal length) that witness(es) this. In Example \ref{exa-list-paths}, the shortest path for $P'(G)$ corresponds to the first path in the table. 

\item {\em No-repeated-node semantics:} 
In this case, $P(G)$ contains all matching paths where each node appears once in the path; such paths are commonly known as \textit{simple paths}. This interpretation makes sense in some practical scenarios; e.g., when finding a route of travel, it is often not desired to have routes that come to the same place more than once. The interaction of this interpretation with RPQs has been studied in depth by the theoretical community \cite{MendelzonW89,ACP12,LM13}. In Example~\ref{exa-list-paths}, only the first path for $P'(G)$ would be selected since others mention a node more than once. 

\item {\em No-repeated-edge semantics:} 
Under this semantics, $P(G)$ contains all matching paths where each edge appears only once in the path.
The Cypher query language of the Neo4j engine currently uses this semantics (see Section 3.4.1. of the Cypher Manual \cite{cypher}). Use-cases for this semantics are similar as for the previous one; e.g., when we want to visit some place more than once, but we do not want to take the same route as before. In Example \ref{exa-list-paths}, the first two paths in $P'(G)$ have no repeated edge, but the other 
paths would not be considered. 
\end{enumerate}
         
\paragraph{\bf Output}  
As hinted at previously, a user may have different types of questions with respect to the paths contained in the evaluation $P(G)$, such as: Does there exist any such path? Is a particular path $\pi$ contained in $P(G)$? What are the pairs of nodes connected by a path in $P(G)$? What are (some of) the paths in $P(G)$? We can categorise such questions by what they return as results:

\begin{itemize} 
	\item
	{\em Boolean:} In some cases, the output of a path query may be a true/false value to ascertain, for example, if $P(G)$ is non-empty, or if there exists a path in $P(G)$ between two particular nodes, etc.
	
	\item 
	{\em Nodes:} In some applications, we are interested in the nodes connected by specific paths (see, e.g., \cite{Woo,B13}). In such cases, we project from $P(G)$ the endpoint nodes: all pairs of nodes $u$ and $v$ linked by some path in $P(G)$. Referring back to Example~\ref{exa-list-paths}, we would project from $P'(G)$ the node pair $(n_1,n_3)$.  
	
	\item
	{\em Paths:} In this case, some or all of the full paths are returned from $P(G)$. For example, if $P(G)$ is applied with a shortest-path semantics, then we would return one or more such shortest paths. In other cases, paths to be returned may be selected based on more complex
conditions, e.g., based on a ranking on paths; this may be useful in, e.g., route finding applications, 
where some top-$k$ ``best'' paths are sought.
	
	\item
	{\em Graphs:} Another solution -- for example under arbitrary path semantics -- is to offer a {\em compact} representation of the output, e.g., in the form of another graph whose paths are precisely the paths in the output of the query \cite{BLLW12}.
\end{itemize} 

While the first two types of answers can be handled under, e.g., a standard relational algebra, there is currently no consensus on how to represent paths as the output of a query. 
In particular, unlike solutions to graph patterns that have a fixed-arity output, paths do not have a fixed-arity, therefore we cannot directly define a mapping from variables to constants as in the case of a bgp match. 
Likewise, although returning graphs as queries is supported in SPARQL~\cite{sparql11} through \texttt{CONSTRUCT}, graph creation is only supported as a final step, where such graphs cannot be manipulated further by other operators. 

\paragraph{\bf Sets vs. bags} In the case of queries that return a boolean value or a graph as a result, there is no distinction between bag or set semantics. Likewise, in the case that full paths -- i.e., the complete sequence of nodes and edges in each path -- are returned, no duplicates can occur and there is no such distinction. However, if nodes are returned, or nodes/edges are projected from a full path, then bag semantics are distinguished from set semantics. In particular, if we consider the case where we are returning end nodes of our path as output, when using set semantics, a pair $(n,n')$ will be returned exactly once when there is at least one path in $P(G)$ connecting $n$ with $n'$, and zero times otherwise; when using bag semantics, this now changes, and a pair $(n,n')$ is returned once for each full path in $P(G)$ connecting $n$ with $n'$. 

Bag semantics combined with arbitrary path semantics is problematic since the set of paths can be infinite; thus this combination is usually not considered in the theoretical literature \cite{Woo,B13}. But even when the number of paths is guaranteed to be finite, there are still several issues with respect to high computational complexity since bag semantics implicitly requires counting paths. For example, it is well-known that counting the number of paths without repeated nodes from node $a$ to node $b$ in a graph $G$ is a \#P-complete problem~\cite{V79}, which implies that it is as difficult as, for example, counting the number of satisfying assignments of a propositional formula, or counting the number of Hamiltonian cycles in a graph.

This high computational complexity has a number of practical consequences. For instance, the initial combination of bag semantics with property paths in drafts of the SPARQL 1.1 standard required that the number of repetitions of a pair of nodes in the answer was equal to the number of paths between them. Thus, a restriction to consider simple paths was added to guarantee finiteness of results. Unfortunately, this gave rise to a path counting problem with a very high complexity~\cite{ACP12,LM13}, which was resolved by imposing a set semantics on property paths of the form \texttt{(p)*} and \texttt{(p)+}, avoiding the counting of paths of unbounded length. On the other hand, Cypher maintains a bag semantics when returning nodes, where a no-repeated-edge semantics is applied by default.

\subsection{Adding paths to basic graph patterns}

Now that we understand how path queries can be used to match paths and how graph patterns can be used to match sub-graphs, we can combine them to produce a powerful query language that allows
to find more flexible matches. In particular, this language allows to express that 
 some edges in a graph pattern should be replaced by a path (satisfying certain conditions) 
 instead of a single edge. 
 
\begin{figure} 
\begin{center}
\resizebox{0.9\linewidth}{!}{
		\begin{tikzpicture}[->,>=stealth',auto,
		thick, scale = 1.0]

		\tikzstyle{rectl}=[rt,minimum width=5mm,minimum height=5mm,inner sep=1ex]
		\tikzstyle{erectl}=[rectl,dashed]
		
		\node [rectl] (x) {\uri{$x$}};
		\node [rectl,right = 5cm of x] (y){\uriq{Kevin Bacon}};
		\node [rectl,left = 5cm of x] (z) {\uriq{Paul Erd\H{o}s}};

		\draw[arrout] (x) to node[ertt,anchor=center] (e1) {\al{80pt}{ \ensuremath{(\uri{acts\_in}\cdot\uri{acts\_in}^-)^+}}} (y);
		
		\draw[arrout] (x) to node[ertt,anchor=center] (e2) {\al{80pt}{\ensuremath{(\uri{author}\cdot\uri{author}^-)^+}}} (z);

		\end{tikzpicture} 
}		
\end{center}
\caption{A query finding the actors with a finite Erd\H{o}s--Bacon number over an edge-labelled graph} \label{fig:baconerdos} 
\end{figure}
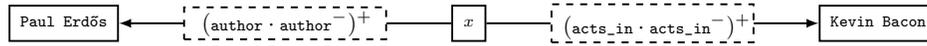

 \begin{example} \label{ex:crpq} 
In Example \ref{ex:rpqs}, we used the query $Q' =  x \xrightarrow{ (\texttt{acts\_in} \cdot \texttt{acts\_in}^-)^+} y$ to find actors that are connected through co-star relations to other actors, and mentioned that this query can be used to find actors with a finite Bacon number. To make our example more challenging, consider now that our movie database from Figure \ref{fig-movies} is extended to also contain bibliographical information about scientific papers and their authors. In such a database, each node is either a movie, a person, or an article. Persons and movies are connected as in Figure \ref{fig-movies}, while a person can also have an \texttt{author} edge connecting it to an article. In such a database we might be interested in finding people with finite Erd\H{o}s--Bacon number, that is, people who are connected to Kevin Bacon through co-stars relations and are connected to Paul Erd\H{o}s through co-authorship relations. This is easily expressed using the query in Figure \ref{fig:baconerdos}, which is a basic graph pattern that permits (two-way) regular path queries on edges. \qed 
\end{example} 

\

Combining path queries with basic graph patterns (bgps) gives rise to {\em navigational graph patterns} (ngps). 
In the case of edge-labelled graphs, ngps are defined similarly as bgps: namely, they are edge-labelled graphs where nodes can be constants or variables, 
and the edge labels can be constants, variables, RPQs\footnote{In the context of ngps we identify the expression defining an RPQ with the RPQ itself.}, 
or the special symbol $*$ denoting an arbitrary path.
Matches are defined as in the case of bgps, but now every edge {\it not} labelled with a variable is mapped to a path. 
That is, if we have $(b,\alpha,c)$ in our ngp, with $\alpha$ either $*$ or a regular expression, 
then our match $h$ must satisfy that $h(b)$ is connected to $h(c)$ by a path in $P(G)$, 
with $P$ being the path query $x \xrightarrow{\alpha} y$. 
Note that in order to keep the arity of matchings bounded by the size of the query, 
we are opting for an {\em existential} interpretation of path expressions in ngps. 
That is, we are considering the boolean output semantics for $P$, which only checks that there is a path in $P(G)$ connecting the nodes $h(b)$ and $h(c)$, but does not return such a path.
Navigational graph patterns for property graphs are defined analogously, but now allowing for elements of property graphs in nodes and edges as per Definition \ref{def-pg}. In particular, if the label $\alpha$ of the edge is $*$ or a regular expression, the end nodes of this edge have to be in the answer to the path query $x \xrightarrow{\alpha} y$ over $G$.

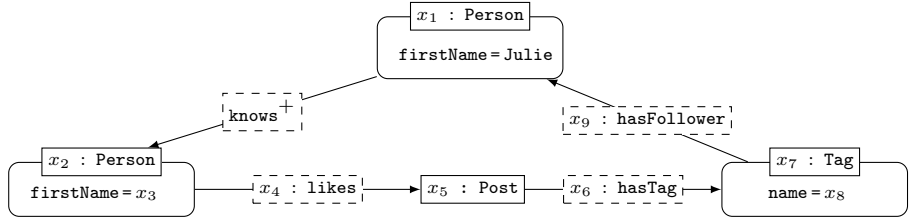
\begin{figure} 
\begin{center}
		\begin{tikzpicture}
		\node[rect] (n1) {
			\alt{60pt}{
				\urie{firstName} & \uri{Julie}}};
		
		\node[rt] (ln1) at (n1.north) {\uri{$x_{1}$ $:$ Person}};
		
		\node[rect, below=1.1cm of n1,xshift=-4.9cm] (n2) {
			\alt{60pt}{ 
				\urie{firstName} & \uri{$x_{3}$}}};
		
		\node[rt] (ln2) at (n2.north) {\uri{$x_{2}$ $:$ Person}};
		
		\draw[arrout] (n1) to 
		node[below, erectw] (e1) {} 
		(ln2);
		
		\node[ert] (le1) at (e1.north) {\uri{knows}$^+$};

		\node[rt,right=3cm of n2] (n4) {\uri{$x_{5}$ : Post}} ;
		
		\draw[arrout] (n2) to
		node[ert] (e5) {\uri{$x_{4}$ $:$ likes}
			}
		(n4);

		\node[rect,right=7cm of n2] (n5) {
			\alt{60pt}{ 
				\urie{name} & \uri{$x_8$}}};
		
		\node[rt] (ln5) at (n5.north) {\uri{$x_{7}$ $:$ Tag}};

		\draw[arrout] (n4) to
		node[ert] (e6) {\uri{$x_{6}$ $:$ hasTag}
			}
		(n5);

		\draw[arrin] (n1) to
		node[ert] (e6) {\uri{$x_{9}$ $:$ hasFollower}
			}
		(n5);

		\end{tikzpicture}
\end{center}
\caption{A navigational graph pattern that characterises the friends of friends of Julie that like a post with a tag she that she follows}
\label{fig:julie} 		
\end{figure}

\begin{example}\label{ex-ngp}
Coming back to the social network from Figure \ref{fig:sn-sample}, we might be interested in finding all friends of friends of Julie that liked a post with a tag that Julie follows. The navigational graph pattern in Figure \ref{fig:julie} 
expresses this query over the property graph of Figure \ref{fig:sn-sample}. \qed
\end{example} 

Navigational graph patterns have received a lot of attention in the theoretical literature
under the name 
{\em conjunctive regular path queries} (CRPQs) \cite{ConsensM90,FLS98,CGLV03,BLR14}. 
A natural extension of ngps is to consider \textit{complex navigational graph patterns} (cngps) by taking the closure of ngps under the relational operations of selection, 
projection, join, union, difference, and optional, as presented in Section \ref{gp}. 
Some other variants and extensions of cngps allowing to compare different paths in a graph have also been considered in the past \cite{BLLW12,BM14,FL15}. As we will see later in Section~\ref{ss:nav_practice}, cngps then form the core of languages such as SPARQL.

\begin{example}\label{ex-cngp}
	To give a brief idea of the expressivity of cngps, consider the ngp of Example~\ref{ex-ngp} and assume we project $x_5$: the ids of the posts liked by friends-of-friends of Julie and that have a tag that she follows. Let's call these results the ``recommended posts'' for Julie. Now consider a copy of the same pattern to find the recommended posts for John. We could use the union of these patterns to find posts recommended for Julie or John
, or intersection to find posts recommended for both, or difference to find posts recommended for Julie but not John, or filter dates to find more recent posts, and so forth. All such queries can then be expressed as cnpgs.\qed
\end{example}

\subsection{Repetition of patterns}
\label{subsec-repetition-patterns}

For the navigational languages we have seen thus far, paths are the only form of recursion allowed, but to express certain types of queries, we may require more expressive forms of recursion. Imagine for instance that as before we wish to check for all pairs of actors in our movie database that are connected by co-stars relations, but we only want to consider actors that have directed a movie (such as Clint Eastwood). We cannot express this query by a regular expression over paths since, aside from finding paths between co-stars, we need to check that each intermediate node in the path has an outgoing edge labelled \urit{directs}. In this section, we present several languages that can express these types of queries, and explain how this can be achieved. 


\paragraph{\bf Nested regular expressions}
The language of {\em nested regular expressions} (NREs) extends RPQs with a {\em branching} or {\em nesting} operator that allows to recursively check other nested RPQs over the nodes of a path. As such, the evaluation of an NRE consists of paths where nodes have a potentially branching path that satisfies the given nested RPQ. Conceptually speaking, NREs thus allow for capturing paths matched by a tree-shaped pattern, offering an increase in expressive power that has been applied in practice, for example, to form the basis of proposed navigational query languages for RDF \cite{PAG10,BPR12}. 

\begin{figure}[t]
	\centering
	\subfigure[Base for repetitions]{
\begin{tikzpicture}[>=stealth]
\filldraw [black] 

(0,0) circle (2pt) 
(1,0) circle (2pt) %
(2,0) circle (2pt) 
(2,1) circle (2pt);%

\draw[->,thick] (0.1,0) -- (0.9,0);
\draw[<-,thick] (1.1,0) -- (1.9,0);
\draw[->,thick] (2,0.1) -- (2,0.9);

%
%
%
\draw (0,0) node[left=1pt] {$x$};
\draw (2,0) node[right=1pt] {$y$};
\end{tikzpicture}
		\label{fig-nre-a}
	}
	\hfill
	\subfigure[Transitive closure of the base pattern]{
		
\begin{tikzpicture}[>=stealth]
\filldraw [black] 

(0,0) circle (2pt) 
(1,0) circle (2pt) %
(2,0) circle (2pt) 
(2,1) circle (2pt)%

(3,0) circle (2pt)%
(4,0) circle (2pt)%
(4,1) circle (2pt)%

(6,0) circle (2pt)%
(7,0) circle (2pt)%
(8,0) circle (2pt)%
(8,1) circle (2pt);

\draw[->,thick] (0.1,0) -- (0.9,0);
\draw[<-,thick] (1.1,0) -- (1.9,0);
\draw[->,thick] (2,0.1) -- (2,0.9);

\draw[->,thick] (2.1,0) -- (2.9,0);
\draw[<-,thick] (3.1,0) -- (3.9,0);
\draw[->,thick] (4,0.1) -- (4,0.9);

\draw[->,thick] (6.1,0) -- (6.9,0);
\draw[<-,thick] (7.1,0) -- (7.9,0);
\draw[->,thick] (8,0.1) -- (8,0.9);

%
%
%
\draw (0,0) node[left=1pt] {$x$};
\draw (8,0) node[right=1pt] {$y$};

\draw (5,0) node {$\cdots$};

\end{tikzpicture}
		
		\label{fig-nre-b}
	}
	\caption{Base of an NRE and the transitive closure over this base. We assume all horizontal edges in the above images to be labelled with \texttt{acts\_in}, and all vertical edges with \texttt{directs}.}
	\label{fig-nre}
\end{figure}
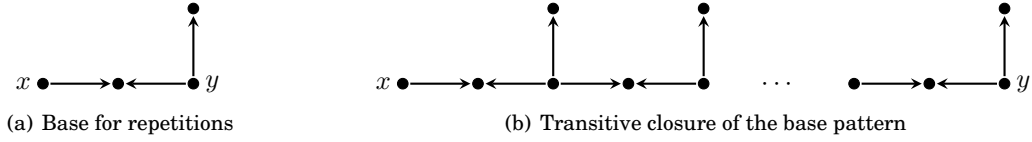

\begin{example}In the language of NREs, we can restrict our co-star paths to only consider directors using the following expression: 
$$x \xrightarrow{\big(\texttt{acts\_in} \cdot \texttt{acts\_in}^- \, [\texttt{directs}]\big)^+}  y\,.$$ 
This query asks for a path whose label belongs to the regular expression (with inverse) 
$(\texttt{acts\_in} \cdot \texttt{acts\_in}^-)^+$, but imposes an additional condition: every intermediate node captured by the sub-expression $\texttt{acts\_in} \cdot \texttt{acts\_in}^-$ must have an outgoing edge labelled $\texttt{directs}$. More generally, the latter bracketed expression is an RPQ that is used as an existential branching test on the preceding sub-expression, checking to ensure that each matched node is connected to some other node by the given bracketed expression. Note that the above pattern does not check that the start node is a director.

This recursive pattern is defined by the structure depicted in Figure~\ref{fig-nre}: one can also think of this structure as taking the base pattern from Figure~\ref{fig-nre-a} and applying it recursively as illustrated in Figure~\ref{fig-nre-b}.\qed
\end{example}



Just as we did with regular path queries, one can consider conjunctions of such patterns to arrive at the language of \emph{conjunctive nested regular expressions} (CRNEs), which has thus far only been studied in theory 
\cite{BPR13,BCOS14}. 
Another direction to extend NREs is to add more expressive features such as negation and unary formulas. By doing so one arrives at a language that is equivalent to applying XPath~\cite{xpath} over graph databases. In fact, as shown by \citeANP{LMV16}~\citeyear{LMV16}, NREs themselves correspond to a positive fragment of XPath.

\paragraph{\bf Regular data path queries}
\label{ss:values} 


While considering NREs, it is perhaps natural to consider how similar such patterns could be applied to property graphs, and in particular, to test the values of various node and edge attributes appearing along the path that one is traversing. To illustrate the issue, consider the following example.

\begin{example}\label{ex:datapaths} Coming back to our social network, recall that in Example \ref{ex:rpqs} we showed how to compute the friend-of-a-friend relation using the expression $\texttt{knows}^+$. Assume we now want to again compute the friend-of-a-friend relation, but we wish to consider only the people who live in the same country: each time we traverse an edge labelled \texttt{knows}, we need to check that the value of the country attribute is equal for both nodes connected by the edge. This can be expressed as follows:
	$$e := ([\texttt{knows}]_{\texttt{start.country=end.country}})^+$$

\noindent
where the filter \texttt{start.country=end.country} checks the aforementioned condition on all pairs of nodes connected by the \texttt{knows} edges that form the path. Note that unlike NREs, here we can express \textit{comparisons} between the values of attributes. Also note that the brackets $[]$ have different meaning in NREs as opposed to RDPQs. Namely, in the former they apply to nodes, and in the latter to the start/end point of a path the expression between the brackets defines.\qed
\end{example}

Expressions such as $e$ above can be formalised by extending the grammar of the ordinary regular expressions with the operator $[exp]_c$, where $exp$ is an expression, and $c$ is a filter of the form $\texttt{start.atr=end.atr}'$, with \texttt{atr} and \texttt{atr}$'$ being attribute names. The full grammar is then given by the following $e := \ a \ ; \ e\cdot e \ ; \ e\,|\,e \ ; \ e^* \ ; \ [e]_c,$ with $c$ a conjunction of expressions of the form $\texttt{start.atr} = \texttt{end.atr}'$ or $\texttt{start.atr}\neq \texttt{end.atr}'$. Allowing any such expression $e$ inside a path query $x \xrightarrow{e} y$ gives rise to {\em regular data path queries} (RDPQs), with the name signifying that paths consider not only navigational aspects, but also reason about the data stored in the graph.

Although queries that allow reasoning about how the attribute values change along paths seem to be relevant in practical applications, they seem to be poorly supported in existing systems. On the other hand, they did receive some attention in the theoretical literature. For instance, the base language for regular data path queries was introduced in \cite{LV12}, and some further extensions allowing first order reasoning over paths \cite{HKBZ13}, or unlimited use of variables \cite{LMV16,BFL13} have also been considered. However, since there is still no clear consensus on the correct language for this task, this seems to be a promising area of future work, both with respect to the theoretical issues, and with respect to the correct techniques for implementing such queries in graph database systems.

\paragraph{\bf Datalog variants}
Thus far all recursive navigational expressions we have considered are based on paths (e.g., RPQs) or trees (e.g., NREs). 
So what happens when we consider more general queries which look for repetitions of arbitrary bgps?
It turns out that such queries can 
typically 
be expressed in \emph{Datalog}-like languages \cite{AHV}, which correspond to powerful recursive languages 
based on rules. 

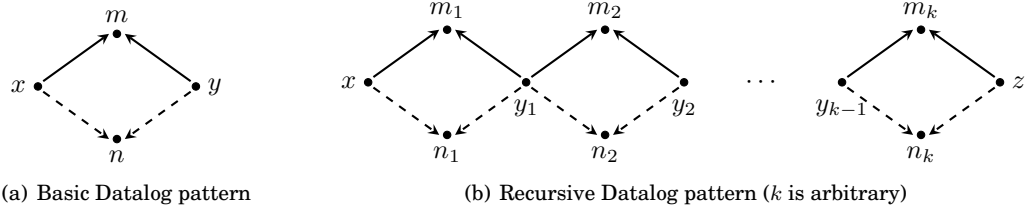
\begin{figure}[t]
	\centering
	\subfigure[Basic Datalog pattern]{
\begin{tikzpicture}[>=stealth, scale=0.7]
\filldraw [black] 

(0,0) circle (2pt) 
(3,0) circle (2pt) 

(1.5,1) circle (2pt) 
(1.5,-1) circle (2pt);

\draw[->,thick] (0.1,0.1) -- (1.35,1);
\draw[->,thick,dashed] (0.1,-0.1) -- (1.35,-1);
\draw[->,thick] (2.9,0.1) -- (1.65,1);
\draw[->,thick,dashed] (2.9,-0.1) -- (1.65,-1);

%
%
%
\draw (0,0) node[left=1pt] {$x$};
\draw (3,0) node[right=1pt] {$y$};
\draw (1.5,1) node[above=1pt] {$m$};
\draw (1.5,-1) node[below=1pt] {$n$};
\end{tikzpicture}
\hspace{0.1cm}
		\label{fig-datalog-a}
	}
	\hfill
	\subfigure[Recursive Datalog pattern ($k$ is arbitrary)]{
		
\begin{tikzpicture}[>=stealth, scale=0.7]
\filldraw [black] 

(0,0) circle (2pt) 
(3,0) circle (2pt) 

(1.5,1) circle (2pt) 
(1.5,-1) circle (2pt)

(6,0) circle (2pt) 

(4.5,1) circle (2pt) 
(4.5,-1) circle (2pt)

(9,0) circle (2pt) 
(12,0) circle (2pt) 

(10.5,1) circle (2pt) 
(10.5,-1) circle (2pt);

\draw[->,thick] (0.1,0.1) -- (1.35,1);
\draw[->,thick,dashed] (0.1,-0.1) -- (1.35,-1);
\draw[->,thick] (2.9,0.1) -- (1.65,1);
\draw[->,thick,dashed] (2.9,-0.1) -- (1.65,-1);

\draw[->,thick] (3.1,0.1) -- (4.35,1);
\draw[->,thick,dashed] (3.1,-0.1) -- (4.35,-1);
\draw[->,thick] (5.9,0.1) -- (4.65,1);
\draw[->,thick,dashed] (5.9,-0.1) -- (4.65,-1);

\draw[->,thick] (9.1,0.1) -- (10.35,1);
\draw[->,thick,dashed] (9.1,-0.1) -- (10.35,-1);
\draw[->,thick] (11.9,0.1) -- (10.65,1);
\draw[->,thick,dashed] (11.9,-0.1) -- (10.65,-1);

%
%
%
\draw (0,0) node[left=1pt] {$x$};
\draw (3,0) node[below=3pt] {$y_1$};
\draw (1.5,1) node[above=1pt] {$m_1$};
\draw (1.5,-1) node[below=1pt] {$n_1$};
\draw (6,0) node[below=3pt] {$y_2$};
\draw (4.5,1) node[above=1pt] {$m_2$};
\draw (4.5,-1) node[below=1pt] {$n_2$};
\draw (9,0) node[below=3pt] {$y_{k-1}$};
\draw (10.5,1) node[above=1pt] {$m_k$};
\draw (10.5,-1) node[below=1pt] {$n_k$};
\draw (12,0) node[right=1pt] {$z$};

\draw (7.5,0) node {$\cdots$};

\end{tikzpicture}
		
		\label{fig-datalog-b}
	}
	\caption{Illustration of the types of patterns a Datalog programs can query. Here all solid edges are labelled \texttt{acts\_in} and all dashed edges are labelled \texttt{directs}.}
	\label{fig-datalog}
\end{figure}

\begin{example} \label{ex:datalog} 
To exemplify how this works, let us focus on edge-labelled graphs (a similar translation can be devised for property graphs). Now instead of considering actors that are connected simply on merit of having co-starred in a movie, let us 
add the constraint that they must additionally direct a movie 
together (possibly a different movie). Let us call a pair of actor--directors connected (directly) in such a fashion ``peers''. 
Taking an edge-labelled graph $G$ (in the style of Figure~\ref{fig:graphdb}), we can 
create a query for peers as follows: $Q = (V,E)$, where $V = \{x,y,m,n\}$ are variables and where 
$E$ contains $(x, \texttt{acts\_in}, m)$, $(x, \texttt{directs}, n)$, 
$(y, \texttt{acts\_in}, m)$ and $(y, \texttt{directs}, n)$.


%
To express this in Datalog we adopt the convention that the relation $\textsf{E}(x,y,z)$ encodes an edge $(x,y,z)$ in an edge-labelled graph $G = (V,E)$. 
We can then represent the original bgp $Q$ as the following Datalog rule:
\begin{align*} 
\mathsf{Q}(x,y) \leftarrow  
\textsf{E}(x, \texttt{acts\_in}, m), \textsf{E}(x, \texttt{directs}, n), 
\textsf{E}(y, \texttt{\texttt{acts\_in}}, m), \textsf{E}(y, \texttt{directs}, n)\,.
\end{align*}


\noindent
Applying this rule 
generates a binary relation $\textsf{Q}$ that contains precisely the matches of bgp $Q$ over $G$; in other words, we can quite easily represent a bgp as a Datalog rule and evaluate it as such.

Let us assume we now wish to find all nodes connected recursively through a peer relation. We can add the following rule:

\begin{center}
	\small%
	$\textsf{Q}(x,z) \leftarrow \textsf{Q}(x,y), \textsf{Q}(y,z)\,.$
\end{center}

\noindent
Applying these two Datalog rules in a recursive fashion 
generates an output $\textsf{Q}$ that contains the transitive closure over peers. 

More importantly -- as illustrated in Figure~\ref{fig-datalog} -- the base pattern of Figure~\ref{fig-datalog-a} is not a path nor a tree, and hence the resulting recursive pattern of Figure~\ref{fig-datalog-b} achieved by these two Datalog rules would not be expressible in any language we discussed earlier: with Datalog, the recursive pattern can be an arbitrary bgp.\qed
\end{example}

In a manner analogous to returning paths for RPQs, one could consider trying to return a similar result for Datalog, but where instead of having sequences of nodes connected by edges in the case of RPQs, we would, intuitively speaking, have something more like sequences of sub-graphs in the case of Datalog. However, since the output of applying Datalog rules is a set of fixed-arity relations, it is not possible to return such a sequence; in fact, how to represent the structures that Datalog navigates is an unexplored area. Instead, Datalog rules can be applied to find pairs of nodes that are connected in such a manner, or to generate a relational representation of a graph that contains all such edges navigated, and so forth. 

There have been attempts to define Datalog-like languages that are specifically tailored to the requirements of graph database applications, in the spirit of the recursive rules 
used in Example \ref{ex:datalog}. The first of these was GraphLog \cite{ConsensM90}, which was designed for querying graphs formed by hypertext documents. More recently, \citeN{RRV15} studied the restriction of Datalog where recursion is only allowed over patterns that output at most two variables; in fact, a number of languages have been proposed in different settings with similar expressive power \cite{belgas,LRV13,RK13,AGP14,BKR15}. 
There have also been attempts to implement query engines that support these languages, specifically over RDF datasets using extensions of SPARQL \cite{RSV15,PSL15}. 
Recently we have also witnessed proposals that combine user-defined functions into Datalog to obtain a graph query language more tailored for graph analytic tasks (see e.g. \cite{socialite}).  

In summary, the use of Datalog-like languages for querying graphs is an active area of research being explored from a number of angles. However, as we discuss in the following section, recursively applying bgps in a declarative manner is not widely supported within the practical query languages we consider in this survey.
 
\subsection{Navigational queries in practice}
\label{ss:nav_practice}
Next we show examples of how navigational queries can be expressed in practical query languages. As before we illustrate this using SPARQL, Cypher and Gremlin.
 
\paragraph{\bf SPARQL} Since version 1.1~\cite{sparql11}, SPARQL permits the use of \textit{property paths}, which are an extended form of regular expression that, beyond usual RPQs, also allow inverses and a limited form of negation~\cite{KRRV15}. As a consequence, we can express any path query from Example~\ref{ex:2rpqs} using SPARQL~1.1.
 
\begin{example}\label{ex:sp_pp}
Consider the RDF graph depicted in Figure~\ref{fig:rdf}. To find all pairs of actors who have finite collaboration distance (i.e. the query $Q'$ from Example \ref{ex:rpqs}) we can use the following SPARQL query:
 	
\begin{center}
\footnotesize
\begin{verbatim}
	SELECT ?x ?y
	WHERE { ?x (:acts_in/^:acts_in)* ?y }
\end{verbatim}
\end{center}

\noindent
Here the symbol `$\verb|/|$' is used to denote concatenation and `$\verb|^|$' to denote the inverse of an edge label. The Kleene closure is given by `$\verb|*|$' as before. Note that if we wanted to extract the actors with a finite Bacon number from our graph database we can just replace the variable \texttt{?x} with the constant \texttt{:Kevin\_Bacon}. \qed 
 \end{example}
 

In one aspect, SPARQL goes beyond RPQs and allows for a (very) limited form of negation called negated property sets~\cite{KRRV15}. This is done by allowing subexpressions of the form $!\{e_1,\ldots,e_n\}$ inside property paths, which will match to all pairs of nodes connected by some edge whose label is not in the set $\{e_1,\ldots,e_n\}$. Apart from ordinary labels, negated property set can also include inverse-edge labels.

%
%

\begin{example}\label{ex:sp_nps}
	Consider the RDF graph depicted in Figure~\ref{fig:rdf} and the following SPARQL query with a negated property-set
	
	\begin{center}
		\footnotesize
		\begin{verbatim}
	SELECT ?y
	WHERE { :Clint_Eastwood (!{:type,:directs})* ?y }
		\end{verbatim}
	\end{center}

\noindent	
This query will match \texttt{:Unforgiven} (the IRI) and \texttt{"Unforgiven"} (the title string)	for \texttt{?y}. Here, \texttt{:Anna\_Levine} is not included since the negated property-set does not include any inverse. However, once any inverse is added, then inverse edges are included:

	\begin{center}
		\footnotesize
		\begin{verbatim}
	SELECT ?y
	WHERE { :Clint_Eastwood (!{:type,:directs,^directs})* ?y }
		\end{verbatim}
	\end{center}

\noindent
This query will additionally return \texttt{:Anna\_Levine} since now inverse edges are also traversed. In a similar manner to the first query, if the negated property set only includes inverses, then only inverse edges are traversed. \qed
\end{example}

Adding this limited form of negation to the RPQ-style features of property paths does not affect the complexity of SPARQL query evaluation~\cite{KRRV15}.
\medskip
 
As aforementioned in the discussion on set vs. bag semantics in Section~\ref{ss-paths}, in a draft of the SPARQL 1.1 standard, the original semantics of property paths was based on simple paths with a bag semantics. However, since it was shown that such a semantics quickly renders query evaluation impractical \cite{ACP12,LM13}, the semantics was changed. Now, in order to evaluate any query containing the transitive closure operator ($\verb|*|$ or $\verb|+|$), SPARQL uses a set semantics, looking for pairs of nodes connected by any path whose label belongs to the language of the regular expression specifying the query. Otherwise, if a property path can be rewritten as a bgp (with projection), SPARQL instead uses the bag semantics defined for bgps (see \cite[\S 9.3]{sparql11} for more details).
 
 
 
 
 %
 
 
Similarly, SPARQL can also express navigational graph patterns (ngps).
 
\begin{example}\label{ex:sp_ngp}
The ngp from Example \ref{ex:crpq} -- find all people with a finite Erd\H{o}s--Bacon number -- can be expressed in SPARQL as:
 	
\begin{center}
\footnotesize
	\begin{verbatim}
	SELECT ?x
	WHERE { ?x (:acts_in/^:acts_in)* :Kevin_Bacon . ?x (:author/^:author)* :Paul_Erdos . }
 	\end{verbatim}
 \end{center}
 	
\noindent
This query is a conjunction of two RPQs, where the symbol $.$ denotes conjunction. \qed
\end{example}
 
Likewise, SPARQL can express complex navigational graph patterns (cngps).

\begin{example}\label{ex:sp_cngp}
Referring back to Example~\ref{ex-cngp}, we can express an RDF version of the query for the posts recommended to Julie but not to John as follows:

\begin{center}
	\footnotesize
	\begin{verbatim}
	SELECT ?x
	WHERE { 
	 {
	  { :Julie :knows+/:likes ?x . ?x :hasTag/:hasFollower :Julie . }
	   MINUS
	  { :John :knows+/:likes ?x . ?x :hasTag/:hasFollower :John . }
	}
	\end{verbatim}
\end{center}

\noindent
This query involves the difference of two cgps, creating a cngp. \qed
\end{example}
 
Finally, although SPARQL cannot express iterations of navigational patterns such as for instance NREs, several extensions capable of doing this have been proposed. These include Datalog-based RDF languages such as RDFox \cite{NenovPMHWB15}, extended property paths \cite{AE14}, or SPARQL extended with a recursion operator using \texttt{CONSTRUCT}
 \cite{RSV15}.
 
\paragraph{\bf Cypher} While not supporting full regular expressions, Cypher still allows transitive closure over a single edge label in a property graph. On the other hand, since it is designed to run over property graphs, Cypher also allows the star to be applied to an edge property/value pair; however, this is again limited to a single repeated label/value. 
 
\begin{example}\label{ex:cy_p}
To compute the friend-of-a-friend relation in Cypher over the graph from Figure~\ref{fig:sn-sample}, we can use the following expression: 
 	
\begin{center}
	\footnotesize
	\begin{verbatim}
	MATCH (x1:Person) -[:knows*]-> (x2:Person)
	RETURN x1,x2 
	\end{verbatim}
\end{center}
 	
\noindent
This expression selects pairs of nodes that are linked by a path completely labelled by \texttt{knows}. To do this, it applies the star operator $*$ over the label \texttt{knows}. \qed
\end{example}
 
Currently Cypher does not allow to apply the recursive operator $*$ over more complex expressions; thus, for example, we are not able to query for actors with a finite Bacon number over the property graph from Figure \ref{fig-movies} (without changing the data to, e.g., give explicit co-star relations). This might change, however, in the near future. 
 
Recall that Cypher uses the no-repeated-edge semantics for cgps; by default, Cypher uses the same semantics for path queries, thus returning all pairs of nodes connected by a path which does not repeat any edges. In fact, Cypher uses a bag semantics, so each pair of nodes will be duplicated for every such path connecting them in the data.

\begin{example}\label{ex:bag_cyp}
Consider the graph from Figure~\ref{fig:sn-sample} and the following query looking for any path (of arbitrary length) between two nodes:
\begin{center}
	\footnotesize
	\begin{verbatim}
	MATCH (x1) -[*]-> (x2)
	RETURN x1,x2 
	\end{verbatim}
\end{center}

\noindent
Here the operator \texttt{*} signifies that the path is of arbitrary length and there is no restriction on edge labels.
The output of this query will contain the pair $(n_1,n_4)$ twice, as there are two distinct paths (that do not repeat an edge) from the node $n_1$ representing \texttt{Julie}, to the node $n_4$ representing the post with the content \texttt{I love U2}. \qed
\end{example} 

However, Cypher also allows for returning a single shortest path connecting two nodes, or all shortest paths connecting them, allowing the user to declaratively change the semantics for evaluating paths within the query.

\begin{example}If we wanted to find friends of friends of Julie in the example above and return only the shortest witnessing path, we could use the following query:
	
	\begin{center}
	\footnotesize
	\begin{verbatim}
	MATCH ( julie:Person {firstname:"Julie"} ), 
	p = shortestPath( (julie) -[:knows*]-> (x:Person) )
	RETURN p
	\end{verbatim}
	\end{center}
	
\noindent
This will return a single shortest witnessing path. If we wanted to return all shortest paths, we could replace ``\texttt{shortestPath}'' with ``\texttt{allShortestPaths}''. \qed
\end{example}

In Section \ref{gp} we have seen how to specify basic graph patterns using Cypher. A restricted form of navigational patterns -- only allowing the star operator on edge labels -- are then supported by allowing path expressions inside basic patterns.
 
\begin{example}\label{ex:cy_ngp}
Coming back to the social network from Figure \ref{fig:sn-sample}, if we want to find all friends-of-friends of Julie that liked a post with a tag that Julie follows, we can use the following Cypher query:
 	
\begin{center}
	\footnotesize
	\begin{verbatim}
	MATCH (x1:Person {firstName:"Julie"}) -[:knows*]-> (x2:Person)
	MATCH (x2) -[:likes]-> () -> [:hasTag] -> (x3)
	MATCH (x3) -[:hasFollower]-> (x1)
	RETURN x2
	\end{verbatim}
\end{center}
 	
\noindent
The first \texttt{MATCH} clause provides a path expression, which when joined with the bgps expressed in the latter two \texttt{MATCH} clauses, forms a navigational graph pattern (ngp). In fact, the query is an abbreviated version of the ngp depicted in Figure~\ref{fig:julie}. \qed
\end{example}
 
Apart from (a restricted form) of RPQs and (c)ngps, Cypher also offers several unique features that make it useful when working with property graphs. First, Cypher allows for specifying the length of the path. For instance, in Example \ref{ex:cy_p} we can change the edge-label constraint \texttt{[:knows*]} to \texttt{[:knows*2..7]} to specify that the path must traverse at least two and at most seven edges. Although this property is syntactic and can be simulated using regular expressions, adding counting to regular expressions is known to improve the succinctness of the language \cite{LM13}. 
\medskip

Another interesting feature available in Cypher is the ability to return paths. 
 
\begin{example} If we wanted to return all friends of friends of Julie in the graph from Figure \ref{fig:sn-sample}, together with a path witnessing the friendship, we can use:
 	
\begin{center}
	\footnotesize
	\begin{verbatim}
	MATCH p = (:Person {name:"Julie"}) -[:knows*]-> (x:Person)
	RETURN x, p
	\end{verbatim}
\end{center}
 	
\noindent
The variable \texttt{p} will be bound by the witnessing path and will return (in Cypher syntax): 
 	
\begin{center}
	\footnotesize
	\vbox{
	\begin{verbatim}
	+---------------------------------------------------------+
	| x       | p                                             |
	+---------------------------------------------------------+
	| Node[2] | [Node[1],:knows[1],Node[2]]                   |
	| Node[1] | [Node[1],:knows[1],Node[2],:knows[2],Node[1]] |
	+---------------------------------------------------------+
	\end{verbatim}
	}
\end{center}
	
\noindent We assume that \texttt{Node[1]} corresponds to $n_1$ (aka.\ John), \texttt{knows[1]} corresponds to $e_1$, and so forth. Each path is a sequence $n_1 e_1 n_2 e_2 n_3 \ldots n_{k-1}e_{k-1}n_k$ as discussed previously. Though not shown, in practice Neo4j will also return all attributes and values on each node and edge. No further paths are returned since they repeat an edge. \qed
\end{example}

Although Cypher does not ``directly'' support features such as NREs or RDPQs, similiar queries -- such as the one from Example \ref{ex:datapaths} -- can be supported through an auxiliary feature called \textit{path unwinding}, which permits the user to return an entire path and iterate over its nodes, all within the query itself.\footnote{This feature is discussed in an appendix that can be found appended to the online version of this paper.} Again, however, all such features are limited by the aforementioned fact that the Kleene closure of paths in Cypher can only be applied over edge labels and not path expressions.
 
 
\paragraph{\bf Gremlin} Gremlin supports navigation by the use of \texttt{repeat}, which enables arbitrary or fixed iteration of any graph traversal. As per SPARQL, Gremlin uses the arbitrary path semantics for navigational queries. However, unlike SPARQL, Gremlin returns bags and not sets of answers. Therefore, when returning nodes, Gremlin might repeat the same pair of nodes multiple (potentially infinite) times, depending on how many paths conforming to the query exist between them, and similarly for paths (which are defined in Gremlin as sequences of nodes).
 
\begin{example}
Recall how we used the following Gremlin expression in Example~\ref{ex:g-ce-ca} to obtain all co-stars of Clint Eastwood:
 	
\begin{center}
\footnotesize
	\begin{verbatim}
	G.V().hasLabel('Person').has('name','Clint Eastwood')
	  .out('acts_in').hasLabel('Movies')
	    .in('acts_in').hasLabel('Person')
	\end{verbatim}
\end{center}
 	
\noindent
For a fixed-length iteration, we can use repeat and specify the number of times the repetition should be performed. For example, the following traversal looks for actors that are linked to Clint Eastwood by a path of length 2: 
 	
\begin{center}
	\footnotesize
	\begin{verbatim}
	G.V().hasLabel('Person').has('name','Clint Eastwood').repeat(
	  out('acts_in').hasLabel('Movies')
	    .in('acts_in').hasLabel('Person')
	).times(2)
 	\end{verbatim}
\end{center}
 	
\noindent
If we want arbitrary traversal we can simply omit the \texttt{times} command; however, this effectively means \textit{iterate an unbounded number of times}, and consequently we may never get anything out of this traversal. For this reason we use the \texttt{emit()} modulator for repeat, which forces the repeat process to output the nodes after each iteration. 
 	
\begin{center}
	\footnotesize
	\begin{verbatim}
	G.V().hasLabel('Person').has('name','Clint Eastwood').repeat(
	  out('acts_in').hasLabel('Movies')
	    .in('acts_in').hasLabel('Person')
 	).emit()
 	\end{verbatim}
\end{center}
This query iterates an unbounded number of times, but at the end of each repetition, the current nodes of the traversal are output for the query. \qed
\end{example}
 
Finally, Gremlin also supports returning complete paths as results.
 
\begin{example}
To find all co-star paths connecting Clint Eastwood to other actors (and himself), we can use the following query:
\begin{center}
	\footnotesize
	\begin{verbatim}
	G.V().hasLabel('Person').has('name','Clint Eastwood').repeat(
	  out('acts_in').hasLabel('Movies')
	    .in('acts_in').hasLabel('Person')
	).emit().path()  
	\end{verbatim}
\end{center}
 	
\noindent
This query will then begin enumerating all paths per the call to \texttt{path()}.\qed
\end{example}
 
There are several other features of repeat that can modify the traversal and output. For example, the \verb+emit()+ command can include conditions, such as \verb+emit(hasLabel('Person'))+ to output only those nodes labelled \verb+'Person'+. Gremlin also includes an \verb+until()+ operator, to provide while-loop-style repetition, for example, to stop when a particular node is reached. Unlike SPARQL and Neo4j, the repeat feature of Gremlin can be combined with the bgp features illustrated in Example~\ref{ex:gremlinBGP} to express arbitrary recursive navigational expressions in the spirit of languages like Datalog. We refer to the documentation for further discussion~\cite{tinkerpop}.

%
%

\subsection{Complexity of evaluating navigational queries}\label{nav-com}
We now discuss the complexity of evaluating increasingly expressive forms of navigational queries, starting with path queries.


\paragraph{\bf Path queries}
We concentrate on the complexity of evaluating RPQs, which has received considerable 
attention in the theoretical literature. 
This is relevant since RPQs form the basis of many path query languages. 
We study the problem with respect to the possible restrictions we mentioned before, focusing on the problems of checking if a path exists, or finding pairs of nodes connected by some path under set semantics:  

\begin{itemize}

\item {\em Arbitrary paths:} 
Determining whether $v$ can be reached from $u$ by a path 
labelled in the regular expression $L$ can be solved in linear time $O(|G| \cdot |L|)$ (see, e.g., 
\cite{Woo,B13}). This bound can be achieved by using folklore algorithms based 
on {\em automata} techniques. 
Such techniques can also be reformulated to compute the set of 
all pairs of nodes that are linked by a path labelled in $L$ in time $O(|G|^2 \cdot |L|)$.  In the special case of an unconstrained path query $Q = x \xrightarrow{*} y$, we can simply perform a directed reachability analysis over $G$. 
This can be done in time $O(|G|)$ for a single pair of nodes, and in $O(|G|^2)$ to compute all pairs of linked nodes.






\item {\em Shortest paths:} 
Applying reachability techniques that return shortest 
paths (e.g., breadth-first search) in combination with the previous automata-based 
algorithms, 
we obtain shortest paths witnessing the constraints stated by RPQs. 
In particular, computing the set of all pairs of nodes that are linked by a path labelled in $L$, and for each such pair a shortest 
path in $G$ witnessing it, can be done in time $O(|G|^2 \cdot |L|)$.  

\item {\em No-repeated-node/edge paths:} Under such semantics, the complexity jumps: evaluation becomes NP-complete even in data complexity \cite{MendelzonW89}. 
Tractable instances of the RPQ evaluation problem under these semantics can be found
by either restricting regular expressions or the class of graph databases \cite{MendelzonW89,BBG13}, but it remains to be seen to what extent 
such restrictions 
are relevant in practice. 
The special case of $Q = x \xrightarrow{*} y$ can still be computed efficiently since any shortest path needs to be simple, and thus finding an unconstrained simple path amounts to finding a shortest path. 
%
\end{itemize} 

In summary, finding nodes connected by arbitrary paths or finding a shortest path satisfying an RPQ can be done in polynomial time, whereas considering simple paths, the problem becomes intractable. An open question then is if there are any practical scenarios in which the (intractable) simple path witness is really justified in terms of computational cost over finding (tractable) witnesses based on shortest paths. 

Please note that the discussion thus far assumes the use of set semantics when returning pairs of nodes or paths. When considering bag semantics in such scenarios, assuming a no-repeated-node/edge semantics, the complexity of the problem is at least that of the problem of counting paths under the chosen semantics~\cite{ACP12,LM13}; in the general case, this leads to a significant leap in complexity for reasons discussed previously.

\paragraph{\bf Navigational graph patterns}
Recall that an ngp is a bgp where the edges can also be labelled by an RPQ, or the special symbol $*$ denoting an arbitrary path. Assuming we adopt a set semantics for paths, evaluating an ngp $Q$ over a graph database $G$ can be implemented as follows:
\begin{enumerate}
\item First, each RPQ $x \xrightarrow{L} y$ that labels an edge of $Q$ is evaluated over the  
graph database $G$, and for each pair $(u,v)$ of nodes that are connected by a path labelled with $L$ we 
add to $G$ a new edge between $u$ and $v$ labelled with $L$. 


\item Second, we evaluate $P$ over the graph we augmented in the first step, but now treating 
$P$ as a bgp (that is, $L$-labelled edges in $P$ must only match to $L$-labelled edges in $G$, and not 
to a pair of nodes connected by a path whose label is in $L$). 


\end{enumerate} 

Therefore, ngp evaluation can be separated into independent phases of path query evaluation (step 1) and graph pattern evaluation (step 2). This helps understand the complexity of evaluating ngps  better. 
\begin{enumerate}
\item
\textit{First, how costly is step 1, i.e., 
building the augmented graph?} Of course this depends on the semantics for path query evaluation we use. If we use a simple path interpretation, this process will be intractable, while if we apply an arbitrary/shortest path interpretation, we can construct the graph 
in time $O(|G|^2 \cdot |Q|)$. 
 \item \textit{Second, how expensive is step 2, namely, evaluating a bgp over a graph?} We know from Section 
 \ref{gp} that this problem is NP-complete in general, but tractable for certain efficient classes of queries and tractable in data complexity. Likewise if we consider c(n)gps, as in the case of SPARQL, the same complexity arguments apply. 
\end{enumerate}  

Note that if we consider a bag semantics for paths, the first step will not succeed since a graph is a set of edges, and duplicate edges will not be preserved in the augmented graph; we would need an alternative strategy to capture such duplicate edges. In any case, the problem of constructing the augmented graph is already intractable in the case of set semantics, and will likewise be intractable in the case of bag semantics.

\paragraph{\bf More expressive queries} When analysing more expressive variants of path queries, the evaluation complexity is deeply connected with the structure of the language. For languages such as NREs or XPath we can find fast evaluation algorithms that are nothing more than extensions 
of the algorithms shown for RPQs. 

Concerning Datalog-based languages, it is well-known that answering unrestricted Datalog queries is \exptime-complete \cite{AHV}. Hence in practical settings, restrictions with lower complexity are sometimes considered. One such restricted language is Linear Datalog \cite{ConsensM90}, for which query evaluation is \pspace-complete. Other languages such as e.g. Regular Queries \cite{RRV15}, may bound the arity of predicates, which returns query evaluation to the same complexity class as ngps: \np-complete.

Finally, with respect to regular data path queries of Section \ref{ss:values}, it can be shown that the base algorithm for RPQs can be modified in order to give a polynomial time evaluation \cite{LMV16}. On the other hand, extending such queries with more expressive features seems difficult, as evaluation quickly becomes intractable \cite{LMV16,BFL13}. Furthermore, implementing these queries using the unwinding operator, as in Cypher, does also not seem to be the best solution, as the operator makes the evaluation \np-hard (see our online appendix for details).

\section{Final Remarks} \label{sec:final}

Graph databases are becoming more and more important in industry, with new graph database engines and query languages being released in recent years. With this emerging variety of systems and languages, understanding the features that each brings, and the fundamental issues that arise as a product of their design choices, is becoming of increasing importance. In this survey, we have provided an overview of the developments in this area, bridging theory and practice in order to develop a categorisation of features that constitute a common core for graph query languages.


\medskip 
\noindent
\textbf{Feature categorisation}. 
We started our review of the core aspects of graph query languages by first presenting two graph database models: the\textit{ edge-labelled graph model}, and the more elaborate \textit{property graph model}. Thereafter we identified the two main core features that are common in all modern graph query languages: \textit{pattern matching} and \textit{navigation}. We think that these two forms of querying are 
at the heart of graph query languages, and thus any reader that is familiar with these two classes of queries -- and the different options that one could  consider with respect to both -- should be qualified to understand the core of any modern graph query language.\footnote{Of course, there are also a number of additional operators which can be considered for graph querying, such as different forms of aggregation, or graph transformations; however, these either do not add anything fundamentally new to the core features we identified, or are implementation specific and not well explored in the literature. We provide a brief overview of such features in the online appendix to our paper.}

To categorise pattern matching features, we identified the class of basic graph patterns (bgps), which should arguably form the core of any graph query language, and are indeed present in all of the practical systems we reviewed. These can be further extended with operators such as projection, union, or optional, among others, giving rise to complex graph patterns (cgps). In terms of navigational queries, following both the research literature and the practical solutions currently available, we identify paths as the core of all navigational queries over graphs, and adopt the well studied notion of regular paths queries (RPQs) as the basis for navigating graphs. These can then be incorporated into bgps giving rise to navigational graph patterns (ngps), which themselves can be further extended with operators such as union, optional, etc., to create the notion of complex navigational graph patterns (cngps).

The choice of the appropriate semantics for each of these forms of queries has proven to be a non-trivial task, and there have been several proposals coming both from practice and from theory. For matching basic graph patterns we classified the main proposals for the semantics into three categories: 
\begin{enumerate}
\item {\em homomorphism-based:} matching the pattern onto a graph with no restrictions.
\item {\em isomorphism-based:} one of the following restrictions is imposed on a match:
\begin{itemize}
\item {\em no-repeated-anything:} no part of a graph is mapped to two different variables,
\item {\em no-repeated-node:} no node in the graph is mapped to two different variables,
\item {\em no-repeated-edge:} no edges in the graph is mapped to two different variables.
\end{itemize}
\item {\em simulation-based:} relaxes the notion of matching an entire query onto a graph, while at the same time preserving local connections.
\end{enumerate}
On the other hand, for path queries one can consider: (a) arbitrary paths; (b) shortest paths only; (c) paths not repeating a node (aka. simple paths); and (d) paths not repeating an edge. For the case of path queries there is also the question of how should their output look like. The options here range from: (i) checking the existence of a path (boolean output); (ii) returning start/end nodes of a path; (iii) returning complete paths; and (iv) returning entire graphs.  In the case of both graph patterns and path queries, one can chose if answers are returned as bags (where duplicate answers are returned per their multiplicity), or sets (only a single copy of each answer is returned).


\renewcommand{\arraystretch}{1.2}
\begin{table}
\tbl{Semantics adopted for pattern matching in SPARQL, Cypher and Gremlin.\\ 
	$*$: All languages support a distinct operator to enable set semantics.\\
	$\dag$: Homomorphism-based semantics can be simulated using multiple 
	MATCH commands; see Example \ref{exa-cypher-match}.\\
	$\ddagger$: Optional can be emulated imperatively.\label{fig-summary-languages-matching}}{%
\begin{tabular}{|c||c|c|}
\hline
\textbf{Language} & \textbf{supported patterns} & \textbf{semantics} \\
\hline
\hline
SPARQL & all complex graph patterns & homomorphism-based, bags$^*$ \\
\hline
Cypher & all complex graph patterns & no-repeated-edges$^\dag$, bags$^*$ \\
\hline
Gremlin & \parbox{4cm}{\centering complex graph patterns \\ without explicit optional$^\ddagger$} & homomorphism-based, bags$^*$ \\ 
\hline
\end{tabular}
}
\end{table}

\begin{table}
	\tbl{Semantics adopted for navigational queries in SPARQL, Cypher and Gremlin.\\
	*: SPARQL adds negated property sets; see Example~\ref{ex:sp_nps}.\\
	$\dagger$: In the case of SPARQL, set semantics applies only when the query can \emph{not} be rewritten as a cgp (e.g., when it uses a $*$ operator); see \cite{sparql11} for details.\\
	$\ddagger$: Cypher also allows to enable shortest-path semantics.\\
	$\S$: A distinct operator is supported to enable set semantics\\
	||: In Gremlin, other semantics can also be enabled or otherwise emulated.
	 \label{fig-summary-languages-navigation} 
	}{%
\begin{tabular}{|c||c|c|c|}
\hline
\textbf{Language} & \textbf{path expressions} & \textbf{semantics} & \textbf{choice of output} \\
\hline\hline
SPARQL & more than RPQs$^*$ & arbitrary paths, sets$^\dagger$ & boolean / nodes \\
\hline
Cypher & fragment of RPQs & no-repeated-edge$^\ddagger$, bags$^\S$ & \parbox{2.8cm}{\centering boolean / nodes / \\ paths / graphs} \\
\hline
Gremlin & more than RPQs & arbitrary paths$^{||}$, bags$^\S$ & nodes / paths \\
\hline
\end{tabular}
}
\end{table}

To exemplify our categorisation, we have reviewed some of the key design choices made for SPARQL, Cypher and Gremlin: three of the currently most popular query languages used in graph database engines. Table \ref{fig-summary-languages-matching} contains a summary of these choices for pattern matching, and Table \ref{fig-summary-languages-navigation} likewise for navigational queries. Of course, all three languages extend upon these core features presented; however, this core offers a good starting point to further formalise, study and understand these languages. 

Throughout, we have also discussed the effects of such design choices on the computational complexity considering various types of semantics and various evaluation problems. With respect to SPARQL, Cypher and Gremlin, we can summarise the following known results in terms of computational complexity of query evaluation, where \textit{PM} refers to Pattern Matching and \textit{NQ} to Navigational Queries.

\begin{itemize}
	\item In terms of complexity, by far the most studied language of the three is SPARQL. 
	\begin{description}
		\item[PM] It is known that the evaluation of bgps with projection is NP-complete and that the evaluation of cgps is PSPACE-complete~\cite{PAG09}. 
		\item[NQ] Evaluating cngps remains within the same complexity class as cgps -- PSPACE-complete -- assuming the set-based semantics of property paths used in the final official version of the SPARQL 1.1 standard~\cite{KRRV15}.
	\end{description}
	\item With respect to Cypher, less is understood. One complication, in particular, is the use of the no-repeated-edge semantics, which has not been well-studied. 
	\begin{description}
		\item[PM]  While evaluating bgps with projection in Cypher directly relates to the subgraph-isomorphism problem (which is NP-complete), there are no results stating how the no-repeated-edge semantics might affect the evaluation of cgps, so it is not clear if the problem is as hard (or perhaps even harder) than in the case of SPARQL: all that we can directly conclude
		is that evaluating cgps under this default semantics in Cypher is NP-hard. However, as per Example~\ref{ex-cyp}, a homomorphism-based semantics can be emulated by using multiple \texttt{MATCH} patterns; assuming such a ``trick'' is used, then SPARQL-like cgps can be modelled and the complexity is PSPACE-hard.
		\item[NQ]  Stating formal results is complicated by the fact that Cypher has a no-repeated-edge semantics (rather than the more well-studied no-repeated-node semantics for simple paths), a bag semantics, and that it does not support full RPQ-style expressions. Little is known of the complexity of such features, however, some lower bounds can be easily inferred. For instance, evaluating path queries is already NP-hard due to the fact that Cypher allows path unwinding (see our online appendix for more details). 
	\end{description}
	\item With respect to Gremlin, the language is Turing-complete. However, if we only consider the core fragments discussed herein, we can make the following conclusions:
	\begin{description}
		\item[PM] As per Table \ref{fig-summary-languages-matching}, we can see that the semantics of Gremlin and SPARQL are almost equivalent; even excluding the imperative features needed to emulate optional patterns in Gremlin, evaluating cgps should still be PSPACE-complete.
		\item[NQ] The study of navigational queries in Gremlin is complicated by the combination of potentially infinite arbitrary paths, the default bag semantics and the presence of features that go beyond RPQs. However, we can note that considering only RPQ-style expressions returning nodes (and not paths) with the default arbitrary-path semantics, the expressivity is equivalent to SPARQL and the complexity of evaluating cngps should thus be the same as for cgps: PSPACE-complete.
	\end{description}
\end{itemize}

\noindent This discussion shows that there are various open questions in terms of the complexity associated with the design choices made, in particular, by the Cypher query language.

 
\medskip 
\noindent
\textbf{Uses of this survey}. First, the categorisation of models, query features, semantics and results covered by this survey offer a useful guide to anyone who wishes to understand a graph query language -- be it an existing such language or one yet to be proposed -- not in terms of superficial issues like query syntax or minor variations in the graph model, but rather in terms of fundamental querying abilities, choice of semantics, and expressivity. Once the core of a language can be understood in this more abstract way, different languages can then be compared and contrasted in a similar, foundational manner. We have provided such a comparison for the languages SPARQL, Cypher and Gremlin. Indeed, even though these languages run over different models and have completely different syntax, etc., by looking at Table~\ref{fig-summary-languages-matching} and Table~\ref{fig-summary-languages-navigation}, one realises that the core of these languages is fundamentally rather similar.
In the same fashion, using this survey as a guide, we could now compare a new proposal for a graph query language by abstracting the pattern matching and navigational capabilities of the language, and asking relevant questions such as: ``\textit{what is the semantics of pattern matching in this language?}''; or ``\textit{what type of navigational features does it include?}''.

This growing diversity of graph database technologies moreover suggests that the time may come for further standardisation of graph query languages. While SPARQL has been formally standardised for RDF databases and has been well studied in the literature, many implementers have opted for custom graph database solutions and engines with custom languages, such as Neo4j with Cypher. Likewise, ad hoc standards like Gremlin have emerged in recent years and have been implemented by multiple vendors. However, unlike SPARQL, the semantics and complexity of languages like Cypher and Gremlin have not been studied. Looking to the future, one can thus expect standardisation efforts to rigorously define and characterise the properties of a general graph query language that takes into consideration the demands of the industry, much like the story for SQL, where core features are abstracted as the relational algebra. Of course, a query language may not always abide by a clean abstraction, as per the case of SQL which goes beyond the relational algebra in various ways, or the languages in Table~\ref{fig-summary-languages-matching} and Table~\ref{fig-summary-languages-navigation} that are annotated with exceptions and support a variety of other features not covered. But yet, the exercise of abstracting languages into core features is a necessary task if one wants to create standards in terms of understanding which features are simply syntactic (i.e., redundant in terms of expressivity), what choice of semantics and features could be considered, and what effects such choices have with respect to achieving desirable computational guarantees in terms of evaluating queries in that language. A notable such example of this were the studies by \citeN{ACP12} and \citeN{LM13} on the complexity of property paths initially proposed in a SPARQL 1.1 draft, which lead to the semantics being changed in the final version. Likewise, we believe that this survey can serve as a useful guide for current and future standardisation processes involving graph query languages.

We also expect that our survey could serve to bridge the theory/practice gap, helping to port theoretical results about abstract languages (such as graph patterns or path queries) into real graph database engines, and also the other way around, helping to state the problems facing current graph engines in a formal manner. Indeed, we have seen multiple times in this survey that a seemingly innocuous change can have a drastic effect on computational complexity upon further examination; for example, we saw how an optional operator in cgps leads to a jump in complexity for query evaluation, or how having a no-repeated-node semantics can render path queries intractable, or (in the aforementioned case of SPARQL 1.1) how the combination of bag semantics and path queries can quickly become problematic. On the other hand, for example, we have also seen that bpgs with projection can be extended with a variety of useful features without a complexity jump in terms of query evaluation, including support for ngps with path expressions. But while the existing theory provides important insights, this survey also reveals gaps in the literature. To name a few examples: the problem of finding tractable subclasses of graph patterns that can be evaluated efficiently over no-repeated-edge semantics is almost unexplored; systems capable of returning paths need a way of representing a set of paths whenever it is infinite; and we have already noted the importance of a more rigorous theoretical formalisation of Cypher and Gremlin in order to determine the exact complexity of evaluating their queries and to understand their expressive power.  In this respect, we believe that our survey can bring research questions from the practical world of graph database engines into theory. 

\medskip 
\noindent
\textbf{Future directions}. 
Our categorisation also opens interesting possibilities for further work in terms of surveying and classifying other important aspects of graph databases 
that are infeasible to cover in this survey with the necessary depth. 
\medskip

The first issue has to do with the implementation and optimisation of modern query languages. In surveying fundamental features, we have only dealt with such issues indirectly. A number of implementations of modern graph query languages using the features in this survey have emerged: 
in terms of some of the most prominent SPARQL engines that have been released, we can name 4store~\cite{4store},  BlazeGraph (formerly BigData~\cite{bigdata}), GraphDB (formally (Big)OWLIM~\cite{owlimswj}), Jena~\cite{jena},  and Virtuoso~\cite{virtuoso}; with respect to property graphs, Neo4j~\cite{cypher} is one of the most popular engines, but one can also cite Titan \cite{titan} and OrientDB \cite{orientdb}. Collectively these engines implement a diverse range of indexing strategies, query planning methods, optimisations and ad hoc heuristics -- with more proposed in the literature -- sometimes borrowing directly from relational databases (e.g., \cite{jena}), others being custom-designed for graphs (e.g., \cite{owlimswj,orientdb}), and others still that intersect the relational and graph worlds (e.g., \cite{virtuoso,ParadiesLB15}). The analysis and classifications of all these implementation strategies is an important task that can benefit tremendously from our framework and 
would make for an interesting complementary survey in the future.
\medskip

The second important line of work has to do with identifying the core of graph analytics and other operations more related to machine learning, or computing statistics over graphs.  
Currently these operations are not commonly compiled into query languages, but instead graph engines normally provide several data-access primitives such that users can implement their own algorithms within a programming environment: a direction in which Gremlin goes, for example. Furthermore, there has recently been a lot of work on domain-specific languages that can take care of particular sets of operations within a certain domain or scenario (see e.g. \cite{greenmarl}). However, in the area of graph analytics, with different tasks ranging from computing weighted shortest paths to computing the PageRank matrix of an entire graph, we see the same diversity problem as with graph query languages: how to abstract the core (possibly declarative) features of such operations? 

More pertinently for this survey, it is not clear where graph query languages start and graph analytics languages end: what is the overlap of features required, how do they complement and/or extend each other, etc. The graph database community has been slow in adopting graph analytics as a problem of study, but as the importance of these operations grow, we expect this to change in the next few years. We believe that the first goal of the community should be identifying a common core of the most widely used operations, just as we have done with graph query languages.
It would also be interesting to understand how classical database-querying tasks compare to engines supporting the so-called vertex-centric programming model, such as 
Apache Giraph \cite{giraph}, GraphX \cite{graphx} or Pregel \cite{pregel}. 
This again goes in the direction of Gremlin, which as we have discussed has many elements that are similar to a declarative query language, but also encapsulates a more imperative style, being supported, for example by, the aforementioned Apache Giraph analytics framework. We have also recently seen the first efforts in designing a more declarative language in this context \cite{graphiql}, and in the following years we expect  more research in this direction. 

\medskip 
\noindent
\textbf{Conclusion}. Recent years have seen the re-emergence of graph databases as an important alternative to their more widely-established relational cousin, bringing with them a variety of new challenges, new demands, and new questions. In this survey, we have provided an overview of the fundamental query features that underlie such databases and have provided a categorisation that generalises much of these recent developments and offers a bridge to known theoretical results while raising some new questions. Of course, there are many open challenges facing graph databases in terms of standardising query languages, implementing and optimising engines for query evaluation, studying the theoretical properties of related problems, as well as evolving graph databases to meet emerging demands for graph analytics. We hope that this survey may serve as a useful guide for those involved in such efforts.

\bibliographystyle{ACM-Reference-Format-Journals}
\bibliography{survey}





\newpage
\elecappendix

\section{Additional features}
\label{sec-bgp-addfeat}
Throughout the survey our main focus is on querying features designed to retrieve nodes, edges, or paths from a graph. However, most practical query languages also include ways to manipulate these results, in particular, aggregating them or transforming them into different structures. While the types of operators offered for the manipulation of results vary significantly amongst different graph query languages, there are some common features in these languages that we explore in this section. In particular we look at aggregation functions, path manipulation and graph-to-graph querying functionalities, and discuss some challenges when implementing these operations over graphs. We wrap up with a brief overview of other extensions proposed for graph query languages in the literature, including domain-specific features.

\subsection{Aggregation and solution modifiers}
\label{sec-agg}
In the development of relational databases, the possibility of grouping values and computing statistics over these groups has been recognised as an important feature. In the case of SQL, the \texttt{GROUP BY} operator allows for grouping values according to some criteria, the \texttt{COUNT} operator allows for counting the number of elements in each such group, and the \texttt{MIN}, \texttt{MAX}, \texttt{SUM} and \texttt{AVG}  operators were included to compute the minimum, maximum, sum and average of the elements in each group, respectively (provided that the group contains values compatible with the operators). These functionalities play such an important role in data analysis that they have been adopted by graph query languages. In what follows, we provide some examples of these features for the practical graph query languages considered in this survey, which will give the reader a clearer idea of how they are used.

\begin{example}\label{exa-aggr}
As a first example, assume we have an edge-labelled graph $G$ storing information about movies and actors, such as the one shown in Figure~\ref{fig:rdf} on page \pageref{fig:rdf}. In order to count the total number of movies in $G$, we can use the following SPARQL query:
\begin{center}
\small
\begin{verbatim}
	SELECT COUNT(?movie) AS ?total
	WHERE { ?movie :type :Movie . }
\end{verbatim}
\end{center}
As explained in Section \ref{gp}, the triple \verb+?movie :type :Movie+ in this query is used to bind the variable \verb+?movie+ to the movies occurring in $G$. The operator \verb+COUNT(?movie)+ is then used to count the number of values for the variable \verb+?movie+, which is stored in the variable 
\verb+?total+ as indicated by the command \verb+COUNT(?movie) AS ?total+. If 
the variable \verb+?movie+ contains repeated values (which could happen in more complicated queries), then by default, duplicates will be counted; to ensure that only distinct values are counted, the command 
\verb+COUNT(?movie)+ can be replaced by \verb+COUNT(DISTINCT ?movie)+. \qed
\end{example}

\begin{example}\label{exa-count-distinct}
As a second example, assume that for each movie we wish to count the number of people acting in it. This query can be formulated as follows in SPARQL:
\begin{center}
\small
\begin{verbatim}
	SELECT ?movie COUNT(DISTINCT ?actor) AS ?number_actors
	WHERE {
		?movie :type :Movie .  
		?actor :acts_in ?movie . ?actor :type :Person .
	}
	GROUP BY ?movie
\end{verbatim}
\end{center}
The three triples inside the \verb+WHERE+ clause are used to indicate that for each pair $b$, $c$ of values assigned to \verb+?movie+ and \verb+?actor+, respectively, $b$ must be a movie and $c$ must be a person who acted in $b$. Then the operator \verb+GROUP BY ?movie+ is used to indicate that a group must be created for each value $b$ in the variable \verb+?movie+, which must contain all values $c$ for the variable \verb+?actor+ such that $b$, $c$ is a valid assignment for \verb+?movie+ and 
\verb+?actor+ according to the triples in the \verb+WHERE+ clause. Finally, for each value $b$ in \verb+?movie+, the operator 
\verb+COUNT(DISTINCT ?actor)+ counts the number of distinct values in the group associated to $b$, which is stored in the variable \verb+?number_actors+ as indicated by the command 
\verb+COUNT(DISTINCT ?actor) AS ?number_actors+. \qed
\end{example}

\begin{example}\label{exa-cypher-max}
Assume now that each movie includes a property that defines its runtime. 
With such information we would like to obtain the longest films in the database. 
This query can be expressed as follows in Cypher:
\begin{center}
\small
\begin{verbatim}
	MATCH (m:Movie) WITH MAX(m.runtime) AS maxTime 
	MATCH (m:Movie) WHERE m.runtime = maxTime 
	RETURN m
\end{verbatim}
\end{center} 

\noindent
The first \texttt{MATCH} clause looks for nodes labelled \texttt{Movie} and stores them in variable \texttt{m}. The list of movies saved in \texttt{m} is explored by the \texttt{WITH} operator to compute the maximum runtime. From this first match clause, only the result of the aggregation (\texttt{maxTime}) can be projected. The second \texttt{MATCH} clause is thus needed to return the movies whose \texttt{runtime} is equal to the \texttt{maxTime} returned by the first \texttt{MATCH}. The filtered list of movies -- movies with the longest runtime -- is returned as the final result of the query. In this case, we say that the pattern initiated by the first \texttt{MATCH} clause is a \textit{sub-query}.\footnote{It may seem counter-intuitive to have the sub-query ``outside'' in Cypher, as in SPARQL the first \texttt{MATCH} corresponds to a sub-query and would rather be written inside; as such, this is an idiosyncrasy of Cypher.}\qed 
 \end{example}
 
All of the above examples can similarly be expressed in Gremlin. 
\medskip

Finally we briefly note that many practical query languages allow for applying \textit{solution modifiers} over results, such as to express a limit for a number of results, or an ordering to apply over results, or an offset that specifies an number of initial results to skip. These solution modifiers can also be embedded within \textit{sub-queries} that project the modified solutions to an outer query.

\begin{example}\label{exa-cypher-max1}
We can achieve a similar result to Example~\ref{exa-cypher-max} by instead using a solution modifier that orders by runtime and selects the first result:
\begin{center}
	\small
	\begin{verbatim}
	MATCH (m:Movie) RETURN m ORDER BY m.runtime LIMIT 1
	\end{verbatim}
\end{center} 
	
\noindent
In this case, we require only one \texttt{MATCH} clause. However, this is not precisely equivalent to Example~\ref{exa-cypher-max}: if we have multiple movies tied for the longest runtime, here we will only return one such movie, while previously we would return all such tied movies. To make the query equivalent, we would instead need a sub-query as follows: 

\begin{center}
	\small
	\begin{verbatim}
	MATCH (m:Movie) WITH m.runtime as maxTime ORDER BY maxTime LIMIT 1 
	MATCH (m:Movie) WHERE m.runtime = maxTime 
	RETURN n
	\end{verbatim}
\end{center} 

\noindent As before, we use a sub-query to match any movie with the longest runtime and then find other movies with the same runtime. Note that unlike Example~\ref{exa-cypher-max} and the \texttt{MAX} aggregate, we could replace \texttt{LIMIT 1} with \texttt{SKIP 2 LIMIT 1} to find movies with the third longest runtime (where we could also replace \texttt{WITH} as \texttt{WITH DISTINCT} to filter ties). 
\qed   
\end{example}

Such solution modifiers are also found in SPARQL and Gremlin. 
\medskip

As one can see, when coupled with basic graph patterns, aggregate operations and solution modifiers have a similar behaviour as in relational databases. On the other hand, when we consider navigational queries, such operations impose some unique challenges not present when dealing with relational data. For example, counting paths or taking the length of individual paths both impose computational challenges when applied in this new context, as were raised in Section~\ref{ss-paths} when discussing the related problem of returning nodes from a path under bags semantics: if the graph database $G$ is cyclic, the number of paths can be infinite and paths may have infinite length; on the other hand, while restrictions such as no-repeated-nodes make the set of paths finite, counting paths is still associated with a high computational complexity~\cite{V79,ACP12,LM13}. Still, languages such as Cypher provide aggregation features that allow for counting such paths or taking their length.

\begin{example}\label{exa-cypher-longest}
Assume a graph database encoding a road network, where the connectivity between five cities ($c_1$, $c_2$, $c_3$, $c_4$ and $c_5$) is given by five (bidirectional) routes ($c_1 \leftrightarrow c_2$, $c_1 \leftrightarrow c_3$, $c_2 \leftrightarrow c_4$, $c_4 \leftrightarrow c_5$ and $c_3 \leftrightarrow c_5$). The longest route between cities $c_1$ and $c_5$ can be expressed in Cypher by the following query:  
\begin{center}
	\small
	\begin{verbatim}
	MATCH p = (a:City {name:"c1"})-[*]->(b:City {name:"c5"})
	WITH MAX(length(p)) AS maxLength
	MATCH p = (a:City {name:"c1"})-[*]->(b:City {name:"c5"})
	WHERE length(p) = maxLength
	RETURN p
	\end{verbatim}
\end{center} 
	
	\noindent
In this example, the \texttt{MATCH} clause is used twice to store all the paths between cities $c_1$ and $c_5$ in variable \texttt{p} (since the sub-query can only return the result of the aggregate; see Example~\ref{exa-cypher-max}).
The \texttt{WITH} clause combines the operators \texttt{MAX} and \texttt{length} to obtain \texttt{maxLength}, i.e., the length of the longest path. The \texttt{WHERE} clause selects paths whose length is equal to \texttt{maxLength}. The final result is the list of the longest paths such that each path is encoded as a collection of nodes and edges, which in this case would be:
\begin{center}
	\small
	\begin{verbatim}
	[{name: c1}, {}, {name: c2}, {}, {name: c4}, {}, {name: c5}] 
	\end{verbatim}
\end{center} 

\noindent	
This is the longest path from city $c_1$ to city $c_5$ without a repeated edge. Note that without the restriction on repeating edges, we could have infinite length paths (for example, subsequently going back and forth between $c_3$ and $c_5$ ad infinitum).\qed
\end{example}

In Cypher, we can also, for example, count all paths.

\begin{example}\label{exa-cypher-count}
Consider a query that counts the number of paths from a source node to a target node in a graph. This query is expressed in Cypher as follows: 
\begin{center}
\small
\begin{verbatim}
	MATCH p = (:A)-[*]->(:B)
	RETURN COUNT(p)
\end{verbatim}
\end{center} 

\noindent
The \texttt{MATCH} clause in this query stores the paths from a node with label \texttt{A} to a node with label \texttt{B} in the variable  \texttt{p}, and the \texttt{COUNT} clause counts the number of paths stored in \texttt{p}; again the no-repeated-edges restriction avoids infinity in the case of cycles. \qed
\end{example}

While in Cypher, the restriction of not repeating edges is offered by default, in Gremlin, a call to \texttt{simplePath()} is required to ensure that nodes are not repeated.
  
\begin{example}
The following Gremlin query computes all paths between nodes with labels \texttt{A} and \texttt{B} such that no path visits the same node twice:  
\begin{center}
\small
\begin{verbatim}
	G.V().hasLabel('A').repeat(out().simplePath()).until(hasLabel('B')).emit().path()
\end{verbatim}
\end{center} 

\noindent
The \texttt{simplePath()} function filters paths that repeat nodes. Interestingly, Gremlin returns paths ordered by ascending length, and thus by keeping only the first answer we can use this query to obtain the shortest path. \qed



\end{example}  


As opposed to the case of relational aggregates, many questions about aggregation functions on paths remain open. In particular, understanding the expressive power of these functions and pinpointing the exact complexity of evaluating them are important open issues that deserve further investigation.

\subsection{Path unwinding}
\label{sec-path-und}

Path unwinding refers to the idea of projecting parts of a path. As previously discussed, SPARQL queries cannot return paths: they can either check for the existence of paths satisfying some conditions, or return the set of start- and/or end-nodes of such paths; thus, SPARQL does not provide any path-ungrouping operator. On the other hand, Cypher provides functions to get path elements independently. 

\begin{example}\label{exa-cypher-ungroup}
Recall the road network described in Example \ref{exa-cypher-longest}.
When travelling between cities $c_1$ and a city $c_5$, we may wish to find two different disjoint routes (visiting disjoint intermediate cities) allowing us to see new scenery on each part of our journey. A query finding two such paths can be expressed in Cypher as follows:
\begin{center}
\small
\begin{verbatim}
	MATCH p1 = (a:City {name:"c1"}) -[*]- (b:City {name:"c5"})
	MATCH p2 = (a:City {name:"c1"}) -[*]- (b:City {name:"c5"})
	WHERE none(x IN nodes(p2) WHERE (x IN nodes(p1) AND x<>a AND x<>b))
	RETURN p1, p2	
\end{verbatim}
\end{center}

\noindent
The variables \verb+a+ and \verb+b+ store the nodes representing the cities with names \texttt{c1} and \texttt{c5}, respectively. The variables \verb+p1+ and \verb+p2+ then store paths between these two cities; these paths are undirected and of arbitrary length as indicated by the expression~\verb+-[*]-+. The expression \texttt{nodes(path)} returns the nodes in the path as a collection, while \texttt{relationships(path)} returns the edges in the path as a collection. The path disjointness condition is defined by using the \texttt{none} operator (which evaluates to \texttt{true} if the condition is false for all elements of a collection). In more detail, the \texttt{WHERE} clause specifies that for two paths \texttt{p1} and \texttt{p2} to be returned, there can be no node \texttt{x} such that:
 (i) \verb+x+ is a node in the path \verb+p2+ as indicated by the condition \verb+x IN nodes(p2)+;
 (ii) \verb+x+ is a node in the path \verb+p1+ as indicated by the condition \verb+x IN nodes(p1)+; and 
 (iii) \verb+x+ is different from \verb+a+ and \verb+b+ as indicated by the condition \verb+x<>a AND x<>b+. 
 In other words, \verb+p1+ and \verb+p2+ are returned only if they do not share any nodes aside from \verb+a+ and \verb+b+.
\qed
\end{example}

Although useful, queries such as the one above are inherently difficult to evaluate. In fact, given a graph $G$ and nodes $b$ and $c$ in $G$, the problem of verifying whether there exist two paths in $G$ between $b$ and $c$ with no nodes in common except for $b$ and $c$ is known to be NP-complete (this problem is referred as the two-disjoint-paths problem in the literature~\cite{GJ}). Hence we see that adding path unwinding to a query language can lead to issues with computational complexity when combined with other features of the language: various well-known hard problems can be trivially expressed using such combinations of features.
\medskip

Gremlin has similar features for path unwinding, where nodes and edges can be extracted from paths and processed with subsequent operators.

\subsection{Graph-to-Graph queries}
Both the input and output of an SQL query are relational tables, so this language is compositional in the sense that the output to a query can be used as the input of another query. Along similar lines, graph query languages provide functionalities that allow to return a graph as the result of a query. 

In the case of SPARQL, the \texttt{SELECT} operator can be replaced by the \texttt{CONSTRUCT} operator in order to produce an RDF graph as the output of a query. More specifically, a SPARQL query of the form \verb+CONSTRUCT {+ $\texttt{t}_\texttt{1}$ $\texttt{t}_\texttt{2}$ \verb+...+ $\texttt{t}_\texttt{n}$ 
\verb+} WHERE { ... }+ produces an RDF graph as output, where each $\texttt{t}_\texttt{i}$ is an RDF triple that can contain variables and constants, and where the \texttt{WHERE} clause is defined as usual. To produce the answer to such a query, first the \texttt{WHERE} clause is evaluated to produce all possible matches. Next, each match is applied to replace the variables occurring in $\texttt{t}_\texttt{1}$, $\texttt{t}_\texttt{2}$, $\ldots$, $\texttt{t}_\texttt{n}$ by constants. A match may not have a value for a variable occurring in a specific triple $\texttt{t}_\texttt{i}$ because of the use of the operators \texttt{OPTIONAL} and \texttt{UNION}; in this case, an output RDF triple is not produced from $\texttt{t}_\texttt{i}$ for that match. Finally, RDF graphs are defined as unordered sets, meaning that duplicates and ordering are not preserved in the output graph.

\begin{example}\label{ex:construct}
Take again the RDF graph in Figure~\ref{fig:rdf} on page \pageref{fig:rdf}, which we denote by $G$. To create an RDF graph $G'$ storing information about people that act together in some movie, we can use the following query:

\begin{center} 
\small
\begin{verbatim}
	CONSTRUCT { ?actor1 :act_together ?actor2 . } 
	WHERE {
	   ?movie :type :Movie .
	   ?actor1 :acts_in ?movie . ?actor2 :acts_in ?movie .
	   FILTER (?actor1 != ?actor2)
	}
\end{verbatim}
\end{center}

\noindent
For each assignment $b$, $c_1$, $c_2$ generated by evaluating the \texttt{WHERE} clause for the variables \verb+?movie+, \verb+?actor1+, \verb+?actor2+, respectively, we have that $c_1$ and $c_2$ act together in the movie $b$, and also that $c_1$ and $c_2$ are distinct actors as indicated by the command \verb+FILTER (?actor1 != ?actor2)+. This assignment replaces \verb+?actor1+ by $c_1$ and \verb+?actor2+ by $c_2$ in the \texttt{CONSTRUCT} clause to produce the 
triple $c_1$ \verb+:act_together+ $c_2$. In the case of Figure~\ref{fig:rdf}, we would thus create a new RDF graph with two edges labelled \texttt{:act\_together} connecting \texttt{:Clint\_Eastwood} to \texttt{:Anna\_Levine}, and vice versa.\qed
\end{example}

In the case of Cypher, 
one can include a \texttt{CREATE} clause inside a query expression to create graph elements (nodes and edges) from the pattern matching step.

\begin{example}
Consider the property graph with movie data from Figure \ref{fig-movies} on page~\pageref{fig-movies}. Similarly to Example \ref{ex:construct}, if we want to construct a graph containing only the pairs of actors that co-starred in a movie, we can use the following Cypher query:
\begin{center}
\small
\begin{verbatim}
	MATCH (a:Person)-[:acts_in]->(:Movie)<-[:acts_in]-(c:Person)
	WHERE a <> c
	CREATE (a)-[r:act_together]->(c)
	RETURN r
\end{verbatim}
\end{center}


\noindent
The \texttt{CREATE} clause will then ``materialise'' the graph containing all pairs of actors that co-starred in the same movie. The \texttt{RETURN} clause then specifies that all of the edges of this graph should be returned. We also add a \texttt{WHERE} clause to distinguish \texttt{a} from \texttt{c}: although Cypher adopts a no-repeated-edge semantics, there may be multiple edges from an actor to a movie, for example, if the actor plays multiple roles in that movie, in which case we would generate vacuous loops on such actors in the output.\qed
\end{example}

A similar mechanism for graph creation is provided by Gremlin.

\begin{example}
Consider now a transportation network that connects two cities if there is a direct bus link between them. Suppose we want to travel, but are only willing to change the bus once. To see our options, we could add an edge labelled $\texttt{twoHoplink}$ between any two cities reachable from each other by at most one change of bus. This can be done using the following Gremlin query:

\begin{center}
\small
\begin{verbatim}
	G.V().as("a").out().out().as("b").addE("twoHoplink").from("a").to("b")
\end{verbatim}
\end{center}

\noindent
In the above expression: 
\texttt{G.V().as("a")} obtains the list of nodes in the graph and store this list in variable \texttt{a};
\texttt{.out().out().as("b")} obtains the nodes \texttt{b} reachable from each node in \texttt{a}, considering a single intermediate node on this path;
\texttt{addE("twoHoplink").from("a").to("b")} creates edges labelled \texttt{twoHoplink} between each pair of connected nodes stored in \texttt{a} and \texttt{b}. 
\qed
\end{example}
 
The graph-to-graph queries illustrated in the examples above are a rather new feature for graph query languages; currently there are few studies about their basic properties and the effects of combining them with other query features. Some work involving the expressive power and the composition of queries using \texttt{CONSTRUCT} in SPARQL has been carried out~\cite{KRU15,PolleresRK16,AU16}. However, the use of these types of queries in Cypher or Gremlin is currently unexplored in the literature, and may be an interesting topic for future research. 

\subsection{Further extensions}

A number of extended features have been proposed and/or included in the SPARQL, Cypher and Gremlin languages. Though the focus of this survey is on the core features of graph matching and navigational queries, we give a brief overview of some of the more prominent extensions, both as included in the respective specifications of the query languages themselves, and as proposed by third parties in the literature.
\medskip

With respect to official extensions, SPARQL 1.1. Update~\cite{sparql11update} defines a specification for making updates to the underlying dataset that the SPARQL engine queries, allowing to add, remove or modify graphs or triples with graphs in a declarative manner; likewise Cypher offers primitives to update nodes, edges and the labels and attributes associated with them~\cite{cypher}, while Gremlin supports updates through use of the Blueprints API that forms part of the TinkerPop framework~\cite{gremlin}. In order to standardise a mechanism for processing queries over data spanning multiple sources, SPARQL 1.1 Federated Query~\cite{sparql11fed} specifies how SPARQL queries can contain nested queries that are sent to and executed by remote SPARQL services, with the results returned to the outer query for further local processing. With respect to reasoning, SPARQL 1.1 Entailment~\cite{sparql11entailment} provides details on how various types of ontological and rule-based entailment regimes can be applied to generate further answers from implicit knowledge during the graph matching process.
\medskip

Aside from official extensions, a wide variety of extensions have been proposed by third parties in the research literature, particularly for the SPARQL language. Various such extensions are concerned with supporting additional meta-information for RDF data: two such proposals are  SPARQL*~\cite{HartigT14} and AnQL~\cite{ZimmermannLPS12}, which both describe general frameworks for \textit{reifying} or \textit{annotating} RDF data (respectively), providing analogous query features in SPARQL. Other general extensions of interest include SPARQL\textsuperscript{AR}~\cite{FrosiniCPW17}, which allows for performing query approximation and relaxation to also return ``near answers''; SPARQLog~\cite{BryFMLLP09}, which extends SPARQL with rules and more flexible forms of quantification, additionally defining fragments that maintain desirable complexity results; XSPARQL~\cite{BishofDKLP12}, which allows for federating queries over SPARQL, XML (through XQuery) and relational databases (through SQL) in a unified manner; as well as work by \citeN{LausenMS08} on using SPARQL (and a proposed extension thereof) to specify relational-like constraints over RDF graphs.

Other proposed extensions of SPARQL target specific domains or types of applications, including tSPARQL~\cite{Hartig09}, which allows for specifying and processing trust annotations in terms of which results can be trusted and why; SciSPARQL~\cite{AndrejevR12}, which provides primitives to deal with numeric arrays of (scientific) information; SPARQL-MM~\cite{KurzSK15}, which proposes user-defined functions to help when querying meta-data about multimedia artefacts; GeoSPARQL~\cite{geosparql,BattleK12}, stSPARQL~\cite{KoubarakisK10} and SPARQL-ST~\cite{PerryJS11}, which propose extensions to support spatial and temporal queries; 
EP-SPARQL~\cite{AnicicFRS11}, C-SPARQL~\cite{BarbieriBCVG10} and Streaming SPARQL~\cite{BollesGJ08}, which deal with processing dynamic information and support, offering event processing, reasoning and querying over windows of streaming data, and so forth.
\medskip


The above discussion suggests that research in the areas of graph querying and analytics is ongoing, with various extended features being continuously proposed. Graph query languages are thus sure to evolve to capture more and more features. Rather than trying to cover all such features in detail, in this survey, we focus on capturing a core set of features that are foundational for querying graphs in a declarative manner and that thus form the common backbone of modern graph query languages.  

%
%
%
%
%
%
%
%


\end{document}